\documentclass[10 pt]{article}

\usepackage[utf8]{inputenc}
\usepackage{dsfont}
\usepackage{amsmath}
\usepackage{amsthm}
\usepackage{amsfonts}
\usepackage{amssymb}
\usepackage{tensor}
\usepackage{mathtools}
\usepackage[T1]{fontenc}
\usepackage{graphicx}
\usepackage{float}
\usepackage{bm}
\usepackage{setspace}
\usepackage{enumitem}
\usepackage{mdwlist}
\usepackage{parskip}
\usepackage{listings}
\usepackage{color}
\usepackage{tikz,datatool}
\usepackage{hyperref}
\usepackage{mathabx}
\usepackage{multicol}
\usepackage{caption}
\usepackage[makeroom]{cancel}
\usepackage{tensor}
\usepackage{relsize}
\usepackage[toc,page]{appendix}
\usepackage{epic}
\usepackage[nottoc,numbib]{tocbibind}
\usepackage{young}
\usepackage{ytableau}
\usepackage{tikz}
\usepackage[bottom]{footmisc}

\usepackage[%margin=3cm,marginparwidth=0pt,
top=3cm,bottom=3cm,width=16.2cm]%[top=3cm,bottom=3cm,left=3.5cm,right=2.5cm]
{geometry}

\usetikzlibrary{decorations.pathreplacing}
\usetikzlibrary{arrows.meta}

\usetikzlibrary{quotes}
\usepackage{siunitx}
\tikzset{every edge quotes/.style =
          { fill = white,
            sloped,
            execute at begin node = $,
            execute at end node   = $  }}

\usetikzlibrary{positioning}

\tikzset{
  pics/diagram/.style 2 args={% #1=i, #2=partition as comma separated list
    code = {
      \def\diagramlabel{}% we build the automatic label i-mu 
      \begin{scope}[scale=0.5, yshift=-5mm]
         \foreach \row [count=\r] in {#2} {
             \ifnum\r=1
               \pgfmathparse{0.5*(\row+1)}
               \xdef\lastcol{\pgfmathresult}
             \fi
             \pgfmathparse{0.5*(\r+1)}
             \xdef\lastrow{\pgfmathresult}
             \xdef\diagramlabel{\diagramlabel\row}
             \foreach \col in {1,...,\row} {
                 \draw[thick](\col,-\r)rectangle++(-1,1);
             }
         }
         \node[rectangle, anchor=north west,
              minimum height=\lastrow cm, minimum width=\lastcol cm]
              (#1-\diagramlabel) at (-0.5,0.5){};
      \end{scope}
    }
  }
}

\newcommand{\hso}{\hspace{.1cm}}

\newcommand{\beq}{\begin{eqnarray}}
\newcommand{\eeq}{\end{eqnarray}}
\newcommand{\beqs}{\begin{equation*}}
\newcommand{\eeqs}{\end{equation*}}

\newcommand{\avg}[1]{\left\langle #1 \right\rangle}
\newcommand{\ket}[1]{\left| #1 \right\rangle}

\newcommand{\bket}[2]{\left\langle #1 \middle| #2 \right\rangle}
\newcommand{\obket}[3]{\left\langle #1 \middle| #2 \middle| #3\right\rangle}
\newcommand{\ldefeq}{\mathrel{\rlap{%
			\raisebox{0.3ex}{$\m@th\cdot$}}%
		\raisebox{-0.3ex}{$\m@th\cdot$}}%
	=}
\makeatletter

\newcommand*{\rdefeq}{\mathrel{\rlap{%
			= 
			\raisebox{0.3ex}{$\m@th\cdot$}}%
		\raisebox{-0.3ex}{$\m@th\cdot$}}%
}
\makeatother
\newcommand{\hg}{\text{ht}}

\newcommand{\tra}{\text{tr}}

\DeclarePairedDelimiter\abs{\lvert}{\rvert}%

\newdimen{\cellsize}

\def\boxformat{}
\newsavebox{\cellcontent}
\def\hidehrule#1#2{\kern-#1%#1=outside thickness, #2=inside thickness
  \hrule height#1 depth#2 \kern-#2 }%
\def\hidevrule#1#2{\kern-#1{\dimen\cellcontent=#1%
    \advance\dimen\cellcontent by#2\vrule width\dimen\cellcontent}\kern-#2 }%
\def\makeblankbox#1#2{\hbox{\lower\dp\cellcontent\vbox{\hidehrule{#1}{#2}%
    \kern-#1 % overlap the rules at the corners
    \hbox to \wd\cellcontent{\hidevrule{#1}{#2}%
      \raise\ht\cellcontent\vbox to #1{}% set the vrule height
      \lower\dp\cellcontent\vtop to #1{}% set the vrule depth
      \hfil\hidevrule{#2}{#1}}%
    \kern-#1\hidehrule{#2}{#1}}}}
\newcommand\cellify[1]{\defaultcell%
\sbox{\cellcontent}{\vbox to \cellsize{%
\vfill%
\hbox to \cellsize{\hfill$\boxformat #1$\hfill}% content of box possibly modifying drawnbox
\vfill}}%
\rlap{\drawnbox}% draw box first
\usebox{\cellcontent}}% then content
\newcommand\tableau[1]{\vtop{\let\\\cr
\baselineskip -16000pt \lineskiplimit 16000pt \lineskip 0pt
\ialign{&\cellify{##}\cr#1\crcr}}}
\newcommand\defaultcell{\gdef\drawnbox{%default def of a box
\makeblankbox{0.2pt}{0.2pt}% 0.4pt is normal thickness of a line
}}
\newcommand\emptycell{\gdef\drawnbox{}}
\newcommand\vdotscell{\gdef\drawnbox{\kern-1.6pt\vbox{\baselineskip=4pt\lineskiplimit=0pt\hbox{}\hbox{.}\hbox{.}\hbox{.}\hbox{}}}}
\newcommand\hdotscell{\gdef\drawnbox{\vbox to \cellsize{\hbox{\kern1pt$\ldotp\ldotp\ldotp$}\vfill}}}
\newcommand\vhdotscell{\gdef\drawnbox{\rlap{\kern-1.6pt\vbox{\baselineskip=4pt\lineskiplimit=0pt\hbox{}\hbox{.}\hbox{.}\hbox{.}\hbox{}}}\vbox to \cellsize{\hbox{\kern1pt$\ldotp\ldotp\ldotp$}\vfill}}}
\newcommand\vcenterbox[1]{\vcenter{\hbox{#1}}}
\newsavebox\vcbox
\newsavebox\vclab
\newdimen\labshift
\newcommand\vcenterboxlabel[2]{\sbox{\vcbox}{#1}%
\sbox{\vclab}{#2}%
\labshift=\ht\vclab\advance\labshift by\dp\vclab%
\advance\labshift by\ht\vcbox\advance\labshift by\dp\vcbox%
\advance\labshift by 5pt%
\raise-0.5\labshift\hbox to 0pt{\hbox to \wd\vcbox{\hfill\usebox{\vclab}\hfill}\hss}%
\vcenter{\hbox{\box\vcbox}}%
}

\makeatletter
\newsavebox{\@brx}
\newcommand{\llangle}[1][]{\savebox{\@brx}{\(\m@th{#1\langle}\)}%
  \mathopen{\copy\@brx\kern-0.5\wd\@brx\usebox{\@brx}}}
\newcommand{\rrangle}[1][]{\savebox{\@brx}{\(\m@th{#1\rangle}\)}%
  \mathclose{\copy\@brx\kern-0.5\wd\@brx\usebox{\@brx}}}
\makeatother

\makeindex

\title{Unitary matrix integrals, symmetric polynomials,\\ and long-range random walks}
\author{Ward L. Vleeshouwers\textsuperscript{1,2*} and Vladimir Gritsev\textsuperscript{1}\\
	$^1$ {\it Institute for Theoretical Physics, Universiteit van Amsterdam}, \\
	{\it Science Park 904, Postbus 94485, 1098 XH Amsterdam, The Netherlands}\\
	$^2$ {\it Institute for Theoretical Physics, Universiteit Utrecht}, \\
	{\it Princetonplein 5, Postbus 80.089, 3584 CC Utrecht, The Netherlands}\\
		*\href{mailto:w.l.vleeshouwers@uva.nl}{w.l.vleeshouwers@uva.nl}}

\begin{document}

\setlength\fboxsep{10pt}

\maketitle

	\begin{center}	
		
		\begin{abstract}
			
			Unitary matrix integrals over symmetric polynomials play an important role in a wide variety of applications, including random matrix theory, gauge theory, number theory, and enumerative combinatorics. We derive novel results on such integrals and apply these and other identities to correlation functions of long-range random walks (LRRW) consisting of hard-core bosons. We generalize an identity due to Diaconis and Shahshahani which computes unitary matrix integrals over products of power sum polynomials. This allows us to derive two expressions for unitary matrix integrals over Schur polynomials, which can be directly applied to LRRW correlation functions. We then demonstrate a duality between distinct LRRW models, which we refer to as \emph{quasi-local particle-hole duality}. We note a relation between the multiplication properties of power sum polynomials of degree $n$ and fermionic particles hopping by $n$ sites. This allows us to compute LRRW correlation functions in terms of auxiliary fermionic rather than hard-core bosonic systems. Inverting this reasoning leads to various results on long-range fermionic models as well. In principle, all results derived in this work can be implemented in experimental setups such as trapped ion systems, where LRRW models appear as an effective description. We further suggest specific correlation functions which may be applied to the benchmarking of such experimental setups.

		\end{abstract}
	\end{center}
	
\tableofcontents
\section{Introduction}
\subsection{Background}
This work presents a study of unitary matrix integrals over symmetric polynomials and their application to correlation function of long-range random walk (LRRW) models. We derive novel results on weighted unitary integrals over symmetric polynomials. These have a wide variety of mathematical and physical applications, including to Random Matrix Ensembles (RME's) and quantum chaos, gauge theories, number theory, and enumerative combinatorics. Connections between RME's and random walks have been explored extensively, see e.g. \cite{9304011}, \cite{Francesco_1995} for early reviews on RME's and \cite{Forrester:1315169}, \cite{Eynard_15} for more recent ones. Random walkers are called vicious when their paths are not allowed to intersect, i.e. when there exists the restriction that no two random walkers may occupy the same site. Vicious random walk models belong to the class of random-turns models when, at each time step, we move a single random walker. On the other hand, there are the lock-step models, where, at each time step, all random walkers are moved. For an early treatment, see \cite{fisherwwwm}, see also e.g. \cite{forrestervrw}, \cite{Forrester_1990},  \cite{Guttmann_1998}, \cite{Baik2000}, \cite{Krattenthaler_2000}, \cite{Nagao_2002}, \cite{Katori_2002}. This work will generally consider long-range versions of the random-turns model. One way in which a relation between RME's and vicious random walks can be seen to arise is from the fact that the joint eigenvalue distribution of unitary RME's is proportional to the squared Vandermonde determinant. This leads to a vanishing probability that any two eigenvalues coincide, which is known as level repulsion in random matrix theory. In physical terms, these non-intersecting random walkers can be interpreted as fermionic or hard-core bosonic particles, see e.g. the review \cite{Okounkov}. This leads to an interpretation of the eigenvalue probability density of models related to classical Lie groups in terms of the ground state density of non-interacting and non-intersecting particles which are subject to confining potential and certain boundary conditions. This language naturally leads to a so-called $\tau$-functions of integrable hierarchies of differential equations with many deep results obtained over the years \cite{Jimbo:1983if}, \cite{Morozov_1994}, \cite{Orlov_2002}, \cite{Wiegmann_06}, \cite{Harnad_2006}, \cite{Harnad_2021}. Connection to integrability is widely used in the recent literature, see \cite{Mironov-95}, \cite{Mironov_2017} and also \cite{Mironov_2022} for the case of unitary matrix models. Further, various correlation functions and important quantities in string theory are expressed in terms of these objects, see e.g. \cite{Aharony_2004}, \cite{Murthy_22} for recent developments. We also mention here applications in algebraic geometry and topology, see \cite{Eynard_15} for a review. 
%The basic building blocks in this story are Schur polynomials.
%functions and partitions of the Young diagrams of the unitary RME

The connection between RME's and random walks appears in a different guise as well. Bogoliubov demonstrated \cite{bogoxx} that the time-dependent correlation functions of the XX0-model are the generating functions of nearest neighbour vicious random walkers, i.e. those which can take only a unit size step to the left or right. Here, the XX0-model refers to the XX-model at zero magnetic field. These correlation functions can then be expressed as certain weighted unitary matrix integrals over Schur polynomials \cite{bogoxx}, \cite{bogpronko}. In general, spin configurations which start with an infinite sequence of particles and end with an infinite sequence of holes correspond uniquely to certain Young diagrams. Time-dependent correlation functions of the XX0-model involving such configurations then take the form of unitary matrix integrals over a product of Schur polynomials associated to the corresponding Young diagrams. In the remainder of this work, we will refer to down spins as particles and up spins as holes. In the nearest neighbour case, the corresponding RME is given by the Gross-Witten-Wadia (GWW) model, where the inverse square of the coupling constant of the GWW model is proportional to the time parameter of the correlation function of the XX0-model. This relation between random walk correlation functions and matrix integrals over Schur polynomials was generalized to the case of LRRW models \cite{pereztierz}, where the particles can hop over greater distances. The particles in question behave as hard-core bosons rather than fermions, as the wave function does not acquire a minus sign when a particle hops over another one. In the long-range case, the weight function of the corresponding matrix model encodes the particular choice of LRRW model. In particular, the $n^\text{th}$ hopping parameter $a_n$, which allows particles to hop a distance of $n$ sites, equals the $n^\text{th}$ Fourier coefficient of the weight function. The hopping parameters are required to decay as $a_n \sim n^{-1-\epsilon} $ for $\epsilon >0$ as $n\to \infty$, corresponding to weight functions which satisfy the strong Szeg\H{o} limit theorem. 

The generalization to LRRW models allows for the application to long-range one-dimensional systems, such as those characterized by dipole-dipole interactions. For more recent examples of such systems, see e.g. \cite{yuzbalt}. Another physically motivated context where long-range systems are of interest is that of Anderson localization in long-range low-dimensional hopping models \cite{Deng_2018}, \cite{Nosov_2019}. These systems can be simulated experimentally with trapped ions, which have been demonstrated to exhibit great tunability in terms of the range of the the relevant hopping amplitude (see e.g. \cite{ions_rev} for a review). Further applications include the dynamics of the Loschmidt echo in 1D systems, already considered in \cite{bogoxx}, see also \cite{Santilli_2020}, \cite{Bogoliubov_2020}. Dynamical phase transition in Loschmidt echo has been recently observed using the matrix models tools in \cite{Tierz_22}. The Loschmidt echo, a measure of chaoticity in quantum systems \cite{Gorin_2006}, can be regarded as a particular realization of the $\tau$-function mentioned above. The Loschmidt echo is also related to other important probes of many-body system dynamics, such as out-of-time-correlation functions. Importantly, the Loschmidt echo is an experimentally measurable quantity, see e.g. \cite{G_rttner_2017}, \cite{Braumuller_2021}, \cite{OTOC1}. In the context of combinatorics, we mention also a relation to the plane partitions, see e.g. \cite{Bogoliubov_2020} for a recent account thereof. 

In this paper, we consider physically relevant quantities, such as Loschmidt echo and various correlation functions in LRRW models, for which we employ identities on Young diagrams and the closely related theory of symmetric functions. We summarize the outline of the paper, including these results, in the remainder of the introduction.
\subsection{Structure of the paper and main results}
This work is structured as follows. We consider unitary matrix integrals over symmetric polynomials, which appear as correlation functions of long-range random walks of hard-core bosons. The basics of symmetric function theory, the relation between correlation functions of LRRW models and weighted unitary integrals over symmetric functions, and the evaluation of these objects are treated in appendices \ref{secsymmpol}, \ref{seccorrev}, and \ref{secmatrint}, respectively. To be precise, we consider correlation functions between configurations which begin with an infinite string of particles in the form of hard-core bosons and end with an infinite string of holes, with a finite region of mixed particles and holes in between. These correlation functions are written as follows,
\beq
F_{\lambda;\mu}(\tau) = \obket{\emptyset}{\sigma_{j_1}^+ \dots \sigma_{j_N}^+ e^{-\tau \hat{H}} \sigma_{l_1}^-  \dots  \sigma_{l_N}^- }{\emptyset}  ~~,~~~ \lambda_r =j_r+r ~,~~ \mu_s = l_s +s ~,
\eeq
%infinite strings of particles on lattice sites $j \leq 0$ and holes on lattice sites $j\geq 1$
where $\ket{\emptyset}$ is a state with holes at all lattice sites, where we remind the reader that we refer to up spins as holes and down spins as particles. The hamiltonian $H$ is a long-range translationally-invariant hamiltonian whose hopping parameters $a_{m,n} = a_{m-n}$ decay faster than $\abs{a_k} \sim \abs{k}^{-1}$, see equation \eqref{tiham}. Defining the matrix model average $\avg{\dots}$ for some weight function $f$ as in \eqref{schuravg}, it was found \cite{bogoxx}, \cite{bogpronko}, \cite{pereztierz} that this can be written (in our notation) as
\beq
\label{fsavg}
\frac{F_{\lambda;\mu}(\tau)}{F_{\emptyset;\emptyset}(\tau)}  = \avg{s_\lambda(U) s_\mu(U^{-1})}~,
\eeq
where the dependence on the particular choice of LRRW model and the generalized time parameter $\tau$ are captured by the weight function $f$. Further, $s_\lambda$ and $ s_\mu$ are Schur polynomials, whose representations $\lambda$ and $\mu$ characterize the particle-hole configurations that occupy a finite interval in between the infinite strings of particles and holes.

Sections \ref{secwick} and \ref{secspincorr} present our results, starting in \ref{subsecgenid} with the generalization of the result due to Diaconis and Shahshahani in \cite{diac1}. Writing $p_\mu = p_{\mu_1}p_{\mu_2}\dots$, where $p_n$ is the $n^\text{th}$ power sum polynomial, our result in \eqref{prpk} reads
\beq
%\label{prpk}
%\boxed{\hspace{.8cm} 
\avg{p_\rho(U)p_\mu(U^{-1})} =\sum_{n=0}^{\tilde{n}} C_n ~. 
%\hspace{.8cm}}~,
\eeq
where $C_n$ and $\tilde{n}$ are the contribution arising from performing $n$ contractions in $\avg{p_\rho p_\mu}$ and the maximum number of such contractions, which are
defined in \eqref{cneq} and \eqref{deftilden}, respectively. Essentially, this result arises from the application of Wick's theorem and the fact that $\avg{p_n(U) p_k(U^{-1}) } - \avg{p_n(U)}   \avg{ p_k(U^{-1})} = n\delta_{n,k}$, which we derived a previous work \cite{tftis}. In sections \ref{seceisp} and \ref{secepsp}, we derive expansions of general correlation functions of the form $\avg{s_\lambda (U) s_\nu(U^{-1})}$ in terms of Schur or power sum polynomials, respectively. Specifically, section \ref{seceisp} establishes a relation between correlation functions $\avg{s_\lambda s_\nu}$ and $\avg{s_{\lambda\setminus \{\eta\} } },\avg{s_{\nu\setminus \{\eta\} } }^\ast$, where $\lambda\setminus \{\eta\}$ is a diagram obtained from $\lambda$ by the removal of border strips $\eta_j$ of sizes $\abs{\eta_j} = \alpha_j$. In section \ref{secepsp}, we express $\avg{s_\lambda s_\nu}$ in terms of $\avg{p_\mu}$, where the expansion coefficients are determined by removing border strips from $\lambda$ and $\nu$ such that the resulting diagram is the same for both $\lambda$ and $\nu$. These results have a direct interpretation in terms of particle-hole configurations which is treated in section \ref{subsecpsbs}, for which reason we state these results with their particle-hole interpretations below.

We first consider the results derived in section \ref{subsecehcorr}, where we apply identities involving elementary and complete homogeneous symmetric polynomials to LRRW correlation functions. One result that arises in this way involves the well-known involution between elementary and homogeneous polynomials corresponding to the transposition of Young diagrams, which exchanges rows and columns. Although this result is mathematically trivial, its physical consequences are quite surprising and profound. Consider any two LRRW models, referred to as model A and B, whose hopping parameters are related by $a_k \to (-1)^{k+1} a_k$, and write their correlation functions as $F_{\lambda;\mu}^{(1)}(\tau)$ and $F_{\lambda;\mu}^{(2)}(\tau)$, respectively. Our result is given by equation \eqref{qlph}, which reads
\beq
\label{qlphintro}
F^{(1)}_{\lambda;\mu} (\tau)  = F^{(2)}_{\lambda^t;\mu^t} (\tau)  ~,
\eeq
where $^t$ refers to transposition of diagrams, which, combined with a parity transformation, implements a particle-hole transformation. Therefore, equation \eqref{qlphintro} states that the correlation functions of models A and B are equal after performing local particle-hole and parity transformations, which only act non-trivially on the interval in between the infinite strings of particles and holes. For this reason, we refer to this result as \emph{Quasi-Local Particle-Hole duality} (QLPH). In this way, a basic property of symmetric polynomials leads to a surprising result on LRRW models. Further such examples, involving e.g. the Pieri formula, are given in equations \eqref{f11e}, \eqref{corrhs}, \eqref{corrhs}, \eqref{corres}, \eqref{corres}.

Section \ref{subsecpsbs} treats the application of identities involving power sum polynomials and border strips, including results derived in section \ref{secschurbs}. We find that the multiplication of power sum polynomials of degree $n$ and the corresponding addition of border strips is related to fermionic particles hopping $n$ sites to the right, whereas removal of such a border strip corresponds to hopping $n$ sites to the left. This allows us to interpret various results in terms of an auxiliary fermionic (rather than hard-core bosonic) system. This reasoning can be applied to the computation of $\chi^\lambda_{~\alpha}$, which are the irreducible characters of the symmetric group, as we explain just above section \ref{subsubsah}. Further, combined with identities we found in \cite{tftis}, it leads immediately to results such as \eqref{traidlong} and \eqref{pnlt}. In section \ref{subsubsah}, we use the relation between border strips and particle hopping to characterize the action of the hamiltonian in terms of Young diagrams. 
%Combined with our result from \cite{tftis}, this leads to various results including \eqref{pnlt}. Further, these results

In section \ref{subsubsecbst}, we use the aforementioned relation between power sums and fermions to derive two results on long-range fermionic models. First of all, from the fact that $\chi^{\lambda/\mu}_{~\alpha}$ does not depend on the order of the entries of $\alpha$ leads us to conclude the following. Consider a one-dimensional fermion configuration corresponding to some diagram $\mu$, and consider all ways of moving not necessarily distinct fermions to the right by $\alpha_1,\alpha_2,\dots$ sites, where $\alpha_j$ are unordered non-negative integers. We find that \emph{the order of the step sizes $\alpha_j$ by which we move fermions has no effect on the outcome of this process}. That is, if we take a fermionic configuration and consecutively move fermions to the right by various step sizes, the outcome depends only on the distribution of the step sizes and not the order in which the steps are taken. Naturally, this statement also holds if we move fermions to the left instead of the right. The second result follows from the fact that $\chi^{\lambda/\rho}_{~\alpha}$ for $\alpha = (n^k)$  is cancellation-free \cite{white}, \cite{jkbook}. Here, $\rho$ is the so-called $n$-core of $\lambda$, which is the unique diagram that remains after removing the maximal number of size $n$ border strips from $\lambda$. In particular, start with a particle-hole configuration, which we call $\lambda$, and consider all ways to move (not necessarily distinct) particles by $n$ sites, where $n$ is any finite positive integer. By taking a particle and moving it by $n$ sites (from site $j$ to $j+n$), it may in the process `jump over' $\leq n-1$ other particles (occupying sites $j+1 $ to $j+n-1$). Our result then states that all ways to go from $\lambda$ to any other configuration $\nu$ involves jumping over \emph{either} an even \emph{or} an odd number of particles, depending only on the choice of $\lambda$ and $\nu$. When applied to a fermionic model where fermions can hop only by $\pm n$ sites for a single choice of $n$, this implies that there is no destructive interference between various ways to arrive at the same fermionic state, allowing for rapid spread through Hilbert space upon time evolution. The same reasoning can be applied to certain LRRW correlation functions, leading to \eqref{pnschur2} and \eqref{pkcf}. 

%{secepsp}
 In sections \ref{specf} and \ref{secpsecf}, we apply our results from section \ref{secschurbs} to LRRW correlation functions. The expansion in \ref{seceisp}, including its particle-hole interpretation in \ref{specf}, given in \eqref{spsavg} and \eqref{spsint}, reads
\begin{align}
\avg{s_\lambda (U)  s_\nu (U^{-1}) }_c  %& \sum_{\alpha} \frac{1}{z_\alpha}  \prod_{j\geq 1} C(n,\alpha_j ;\lambda)C(n,\alpha_j;\nu) \notag \\
& =    \sum_{\alpha} \frac{1}{z_\alpha} \left( \sum_{\{\eta\}} (-1)^{\hg(T_{\eta})} s_{\lambda\setminus \{\eta\} }(y) \right)\times \left( { y \to x \atop \lambda \to \nu } \right) \\
 \!\!\!\!\!\!\!\!\!\!\!\!\!\!\!\!\!\!\!\!\! =   \sum_\alpha \frac{1}{z_\alpha} & \left(  (-1)^{P} s_{\lambda\setminus \{\eta\} }(y) \left\{ \begin{tabular}{lll}
		  Distinct ways to take particles in $\lambda$\\ and move them $\alpha_1, \alpha_2,\dots$ sites to the\\
		left, thereby hopping over $P$ particles
   \end{tabular}
    \right\} \right)\times \left( { y \to x \atop \lambda \to \nu } \right) ~. \notag
\end{align}
 %
 %In the above expression, $\{ \eta \} = \{ \eta_1 ,\eta_2,\dots\}$ are border strips of sizes $\abs{\eta_j}  = \alpha_j$, and $\lambda \setminus \{\eta\}$ is the diagram obtained from $\lambda$ after the removal of these border strips. 
 The expansion in terms of power sum polynomials derived in section \ref{secepsp}, including its particle-hole interpretation in \ref{secpsecf}, is given by equations \eqref{corrpp} and \eqref{pspint} as
 \begin{align}
 \label{summpsp}
\avg{ s_\lambda (U) s_\nu(U^{-1})} = & \sum_{\omega, \gamma} \frac{p_\omega(y)p_\gamma(x)}{z_\omega z_\gamma} \sum_{\{\eta\} , \{ \xi\} } (-1)^{\hg(T_\eta) +\hg(T_\xi)} \delta_{\lambda\setminus \{\eta\}, \nu \setminus \{ \xi\}} \notag\\
 = & \sum_{\omega, \gamma} \frac{p_\omega(y)p_\gamma(x)}{z_\omega z_\gamma} (-1)^{P}
\left\{ \begin{tabular}{llll}
		  Distinct ways to move particles in $\lambda$ and $\nu$\\ to the left by $\gamma_1, \gamma_2,\dots$  and $\omega_1, \omega_2,\dots$ sites,\\
		   respectively, so hopping over $P$ other particles\\
		   and ending up in the same configuration.
   \end{tabular}
    \right\}  ~. 
 \end{align}  
 The expansion in \eqref{summpsp} is particularly convenient in applications where one has access to the power sum polynomials, such as for LRRW models, where they are given by $p_k =  k \tau a_{\pm k}$, in terms of the hopping parameters $a_k$ and the generalized time parameter $\tau$. Then, in section \ref{sec:Benchmarking} we suggest applications of some of our results to experimental benchmarking. We finish with our conclusions, as well as possible generalizations and applications of our results, which include both theoretical and experimental suggestions. 
\section{Unitary matrix integrals and Wick's theorem}
\label{secwick}
We now proceed with the derivation of our results. In this section, we will derive expansions for unitary matrix integrals over symmetric polynomials. The reader may consult appendix \ref{secmatrint} for more information on these objects. Taking the limit $N\to \infty$ allows us to apply an expression we found in \cite{tftis}. For $k,n \in \mathds{N}^+ $, and $f(z)=H(x;z)H(y;z^{-1})$ satisfying the strong Szeg\H{o} limit theorem, we demonstrated that 
\beq
\label{trakn1}
\avg{\tra U^n \tra U^{-k} } = n\delta_{n,k} + p_n(y) p_k (x) ~.
\eeq
For the CUE, $x_j =0 =y_j$ for all $j$, so that $p_n(x)=0$ for all $n$. Therefore, $\avg{\tra U^n \tra U^{-k} }_{\text{CUE}} = n\delta_{n,k}$ \cite{diac2}. We briefly summarize the derivation of our more general result in \eqref{trakn1} as we will be using similar arguments throughout the remainder of this work. We repeat that, for $n \in \mathds{N}^+$,
	\begin{equation}
	\tra U^n =  p_n(e^{i\theta_j}) = \sum_{r = 0}^{n-1} (-1)^r s_{(n-r,1^r)}(U) ~.
	\end{equation}
It is clear that this also holds when we replace $U$ with $U^{-1}$. Using \eqref{singleschur}, we have
	\beq
	\avg{\tra U^n}=%\sum_{r = 0}^{n-1} (-1)^r s_{(n-r,1^r)}(y) = 
	p_n(y)~~, ~~~ \avg{\tra U^{-n}}= 	p_n(x)~.
	\eeq
	Then, 
	\beq
	\label{trakn2}
\avg{\tra U^n \tra U^{-k} }  =\sum_\nu \sum_{r = 0}^{n-1}  (-1)^{r} s_{(n-r,1^r)/\nu} (y)   \sum_{s = 0}^{k-1}  (-1)^{s} s_{(k-s,1^s)/\nu} (x)  ~.
	\eeq
 The first sum on the right hand side of runs over all representations $\nu$ satisfying $\nu \subseteq (n-r,1^r)$ as well as $\nu \subseteq (k-s,1^s)$. We list below the contributions arising for various choices of $\nu$. 	We will often omit the variables $x$ and $y$ henceforth.
	\begin{enumerate}
		\item If $\nu$ is the empty partition $\nu = \emptyset$, $s_{\lambda/\nu} =s_\lambda$ i.e. the skew Schur polynomials reduces to the usual (non-skew) Schur polynomial. This contributes 
			\beq
			\sum_{r = 0}^{n-1}  (-1)^{r} s_{(n-r,1^r)}(y)
  \sum_{s = 0}^{k-1}  (-1)^{s} s_{(k-s,1^s)}(x)=  p_n(y) p_k(x) ~. 
	\eeq
		\item
		If $\nu = \lambda$, the skew Schur polynomial $s_{\lambda/\nu } = s_{\lambda/\lambda} = 1$. For $\lambda=(n-r,1^r)$ and $\mu=(k-s,1^s)$, one can only have $\lambda=\nu =\mu$ if $k=n$ and $r=s$. For $k=n$ and $\nu = (n-r,1^r)$ and we sum over $r$, we get a contribution of the following form
		\beq
		\sum_{r=0}^{n-1} (s_\emptyset)^2 = n~.
		\eeq
		This is where the term $ n\delta_{n,k}$ in \eqref{trakn1} originates.
		\item Take $k \leq n$ without loss of generality. The only remaining choice for $\nu$ is $\nu \neq \emptyset$ and $\nu \neq (n-r,1^r)$. For $\lambda = (a,1^b)$ and $\nu = (c,1^d)$ such that $\nu \subseteq \lambda$, we have $\lambda /\nu = (a-c) \otimes (1^{b-d})$, e.g. for $\lambda = (4,1^2)$ and $\nu = (2,1)$, we have the following.
		%
		%	\begin{center}
		\beq
		\hbox{
			\ydiagram{4,1,1} {  \huge{/ }}\ydiagram{2,1}  \hspace{.1cm} { \LARGE   =  }   \ydiagram{1}  \hspace{.1cm} \ydiagram{2} 
			}
			\eeq
	%	\end{center}
		\vspace{.3cm}
	Fixing a single such $\nu=(a,1^b)$ and considering only the sum over $r$ in \eqref{trakn2}, we have
	\beq 
	\label{hezero}
	\sum_{r = 0}^{n-1}  (-1)^{r} s_{(n-r,1^r)/(a,1^b)} = 	
	\sum_{r=b}^{n-a}(-1)^r h_{n-r-a}e_{r-b} = 0 ~,
	\eeq
	where we applied \eqref{ehzero}. That is, we get zero contribution for all $\nu \neq \emptyset, (n-r,1^r)$.
		\end{enumerate}
		Combining the above arguments leads to equation \eqref{trakn1}. Of course, the same reasoning can be applied to other expectation values involving $\tra U^n$, such as 
	\beq
	\label{trantral}
	\avg{\tra U^{-n} s_\lambda (U)}_c = \sum_{\nu \neq \emptyset}  \sum_{r=0}^{n-1} (-1)^r s_{(n-r,1^r)/\nu }   ~ s_{\lambda/\nu} ~.
	\eeq
As mentioned in appendix \ref{secmatrint}, the connected expectation value $\avg{\dots}_c$ in the above equation is given by $\avg{AB}_c = \avg{AB}-\avg{A}\avg{B}$ . The sum in \eqref{trantral} is over all $\nu \neq \emptyset$ such that $\nu\subseteq \lambda, (n-r,1^r)$. Fixing any $\nu \subseteq (n-r,1^r)$ in \eqref{trantral} with $\nu\neq (n-r,1^r)$ gives zero when summing over $r$ by application of \eqref{hezero}. Therefore, we only get a non-zero answer for terms for which $\nu = (n-1,1^r) \subseteq \lambda$. That is,
	\beq
	\label{trutrl}
	\avg{\tra U^{-n}  s_\lambda (U)}_c   =   \sum_r%=\min(0,  n-\lambda_1)}^{\min( n-1,\lambda_1^t+1)} 
	(-1)^r s_{\lambda/(n-r,1^r)}   ~,
	\eeq
	where the sum is over $\min(0,  n-\lambda_1) \leq r \leq \min( n-1,\lambda_1^t+1)$. Using \eqref{pnskew1}, we can express this as
	\beq
	\label{pnskew2}
		\avg{\tra U^{-n}  s_\lambda (U)}_c   = \sum_{\nu}(-1)^{\text{ht} (\lambda/\nu) } s_\nu~,
		\eeq
		where the sum is over all $\nu$ such that $\lambda/\nu$ is a border strip $\eta$ of size $n$, which we also write as $ \nu = \lambda \setminus \eta$. More information on border strips can be found in appendix \ref{secsymmpol}, see equation \eqref{bseqn} and especially \eqref{schurprodexp} and below.  As an example, take $\lambda =(6,4,3^2,2) $ and $n=4$. We show the diagrams $\nu$ appearing in \eqref{pnskew2} below, where the cells that are removed are again given in black. 
		
		 \vspace{.1cm}
		%
	%	\begin{center}
		\beq
		\label{bsremxmp}
		\hbox{
		{\large  $\mathrm{-}$ }  \begin{ytableau}
 *(white) & *(white) & *(white) & *(black)  & *(black)  & *(black)\\ 
*(white)  & *(white)    & *(white)  & *(black) \\
*(white)  & *(white)  & *(white) \\
*(white) & *(white) & *(white)  \\
*(white)  & *(white) 
\end{ytableau} 
	\hspace{.05cm} {\large   + }  	
  \begin{ytableau}
 *(white) & *(white) & *(white) &  *(white) & *(white) & *(white) \\ 
*(white)  & *(white)   & *(black)  & *(black) \\
*(white)  & *(white)    & *(black)  \\
*(white) & *(white)   & *(black)  \\
*(white)  & *(white) 
\end{ytableau} 
\hspace{.05cm}{\large  $\mathrm{-}$ }    
    \begin{ytableau}
 *(white) & *(white) & *(white) &  *(white) & *(white) & *(white) \\ 
*(white)  & *(white) & *(white)  & *(white) \\
*(white)  & *(white)    & *(white) \\
*(white)   & *(black)  & *(black)   \\
  *(black)  & *(black) 
\end{ytableau} 
}
\eeq

%\end{center} 
%\vspace{.2cm}
%

\subsection{Applying Wick's theorem}
We will now combine equation \eqref{trakn1} with Wick's theorem to compute more complicated objects. In particular, for terms of the form $\avg{ \tra U^{n_1} \tra U^{n_2} \dots \tra U^{-k_1} \tra U^{-k_2}\dots}$, we sum over all ways to contract between copies of $\tra U^{n_j}$ and $\tra U^{k_m}$. This contraction is done using the connected version of equation \eqref{trakn1}, which equals the CUE result,
\beq
\avg{\tra U^n \tra U^{-k}}_c = n\delta_{n,k}~.
\eeq
We first consider how Wick's theorem arises from the properties of Young diagrams for some relatively simple cases. For example, we have
\beq
\label{wickid1}
\avg{(\tra U^{2})^2 \tra U^{-1} }_c = 2 \avg{\tra U \tra U^{-2}}_c \avg{ \tra U^{-2}}=0~,
\eeq
where the factor two in the second expression arises from the fact that there are two ways to contract between $(\tra U^{2})^2$ and $ \tra U^{-1}$, which both contribute a term proportional to $\avg{\tra U \tra U^{-2}}_c=0$. We will now express the same equation using Young diagrams. Using \eqref{powersumschur}, the diagrams corresponding to $\tra U^2$ are as follows:
%
%\begin{center}
	\beq
		\hbox{
   \ydiagram{2}  \hso  { \large   $\mathrm{-}$ } \ydiagram{1,1} %\hso {\large   $+   $  $2$ }   \ydiagram{2,2} \hso { \large $\mathrm{-}$ } \ydiagram{2,1,1} \hso { \large + } \ydiagram{1,1,1,1}  
   }
   \eeq
%\end{center}
%\vspace{.2cm}
%
We take the square of the above expression and apply equation \eqref{schurpowerprod} (or, in this case, the Pieri formulas \eqref{hpieri} and \eqref{epieri}) to find the diagrams contributing to $(\tra U^2)^2$. These are as follows.
 \vspace{.1cm}
	\beq
		\hbox{
%\begin{center}
   \ydiagram{4}  \hso  { \large   $\mathrm{-}$ } \ydiagram{3,1} \hso {\large   $+   $  $2$ }   \ydiagram{2,2} \hso { \large $\mathrm{-}$ } \ydiagram{2,1,1} \hso { \large + } \ydiagram{1,1,1,1}  
   }
   \eeq
%\end{center}
%\vspace{.2cm}
%
We will denote the above sum over diagrams as $ \sum_\lambda b_\lambda  s_\lambda$, i.e. $b_{(4)} = 1 ,~ b_{(3,1)}=-1 ,~b_{(2,2)}=2,~ b_{(2,1,1)}=-1 ,~ b_{(1^4)}=1 $. Applying \eqref{largen}, we have
\beq
\label{wickid1b}
\avg{(\tra U^{2})^2 \tra U^{-1} }_c =  \sum_\lambda b_\lambda  s_{\lambda/{\small \Box} }
\eeq
That is, we take $  \sum_\lambda b_\lambda  s_\lambda$ and find the skew diagrams $\lambda/ {\small \Box}$, found by removing a single cell from $\lambda$ which has no cells to its right or below it. From these constraints, it follows that the resulting object after removing an cell is still a valid, non-skew diagram. This gives the following sum over diagrams, which evidently mutually cancel.%\\
		\begin{align}
		\label{wickidxmp}
	  \begin{ytableau}
*(white) & *(white) & *(white) & *(black)  
\end{ytableau} 
	\hspace{.05cm} ~  & - ~ 
\begin{ytableau}
*(white) & *(white) & *(white) \\ 
*(black) 
\end{ytableau} 
\hspace{.05cm} ~  -  ~	
 \begin{ytableau}
 *(white) & *(white) & *(black)  \\ 
*(white)  
\end{ytableau} 
 \hspace{.05cm} ~  + ~  2 ~~
  \begin{ytableau}
 *(white) & *(white) \\ 
*(white)   & *(black) 
\end{ytableau} 
  \notag\\[15pt]
	\hspace{.05cm}  & - ~ 	
\begin{ytableau}
*(white) & *(white) \\ 
 *(white) \\
*(black) 
\end{ytableau}
\hspace{.05cm} ~  - ~  	
\begin{ytableau}
*(white) & *(black) \\ 
 *(white) \\
  *(white)
\end{ytableau} 
\hspace{.05cm}  ~  + ~	
\begin{ytableau}
*(white)  \\ 
 *(white) \\
  *(white) \\
*(black) 
\end{ytableau} 
\hso  ~  = ~ 0
%\end{center}
\end{align}

\vspace{-.3cm}
We thus explicitly confirm that Wick's theorem is satisfied for the case of equation \eqref{wickid1}. Consider now $\avg{(\tra U^2)^2 \tra U^{-2}}_c $. Using \eqref{trakn1}, this should give
\beq
\label{wickid2}
\avg{(\tra U^2)^2 \tra U^{-2}}_c=2\avg{\tra U^2 \tra U^{-2}}_c\avg{\tra U^2}= 4 \avg{\tra U^2}
\eeq
We will check this explicitly as well. Applying \eqref{pnskew2} to $ \sum_\lambda b_\lambda  s_\lambda $ gives
\beq
\label{wickid2b}
 \avg{(\tra U^2)^2 \tra U^{-2}}_c=\sum_\lambda b_\lambda  \left[  s_{\lambda / (2)} -s_{\lambda/(1^2)} \right] = \sum_\lambda b_\lambda 
 \sum_\nu (-1)^{\text{ht}(\lambda/\nu)} s_\nu~,
 \eeq
where the sum is over all $\nu$ such that $\lambda/\nu $ is a border strip of size 2, i.e. $\lambda/\nu = $ {\tiny \ydiagram{2} } or $\lambda/\nu = $ {\tiny  \ydiagram{1,1} } . In terms of diagrams, this is given below. The first three diagrams corresponds to $\lambda/\nu = $ {\tiny \ydiagram{2} } whereas the latter three correspond to $\lambda/\nu = $ {\tiny  \ydiagram{1,1} } , which appear with a minus sign due to the factor $(-1)^{\text{ht}(\lambda/\nu)}$.\\
	\beq
	\label{tr2wickxmp}
	{ \small
	  \begin{ytableau}
*(white) & *(white)  & *(black)  & *(black)  
\end{ytableau} 
~
\hspace{.05cm} {\large   \mathrm{-} }  	
~
 \begin{ytableau}
 *(white)  & *(black) & *(black)  \\ 
*(white)  
\end{ytableau} 
~
 \hspace{.05cm}{\large   +  ~   2 } 
 ~
  \begin{ytableau}
 *(white) & *(white) \\ 
*(black)   & *(black) 
\end{ytableau} 
~
 \hspace{.05cm}{\large   -     ~2~ } 
  \begin{ytableau}
 *(white) & *(black) \\ 
*(white)   & *(black) 
\end{ytableau} 
~
	\hspace{.05cm} {\large   +}  
	~
\begin{ytableau}
*(white) & *(white) \\ 
 *(black) \\
*(black) 
\end{ytableau} 
~
\hspace{.05cm} {\large   \mathrm{-} }  	
~
\begin{ytableau}
*(white)  \\ 
 *(white) \\
*(black)\\
*(black) 
\end{ytableau} 
~
\hspace{.05cm} {\large   =  ~ 4 } ~ \begin{ytableau}
*(white)  & 
 *(white) \end{ytableau}
 ~
 \hspace{.05cm}{\large  \mathrm{-} ~ 4 }
 ~
 \begin{ytableau}
*(white)  \\ 
 *(white) 
 \end{ytableau}}
 \eeq
%\end{center} 
\vspace{.2cm}
We see that $\avg{(\tra U^2)^2 \tra U^{-2}}_c=2\avg{\tra U^2 \tra U^{-2}}_c\avg{\tra U^2} = 4 $ {\tiny \ydiagram{2} } $- $ $4$  {\tiny \ydiagram{1,1} } $ = 4 \avg{\tra U^2}$, thus confirming \eqref{wickid2}. Lastly, we will briefly check
\beq
\label{wickid3}
\avg{(\tra U^2)^2 (\tra U^{-1})^2}_c = 0~.
\eeq
We have
	\beq
		\hbox{
%\begin{center}
    \ydiagram{1} \hso {\Large$ \otimes$} \ydiagram{1} \hso { \large   = } 
\hso \ydiagram{2} \hso {\large +} \ydiagram{1,1}\hso 
}
\eeq
%\end{center}

Note that  {\tiny \ydiagram{1,1} }  appears with a positive sign, instead of with a minus sign as it does for $\tra U^2$. Then, 
\vspace{-.2cm}
\beq
\label{trtrt}
\avg{(\tra U^2)^2 (\tra U^{-1})^2}_c=s_{\tiny \Box} \cancelto{~0}{\sum_\lambda b_\lambda   s_{\lambda / \Box}} + \sum_\lambda b_\lambda  \left[  s_{\lambda / (2)} + s_{\lambda/(1^2)} \right] ~,
\eeq
where we applied the fact that $ \sum_\lambda b_\lambda   s_{\lambda / \Box} =0 $, see the diagrams in equation \eqref{wickidxmp}. Note that \eqref{trtrt} gives an equation very similar to \eqref{wickid2b}, but now $ s_{\lambda/(1^2)}$ carries a positive sign. This gives the same six diagrams as in the top line of equation \eqref{tr2wickxmp}, except that the last three diagrams are multiplied by $-1$, as can be seen below.   \vspace{.1cm}
%
%\begin{center}
	\beq
		\hbox{
	{ \small
	  \begin{ytableau}
*(white) & *(white)  & *(black)  & *(black)  
\end{ytableau}
\hspace{.05cm} {\large   $\mathrm{-}$ }  	
 \begin{ytableau}
 *(white)  & *(black) & *(black)  \\ 
*(white)  
\end{ytableau} 
 \hspace{.05cm}{\large   $+   $  $2$ } 
  \begin{ytableau}
 *(white) & *(white) \\ 
*(black)   & *(black) 
\end{ytableau} 
 \hspace{.05cm}{\large  $+$     $2$ } 
  \begin{ytableau}
 *(white) & *(black) \\ 
*(white)   & *(black) 
\end{ytableau} 
	\hspace{.05cm} {\large     $\mathrm{-}$ }  	
\begin{ytableau}
*(white) & *(white) \\ 
 *(black) \\
*(black) 
\end{ytableau} 
\hspace{.05cm} {\large   $+$ }  	
\begin{ytableau}
*(white)  \\ 
 *(white) \\
*(black)\\
*(black) 
\end{ytableau} 
\hspace{.05cm} {\large   = 0 }}
}
\eeq
%\end{center}
%\vspace{.2cm}
%
This confirms equation \eqref{wickid3}. From these relatively simple examples, one can explicitly see how Wick's theorem arises from equation \eqref{largen} and the multiplication rules for Young diagrams given in appendix \ref{secsymmpol}. We proceed to apply Wick's theorem to the computation of more general objects in the remainder of this section.
\subsection{Generalization of an identity due to Diaconis and Shahshahani}
\label{subsecgenid}
We use Wick's theorem and equation \eqref{trakn1} to generalize an identity due to Diaconis and Shahshahani \cite{diac1}, see also \cite{diac2}. Writing $p_\rho=p_{\rho_1}p_{\rho_2}\dots p_{\rho_{\ell(\rho)}} $  and $ m_j(\rho)=\text{Card} \{k:\rho_k=j\} $, as in equations \eqref{generalizedsympol} and \eqref{zmf}, respectively, we wish to calculate $\avg{p_\rho (U) p_\mu(U^{-1})}$. This is written out as follows,
\beq
\label{pavg}
\avg{p_\rho (U) p_\mu(U^{-1})} =\avg{ (\tra U^{j_1})^{m_{j_1}(\rho)}  (\tra U^{j_2})^{m_{j_2}(\rho)}\dots  (\tra U^{-k_1})^{m_{k_1}(\mu)} (\tra U^{-k_2})^{m_{k_2}(\mu)}\dots } ~.
\eeq
We start with a simpler object that we can easily apply Wick's theorem to. We see that performing $n$ contractions on $\avg{(\tra U^j)^a (\tra U^{-j})^b} $ leads to the following expression
\beq
\label{fnjeq}
C_{n,j} \coloneqq \frac{a!b!j^n}{(a-n)!(b-n)!n!}p_j(x)^{b-n} p_j(y)^{a-n}%~~;~~~C_{n,j}=0 
~,~~ n\leq  \text{Min}(a,b) ~.
\eeq 
Further, we have $C_{n,j}=0$ for $n \geq   \text{Min}(a,b) +1$. Equation \eqref{fnjeq} arises as follows. There are $\frac{a!b!}{(a-n)!(b-n)!n!}$ ways to perform $n$ contractions between  $(\tra U^j)^a $ and $(\tra U^{-j})^b$, and the contracted terms give a contribution equal to $\left(\avg{\tra U^j \tra U^{-j}}_c\right)^n= j^n $. The $a+b-2n$ uncontracted traces contribute $\avg{\tra U^j}^{a-n}\avg{\tra U^{-j}}^{a-n} =p_j(x)^{b-n} p_j(y)^{a-n}$. We now consider all possible ways to perform $n$ contractions between $p_\rho(U)$ and $p_\mu(U^{-1})$. This leads to a sum over $\alpha= (\alpha_1,\alpha_2,\dots) $ which are partitions of $n$, which specify the contractions that are performed. In particular, $m_j(\alpha)$ gives the number of $\tra U^j$ and $\tra U^{-j}$ which are contracted. The contribution coming from $n$ contractions in $\avg{p_\rho(U)p_\mu(U)^{-1}}$ can then be written as
\beq
\label{cneq}
C_n =  \sum_\alpha \prod_j C_{m_j(\alpha),j}~,%\sum_{\substack{ m =\text{Co}(n)  \\  \{ \text{distinct } j\} }}% \sum_{m =\text{Co}(n)} C_{m_1,j_1}C_{m_2,j_2}\dots C_{m_k,j_k}~,
\eeq
%where $(m)\vdash n$ means that $(m)=(m_1,m_2,\dots)$ is a partition of $n$, i.e. $\sum_k m_k =n$. Further,As indicated, the $j$'s we sum over are required to all be distinct. This restriction arises from the fact that (generally) $C_{n,j} \neq C_{m,j}C_{r,j}$ for $m+r = n$, so we have to have to ensure that each $j_k$ appears only a single time.
where the sum is over $\alpha$ that are partitions of $n$. We denote by $\tilde{n}$ is the maximal number of contractions one can perform, which is given by
\beq
\label{deftilden}
 \tilde{n}= \text{Max}(n) = \sum_{j\geq 1} \text{Min}(m_j(\rho),m_j(\mu))~.
 \eeq
 Summing over all possible contractions and applying \eqref{fnjeq} and \eqref{cneq}, we arrive at our result
\beq
\label{prpk}
\boxed{\hspace{.2cm} 
\avg{p_\rho(U)p_\mu(U^{-1})} =\sum_{n=0}^{\tilde{n}} C_n ~.  \hspace{.2cm}
}
\eeq
As mentioned before, this is a generalization of a result in \cite{diac1}, which considered the CUE, where $f=1$ so that $p_j(x) =0$ for all $j \neq 0$. Therefore, in the CUE case, one only gets a non-zero result when $ \rho=\mu$. In our notation, their result reads
\beq
\label{avgtrtr}
 \avg{p_\rho(U)p_\mu(U^{-1})}_{\text{CUE}} = z_\rho \delta_{\rho ,\mu } ~,
\eeq
where $\delta_{\rho ,\mu } =1 $ for $\rho=\mu$ and zero otherwise. Note that \eqref{avgtrtr} is only the last term in the full expansion in \eqref{prpk}, corresponding to $n=\tilde{n} = \sum_j m_j(\rho)$, so that all power sums in $p_\rho(U) $ and $p_\mu(U^{-1})$ are contracted.
	\subsection{Border strips and their application to unitary matrix integrals over Schur polynomials }
	\label{secschurbs}
We will derive two expressions for general $\avg{s_\lambda(U)s_\nu(U^{-1})}$ which rely on removing border strips from $\lambda $ and $\nu$. The first of these, equation \eqref{spsavg}, relates $\avg{s_\lambda(U)s_\nu(U^{-1})}$ to sums over $\avg{s_\mu (U)}\avg{s_\rho (U^{- 1})} = s_\mu(y)s_\rho(x)$, where $\mu$ and $\rho$ are related to $\lambda$ and $\nu$ by the removal of border strips, respectively. The second expression, in equation \eqref{corrpp}, provides an expansion of $\avg{s_\lambda(U)s_\nu(U^{-1})}$ in terms of the power sums $p_k(x)$ and $p_k(y)$. The latter expression appears to be particularly useful, as power sums are simpler objects than general Schur polynomials. In the context of LRRW models, $p_k(x)$ and $p_k(y)$ are given by $\pm \tau k a_{\pm k}$, where $a_{\pm k}$ are the hopping parameters in \eqref{tiham}. Equation \eqref{corrpp} therefore provides an expansion in $\tau$, where the expansion coefficients depend on $a_k$ which can be read off from the hamiltonian. We will treat the application of these formulas to LRRW models in sections \ref{specf} and \ref{secpsecf}. 
\subsubsection{Expansion in Schur polynomials}
\label{seceisp}
From \eqref{pnskew1}, \eqref{schurprodexp}, and \eqref{pnskew2}, we have
\beq
\label{spavg}
\avg{s_\lambda p_n}_c = \sum_\eta (-1)^{\hg (\eta)} \avg{s_{\lambda \setminus \eta}} = \sum_{\mu} \frac{\chi^\lambda_{~\mu}}{z_\mu } \avg{p_{\mu-(n)}} n m_n(\mu)~,
\eeq
where the sum is over all $\mu$ containing a row of size $n$ and $\mu-(n)$ is the remainder of $\mu$ after removing a row of size $n$. We remind the reader that $\lambda \setminus \eta$ is the diagram that results from $\lambda$ after removing border strip $\eta$ with (in this case) $\abs{\eta} = n$. We can also use the recursive definition of $\chi^\lambda_{~\mu}$ in equation \eqref{chirec} to see that the second equality in \eqref{spavg} should hold. From \eqref{zmf}, it is clear that
\beq
z_\mu = z_{\mu - (n)}\hspace{.02cm} n m_n(\mu)~.
\eeq
Plugging this into the rightmost expression in \eqref{spavg} and applying \eqref{chirec} leads to
\beq
\avg{s_\lambda p_n}_c = \sum_\mu \sum_\eta (-1)^{\hg (\eta)} \frac{\chi^{\lambda \setminus \eta}_{~\mu-(n)}}{z_{\mu - (n)}} \avg{p_{\mu - (n)}}~. 
\eeq
We can then apply equation \eqref{schurprodexp}, which leads to the second equality in \eqref{spavg}. If we insert two identical power sums, we find the following% gives
\begin{align}
\label{sppavg}
\avg{s_\lambda p_n^2}_c = & \sum_{ \eta ,\zeta}  (-1)^{\hg (\eta)+\hg (\zeta)} \avg{s_{(\lambda \setminus \eta ) \setminus  \zeta }} +  2\avg{p_n}\avg{s_\lambda p_n}_c   \notag \\
= & \sum_{\mu} \frac{\chi^\lambda_{~\mu}}{z_\mu } \avg{p_{\mu-(n^2)}} 2 n^2 m_n(\mu)(m_n(\mu)-1)+     2\avg{p_n}\avg{s_\lambda p_n}_c   ~,
\end{align}
where we consecutively remove border strips $\eta $ and $\zeta$ satisfying $\abs{\eta} = n = \abs{\zeta}$, resulting in the partition $(\lambda \setminus \eta ) \setminus  \zeta$. The sum on the right hand side runs over all $\mu$ containing (at least) two rows of length $n$, and $\mu-(n^2)$ is the remainder of $\mu$ after removing two such rows. The term $ 2\avg{p_n}\avg{s_\lambda p_n}_c $ arises from a single contraction between $p_n$ and $s_\lambda$. We can again apply equation \eqref{chirec} to demonstrate the second equality in \eqref{sppavg}, where we focus on the term arising from two contractions. Plugging %We apply equation \eqref{zmun} again to find 
\beq
z_\mu = z_{\mu - (n^2)}\hspace{.02cm} n^2  m_n(\mu) ( m_n(\mu) -1 ) ~.
\eeq
%
%Plugging this
into the first term on the right of \eqref{sppavg} leads to 
\beq
\sum_\mu \sum_{\zeta,\eta} (-1)^{\hg(\zeta) + \hg ( \eta) } \frac{\chi^{(\lambda \setminus \eta ) \setminus \zeta }_{~ \mu - (n^2)} }{ z_{\mu - (n^2)} } \avg{ p_{\mu - (n^2)}} ~,
\eeq
which, by \eqref{schurprodexp}, recovers the top line of \eqref{sppavg}. 

We then take $\avg{s_\lambda (p_n )^k} $ and perform all $k$ contractions. This gives an object we denote by $C(n,k;\lambda)$, which equals
\beq
\label{pscontr}
C(n,k;\lambda) = \sum_{\mu} \frac{\chi^\lambda_{~\mu}}{z_\mu } \avg{p_{\mu-(n^k)}} k! n^k \frac{m_n(\mu)!}{(m_n(\mu)-k)!} = \sum_{\{\eta\}} (-1)^{\hg(T_{\eta})} \avg{s_{\lambda\setminus \{\eta\} }} ~.
\eeq
The sum on the left is over all $\mu$ containing at least $k$ rows of length $n$. The sum on the right is over $k$ border strips $\{\eta\} = \{\eta_1 ,\dots \eta_k\}$ satisfying $\abs{\eta_j} = n$ for all $j$, where $\lambda\setminus \{\eta\} $ is the diagram obtained after removing all $\eta_j$ from $\lambda$, and $T_{\eta}$ is the BST consisting of the union of $\eta_1,\dots,\eta_k$. It follows that the term in \eqref{pscontr} gives zero if it is not possible to construct a subdiagram of $\lambda$ with $k$ border strips of size $n$, e.g. simply if $\abs{\lambda} < nk$. Note that $C(n,1;\lambda) = \avg{s_\lambda p_n}_c $ and $C(n,2;\lambda) = \avg{s_\lambda p_n^2 }_c - 2\avg{p_n}\avg{s_\lambda p_n}_c $.

We now consider
\beq
\label{slsnavg}
\avg{s_\lambda(U)s_\nu(U^{-1})}_c = \sum_{\mu,\rho} \frac{\chi_{~\mu}^\lambda}{z_\mu}\frac{\chi_{~\rho}^\nu}{z_\rho} \avg{p_\mu(U) p_\rho(U^{-1})} ~.
\eeq
When we consider those $\mu$ and $\rho$ that contain a row of size $n$ and we contract a single copy of $p_n$ between $p_\mu(U)$ and $ p_\rho(U^{-1})$, we get an object we denote by $A (n,1; \lambda , \nu  )$, which is given by %the following,
\beq
A (n,1; \lambda , \nu  ) = \sum_{\mu,\rho} \frac{\chi_{~\mu}^\lambda}{z_\mu}\frac{\chi_{~\rho}^\nu}{z_\rho} n m_n(\mu) m_n(\rho) \avg{p_{\mu-(n)}(U)} \avg{ p_{\rho-(n)}(U^{-1}) }  = \frac{1}{n} \avg{s_\lambda p_n}_c \avg{s_\nu p_n}_c  ~.
\eeq
Using \eqref{spavg} gives
\beq
A (n,1;\lambda , \nu  ) = \frac{1}{n } C(n,1;\lambda)C(n,1;\nu)  = \frac{1}{n} \left(  \sum_\eta (-1)^{\hg (\eta)} \avg{s_{\lambda \setminus \eta}} \right)  \left(  \sum_\zeta (-1)^{\hg (\zeta)} \avg{s_{\nu \setminus \zeta}} \right)   ~,
%\avg{s_\lambda p_n}_c 
\eeq
where the sums are again over border strips satisfying  $\abs{\eta} = n = \abs{\zeta}$. Consider now $\mu$ and $\rho$ that contain at least two rows of length $n$ and contract two copies of $p_n$ between $p_\mu(U)$ and $ p_\rho(U^{-1})$, 
\begin{align}
A (n,2 ; \lambda , \nu  ) = & \sum_{\mu,\rho} \frac{\chi_{~\mu}^\lambda}{z_\mu}\frac{\chi_{~\rho}^\nu}{z_\rho} \frac{n^2}{2} m_n(\mu)(m_n(\mu)-1) m_n(\rho)(m_n(\rho)-1) \avg{p_{\mu-(n)}(U)} \avg{ p_{\rho-(n)}(U^{-1}) } \notag \\
= &  \frac{1}{2n^2} C(n,2;\lambda)C(n,2;\nu)
~.
\end{align}
More generally, performing $k$ contractions between $ (\tra U^n)^k$ and $(\tra U^{-n})^k$ results in
\beq
A (n,k ; \lambda , \nu  ) = \frac{1}{k!n^k}C(n,k;\lambda)C(n,k;\nu)~.
\eeq
Consider a partition $\alpha$  and, as above, contract over $\alpha_1 $ copies of $p_1(U)$ and $p_1(U^{-1})$, $\alpha_2 $ copies of $p_2(U)$ and $p_2(U^{-1})$, etc. This gives the following.
\beq
A (n,\alpha ; \lambda , \nu  )= \frac{1}{z_\alpha}  \prod_{j\geq 1} C(n,\alpha_j ;\lambda)C(n,\alpha_j;\nu)~.
\eeq
We can apply the above expression and \eqref{pscontr}, leading to
%
%\begin{align}
\beq
\label{spsavgg}
\avg{s_\lambda s_\nu}_c =  \sum_{\alpha} \frac{1}{z_\alpha}  \prod_{j\geq 1} C(n,\alpha_j ;\lambda)C(n,\alpha_j;\nu)~, % \notag \\
\eeq
which can be written out as
\beq 
\label{spsavg}
\boxed{~~  \avg{s_\lambda(U)  s_\nu (U^{-1})}_c =    \sum_{\alpha} \frac{1}{z_\alpha} \left( \sum_{\{\eta \} } (-1)^{\hg(T_{\eta})} s_{\lambda\setminus \{\eta\} }(y) \right)\times \left( { y \to x \atop \lambda \to \nu } \right) ~. ~~ }
\eeq
The second sum above is over all border strips $\eta_j$ satisfying $\abs{\eta_j} = \alpha_j$. The above equation can be interpreted as follows. For any $\alpha$, consider all ways to remove $\alpha_1$ border strips of unit size (single cells) from $\lambda $ and $\nu$, $\alpha_2$ border strips of size 2, $\alpha_3 $ of size 3, and so on. The resulting diagrams are written as $\lambda\setminus \{\eta\}$ and idem for $\lambda \to \nu$. Remember from equation \eqref{chirec} and the comments below that the resulting diagrams do not depend on the order of the entries of $\alpha$, that is, on the order in which we remove border strips of various sizes. Indeed, the diagrams resulting from this procedure only depend on $\lambda$ and the set of cardinalities $m_j(\alpha)= \text{Card} \{k:\lambda_k=j\}$.

Equation \eqref{spsavg} expresses general expectation values $\avg{s_\lambda s_\nu}_c $ in terms of the non-skew Schur polynomials corresponding to $\lambda\setminus \{\eta \}$ and $\nu \setminus \{\eta \}$. On the other hand, equation \eqref{largen}, which is our starting point, gives an expansion in terms of skew Schur polynomials. There are various applications where an expression in terms of non-skew Schur polynomials is desirable. In general, this may be due to the fact that there are many more skew diagrams than non-skew ones, so that an expansion in non-skew diagrams may reveal underlying structures that are otherwise difficult to discern. This is also the case for the LRRW correlation functions we will be considering in the next section, where equation \eqref{spsavg} will reveal relations between various correlation functions. 

We consider now the special case where one can form BST's of shapes $\lambda$ and $\nu$ from $ \alpha_j$ border strips of size $j$, summed over $j$, such that we can fully contract between $p_\mu(U)$ and $p_\rho(U^{-1})$. That is, we consider the case where we can completely tile both $\lambda$ and $\nu$ with $\alpha_1$ single cells, $ \alpha_2 $ border strips of size 2 (dominoes), and so on, for the same choice of $\alpha$. This clearly requires $\abs{\lambda} = \abs{\nu} = \sum_{j\geq 1} j \alpha_j$, which is a necessary (but far from sufficient) condition for $\chi^\lambda_{~\alpha} , \chi^\nu_{~\alpha}  \neq 0$. Consider the CUE, where $\avg{s_{\lambda/\mu}(U^{\pm 1})}=0$ for any $\lambda/\mu \neq \emptyset$. Applying the Murnaghan-Nakayama formula given in equation \eqref{mnrule},
 \beq
 \sum_{\alpha_1,\dots,\alpha_k} (-1)^{\hg(T_{\alpha})}  = \chi^\lambda_{~\alpha}~,
 \eeq
we then find that
\beq
\label{slsncue}
\avg{s_\lambda s_\nu}_{\text{CUE}} = \sum_\alpha z_\alpha^{-1} \chi^\lambda_{~\alpha} \chi^\nu_{~\alpha }= \delta_{\lambda , \nu}~.
\eeq
This is just the orthonormality property of symmetric group characters, see e.g. [Proposition 7.17.6b, \cite{stanley}]. Another way to arrive at the same expression is to directly plug equation \eqref{avgtrtr},
\beq
\avg{p_\mu p_\rho}_{\text{CUE}} =z_\mu \delta_{\mu,\rho}~,
\eeq
 into equation \eqref{slsnavg} to find the orthonormality relation in \eqref{slsncue}.

We work out the explicit example for $\lambda = (3,2)$, which is a sufficiently small partition that we can still apply \eqref{largen} for comparison. Indeed, applying \eqref{largen} to $\avg{s_{(3,2)} s_{(3,2)} }_c= \sum_\nu s_{(3,2)/\nu} s_{(3,2)/\nu}   $ gives the following diagrams $(3,2)/\nu$.
%
%	\begin{center}
	\beq
		\hbox{
\begin{ytableau}
\none & *(white) & *(white)  \\ 
*(white)  & *(white)     
\end{ytableau}
\hso ,  \hso \hso	
%\hspace{.05cm}  
%{\large   + }  	
  ~\begin{ytableau}
 *(white) & *(white)  \\ 
 *(white)     
\end{ytableau}
\hso ,  \hso \hso
%\hso \hso	
%\hspace{.05cm}{\large  + }   
 \begin{ytableau}
*(white)  & *(white)     
\end{ytableau} 
%\hspace{.02cm}
\begin{ytableau}
*(white) 
\end{ytableau}
\hso ,  \hso \hso
%\hspace{.02cm}
\begin{ytableau}
*(white) & *(white)
\end{ytableau}
\hso ,  \hso \hso	 
\begin{ytableau}
*(white) 
\end{ytableau}
\begin{ytableau}
*(white) 
\end{ytableau}
\hso ,  \hso \hso
 2 
\begin{ytableau}
*(white) 
\end{ytableau}
\hso ,  \hso \hso
{ \Large $\emptyset$}
}
\eeq
%\end{center} 
%\vspace{.2cm}
 \vspace{-.4cm}
 
Note that {\tiny \ydiagram{1}} appears with multiplicity 2 as it arises from $\nu = (3,1)$ and $\nu = (2^2)$. By applying \eqref{pnskew1} for $n=1$ (or, equivalently, \eqref{piericorh} or \eqref{piericore}) to the leftmost diagram, we find the following.
 \vspace{.1cm}
%
%	\begin{center}
	\beq
		\hbox{
\begin{ytableau}
\none & *(white) & *(white)  \\ 
*(white)  & *(white)     
\end{ytableau}
{ \large   =  } 
\begin{ytableau}
 *(white) & *(white)  \\ 
*(white)  & *(white)     
\end{ytableau}
\hspace{.05cm}{\large  + }   
\begin{ytableau}
 *(white) & *(white) & *(white)  \\ 
*(white)   
\end{ytableau}
}
\eeq
%\end{center} 
%\vspace{.2cm}
%
%
This leads to
\beq
\label{s32ln}
\avg{s_{(3,2)} s_{(3,2)} }_c  = \left(s_{(2,2)} + s_{(3,1)} \right)^2 + \left( s_{(3)}+  s_{(2,1)}\right)^2 + s_{(2,1)}^2  + \left( s_{(2)}+  s_{(1^2)}\right)^2 + s_{(2)}^2 + 2 s_{(1)}^2 +1  ~. ~~~
\eeq
Note that the above equation is strictly speaking only correct when $x=y$. For $x\neq y$, we have e.g. $\left(s_{(2,2)}(x) + s_{(3,1)}(x) \right)\left(s_{(2,2)}(y) + s_{(3,1)}(y) \right)$ instead of $\left(s_{(2,2)} + s_{(3,1)} \right)^2$, but we write it as above to avoid clutter.

%
%\vspace{-.3cm}
%\beq
%\begin{tikzpicture}
%\begin{scope}%[every node/.style={circle,thick,draw}]
 %   \node (A) at (0,0) {$\mathlarger{\mathlarger{\lambda}}$};
 %   \node (B) at (1.9,0) {$\mathlarger{\mathlarger{\mu}}$};
%\end{scope}
%
%\begin{scope}[>={Stealth[black]},
 %             every node/.style={fill=white,circle},
  %            every edge/.style={draw=black,thick}]
   % \path [->] (A) edge node {$p_j$} (B);
%\end{scope}
%\end{tikzpicture} 
%\eeq
%\vspace{-1cm}
%
%

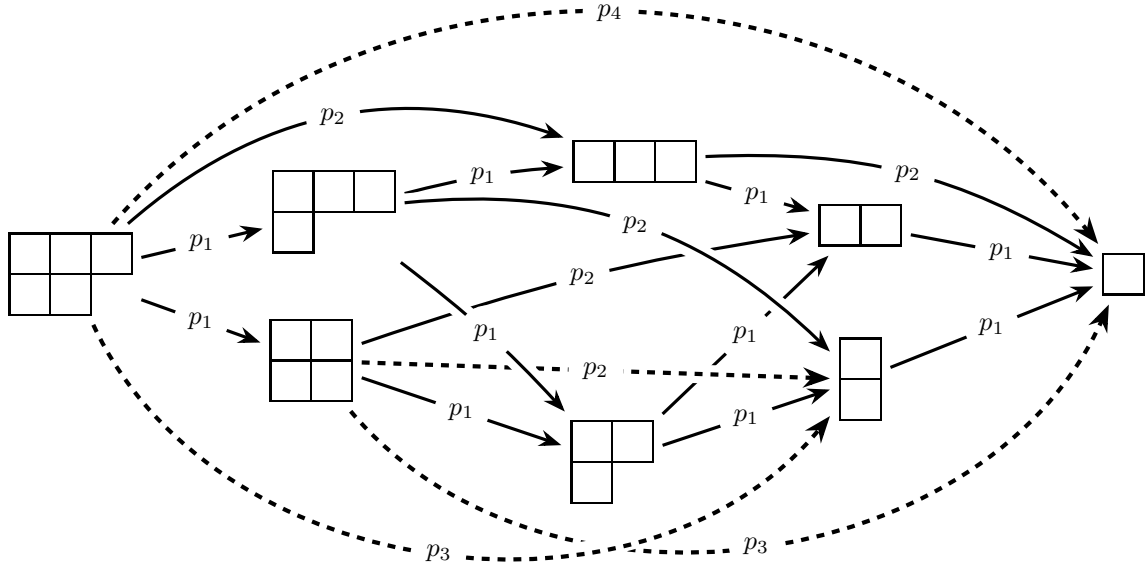
\begin{figure}[ht]
%
 %&
 \begin{tikzpicture}[on top/.style={preaction={draw=white,-,line width=#1}},
on top/.default=7pt]
\begin{scope}%[every node/.style={circle,thick,draw}]
    \node (A) at (0,0) {$\ydiagram{3,2}$};
    \node (B) at (3.5,.83) {$\ydiagram{3,1}$};
    \node (C) at (3.2,-1.15) {$\ydiagram{2,2}$};  
    \node (D) at (7.5,1.5) {$\ydiagram{3}$};
    \node (E) at (7.2,-2.5) {$\ydiagram{2,1}$};
    \node (F) at (10.5,.65) {$\ydiagram{2}$};
    \node (G) at (10.5,-1.4) {$\ydiagram{1,1}$}; 
    \node (H) at (14,0) {$\ydiagram{1}$};    
   % \node (I) at (13,0) {$\mathlarger{\mathlarger{\mathlarger{\mathlarger{\emptyset}}}}$};      
\end{scope}
\begin{scope}[>={Stealth[black]},
              every node/.style={fill=white,circle},
              every edge/.style={draw=black,very thick}]
    \path [->] (A) edge node {$p_1$} (B);
    \path [->] (A) edge node {$p_1$} (C);
    \path [->] (B)[bend left=5] edge node {$p_1$} (D);
    \path [->] (C) edge node {$p_1$} (E); 
    %\path [->] (C) edge[bend right=58] node {$p_3$} (H); 
    \path [->] (C) edge[bend left=5]
    node {$p_2$} (F); 
    \path [->] (D) edge node {$p_1$} (F);
    \path [->] (E) edge node {$p_1$} (G);   
    \path [->] (F) edge node {$p_1$} (H);
    \path [->] (E) edge node {$p_1$} (F);     
    \path [->] (G) edge node {$p_1$} (H);    
    \path [->] (A) edge[bend left=30] node {$p_2$} (D); 
    \path [->] (D) edge[bend left=17] node {$p_2$} (H);
    %\path [->] (A) edge[bend left=50] node {$p_4$} (H); 
\end{scope}
\begin{scope}[>={Stealth[black]},
              every node/.style={fill=white,circle},
              every edge/.style={draw=black,ultra thick,dashed}]
   \path [->] (A) edge[bend left=50] node {$p_4$} (H); 
   \path [->] (C) edge[bend right=58] node {$p_3$} (H); 
   \path [->] (A) edge[on top, bend right=58] node {$p_3$} (G);
   \path [->] (C) edge[on top]%[bend right=3]
                 node {$p_2$} (G); 
\end{scope}    
\begin{scope}[>={Stealth[black]},
              every node/.style={fill=white,circle},
              every edge/.style={draw=black,very thick}]
    \path [->] (B) edge[on top, bend left=5] node {$p_1$} (E);
    \path [->] (B) edge[on top, bend left=25] node {$p_2$} (G);
        % \path [->] (A) edge[on top, bend right=55] node {$p_3$} (G);
     \end{scope}    
\end{tikzpicture}
\vspace{-.7cm}
\caption{The diagrams for $\lambda = (3,2)$ and those obtained by removal of border strips of size $j$, which is indicated by $p_j$. The solid and dashed lines again indicate the removal of border strips of even and odd heights, respectively. This graph indicates all ways to remove border strips from $(3,2)$ and the resulting diagrams, except for those cases where the removal of a border strip leads to the empty diagram.   \label{y32graph}}
\end{figure}

We now apply \eqref{spsavg} to compute $\avg{s_{(3,2)} s_{(3,2)}}_c$. To do so, we successively remove border strips from $(3,2)$ to find the various partitions $(3,2) \setminus \{\eta\}$ in \eqref{spsavg}. This is illustrated below, where two diagrams connected by an arrow as $\overset{p_j}{\longrightarrow}$ again indicates that partitions $\lambda$ and $\mu$ are related by the removal of a border strip of size $j$. The graph below contains all information about the removal of border strips from $(3,2)$, except for those cases where the removal of a border strip leads to the empty diagram. All diagrams in figure \ref{y32graph} except for $(3,2)$ and $(2,2)$ are hook shapes and therefore border strips, so that they are related to $\emptyset$ by the removal of a single border strip.
%\vspace{-.5cm}
%
%
%\begin{center}
%\begin{align}

The number of ways to arrive at a diagram by following the arrows in figure \ref{y32graph} gives the multiplicity of that diagram in \eqref{spsavg}. For example, for $~{\tiny \ydiagram{3,2}}$ , the diagram $~{\tiny \ydiagram{1}}$ has multiplicity 5 for $\alpha = (1^5)$, as there are 5 distinct ways to arrive at $~{\tiny \ydiagram{1}}$ following arrows indicated by $p_1$ in figure \ref{y32graph}. For $\alpha$ containing elements of different sizes, a single ordering should be fixed to find the correct multiplicity i.e. we do not sum over different compositions of the same cycle type. Applying equation \eqref{spsavg} then leads to \eqref{sps32}, where we indicate the compositions $\alpha$ of the power sum $p_\alpha$ over which we contract. That is, to arrive at a term indicated by some $\alpha$, one should start from { \tiny \ydiagram{3,2}} and follow arrows indicated by the $\alpha_j$ in any order, keeping in mind the sign given by $(-1)^{\hg(\eta_j)}$ with $\abs{\eta_j} = \alpha_j$. This gives the following.
\begin{align}
\label{sps32}
    \avg{s_{(3,2)} s_{(3,2)} }_c =  &  \overset{(1)}{\left(s_{(2,2)} + s_{(3,1)} \right)^2} + \overset{~~(2)}{\frac{1}{2} s_{(3)}^2}  +  \overset{(1^2)}{\frac{1}{2} \left( 2 s_{(2,1)} + s_{(3)} \right)^2   } + \overset{~~~~(3)}{+ \frac{1}{3} s_{(1^2)}^2 } + \overset{~~~(2,1)}{\frac{1}{2} s_{(2)}^2 } + \notag \\
     &  + \underset{(1^3)}{\frac{1}{3!} \left(3 s_{(2)} + 2 s_{(1^2)}\right)^2 }
    +  % \notag \\      &   +
  \left(\frac{1}{3}+ \frac{1}{8} + \frac{1}{4} + \frac{1}{4} + \frac{5^2}{4!}   \right) s_{(1)} + 1 ~.
\end{align}
The last two terms on the bottom line of the expression above arise from $\abs{\alpha} = 4,5$. In the latter case, following arrows indicated by $p_{\alpha_j}$ leads to $\emptyset$.  One can see that there is only a single way to start at $~{\tiny \ydiagram{3,2}}$ and arrive at $\emptyset $ for $\alpha = (4,1),(3,2),(3,1^2),(2^2,1), (2,1^3)$ by following the corresponding arrows in the graph above. However, we already noted that there are 5 distinct ways $~{\tiny \ydiagram{3,2}}$ and arrive at $~{\tiny \ydiagram{1}}$ following arrows with $p_1$. This implies there are also 5 ways to arrive at $\emptyset$ by removing single elements, simply by taking the additional step $~{\tiny \ydiagram{1}} \overset{p_1}{\longrightarrow} \emptyset$. This gives
\beq
\frac{1}{z_{(4,1)}} + \frac{1}{z_{(3,2)}} +  \frac{1}{z_{(3,1^2)}} + \frac{1}{z_{(2,1^3)}} + \frac{1}{z_{(2^2,1)}} +  \frac{5^2}{z_{(1^5)}} = \frac{1}{4} + \frac{1}{6} + \frac{1}{6} + \frac{1}{12} +\frac{1}{8 } + \frac{5^2}{5!} = 1~,
\eeq
leading to the unit contribution in equation \eqref{sps32}. Via this reasoning, one may check that expression \eqref{sps32} and \eqref{s32ln} are identical. The above example may not appear to give a very convincing argument in favor of equation \eqref{spsavg} over \eqref{largen}, as the application of \eqref{spsavg} appears to be more complicated than \eqref{largen} for the case of $\lambda = (3,2)$. Indeed, for partitions containing few cells, such as $\lambda = (3,2)$, it is convenient to use \eqref{largen}, as the skew partitions $\lambda/\nu$ can easily be related to non-skew partitions, as in \eqref{s32ln}. However, for larger partitions, this is no longer the case, as \eqref{piericorh}, \eqref{piericore}, and \eqref{pnskew1} can no longer be applied. When considering larger partitions, then, equation \eqref{spsavg} can still be used to express general objects $\avg{s_\lambda s_\nu }_c$ in terms of non-skew Schur polynomials. This will allow us to express complicated correlation functions in terms of simpler ones in section \ref{specf}, but can be used more generally in situations where an expression for $\avg{s_\lambda s_\nu}$ in terms of non-skew Schur polynomials is desirable. 
\subsubsection{Expansion in power sum symmetric polynomials}
\label{secepsp}
For certain applications, such as the LRRW models we will be considering in the next section, expanding $\avg{s_\lambda(U) s_\nu (U^{-1})}$ in terms of power sums $p_k(x)$ and $p_k(y)$ may be particularly useful. One way to do so is to use the expansion in \eqref{schurprodexp}, calculate all $\chi^\lambda_{~\alpha},\chi^\nu_{~\alpha}$, and contract power sums using \eqref{trakn1}. However, this is rather inconvenient as the computation of all the symmetric group characters is rapidly becomes more complicated for larger $\lambda,\nu$. Instead, it would be more effective to once more use equation \eqref{chirec} to find a recursive expansion in terms of power sum polynomials. We will do so here, ultimately leading to equation \eqref{corrpp}. For comparison, we will also consider the more complicated method that involves the calculation of all $\chi^\lambda_{~\alpha},\chi^\nu_{~\alpha}$ at the end of this subsection to demonstrate its inconvenience compared to equation \eqref{corrpp}.

The expression that we will be deriving provides an iterative method for expanding in $p_\gamma (x)$ and $p_\omega(y)$ that does not require us to find $\chi^\lambda_{~\alpha}$ and $\chi^\nu_{~\alpha}$ for all $\alpha$ and then contract over all combinations of power sum polynomials. Essentially, we apply equations \eqref{prpk} and \eqref{schurprodexp} to $\avg{s_\lambda(U) s_\nu(U^{-1})}$ and make use of the orthogonality properties of the symmetric group characters. In particular, we will revert the order of expansion in section \ref{seceisp} and start from the term with all $p_\mu$ contracted, see equation \eqref{slsncue}, then consider the term where a single $p_j$ is not contracted, then two uncontracted power sums, and so on. For simplicity, we start by considering autocorrelation function up to the subleading term, which gives
%
%\begin{align}
\beq
\label{llap}
\avg{s_\lambda (U) s_\lambda(U^{-1})} =  \sum_{\mu ,\rho}  \frac{\chi_{~\mu}^\lambda}{z_\mu}\frac{\chi_{~\rho}^\lambda}{z_\rho} \avg{p_\mu(U) p_\rho(U^{-1})} %\notag \\
 = 1+  \sum_{j \geq 1}  \sum_\mu  \frac{(\chi_{~\mu}^\lambda)^2 }{z_\mu }  \frac{m_j(\mu)}{j}p_j(x)p_j(y) + \dots
\eeq
where sum is over all $\mu$ containing a row of size $j$. We have $\mu =\rho$ since this is the only way to contract all of $p_\mu$ with $p_\rho$ except for two copies of $p_j$ (one from $p_\mu$ and the other from $p_\rho$). Using the recursive formula for $\chi^\lambda_{~\mu}$ in \eqref{chirec} and the orthogonality of symmetric group characters in \eqref{slsncue}, the rightmost term in equation \eqref{llap} then gives
\beq
\label{subl}
 \sum_j \frac{1}{j^2} \sum_\mu  \frac{ \left( \sum_\eta (-1)^{\hg(\eta)} \chi^{\lambda \setminus \eta}_{~\mu - (j)} \right)  \left( \sum_\zeta (-1)^{\hg(\zeta)} \chi^{\lambda \setminus \zeta}_{~\mu - (j)} \right)}{z_{\mu-(j)}  }  p_j(x)p_j(y) = \sum_{j\geq 1} \frac{p_j(x)p_j(y)}{j^2} \sum_\eta 1~,
\eeq
where $\eta $ and $\zeta $ are border strips of size $j$. Equation \eqref{subl} then tells us that the coefficient $p_j(x)p_j(y)$ in the power sum expansion of $\avg{s_\lambda (U) s_\lambda(U^{-1})}$ is given by $\frac{1}{j^2}$ times the number of ways to remove a border strip of size $j$ from $\lambda$.  Consider another term in this expansion, where $\mu = (j,\alpha)$, $\rho=(k^2 ,\alpha)$ with $j=2k$, and contract over $p_\alpha$ leading to the term proportional to $p_j(x)p_k(y)^2$. Via the same reasoning as above, this is given by
\beq
\label{pjpkkcorr}
\frac{p_j (x) p_k(y)^2}{2 jk^2} \sum_{\eta ,\zeta,\xi} (-1)^{\hg(\eta)+  \hg(\zeta)+\hg(\xi)} \delta_{\lambda\setminus \eta,(\lambda\setminus \zeta)\setminus \xi}~,
\eeq
where the sums are over all $\eta$, border strips of size $j$, and $\zeta$ and $ \xi$, border strips of size $k$. We see that the terms appearing in the power sum expansion of $\avg{s_\lambda (U) s_\lambda(U^{-1})}$ are found by removing border strips $\{\eta\}$ and $\{\zeta\}$ from $\lambda$ such that $\lambda \setminus \{\eta\} =\lambda \setminus \{\zeta\}$. Continuing this procedure gives the following. Consider not contracting $p_\omega (U)$ and $p_\gamma(U^{-1})$ for some $\omega$ and $\gamma$, leading to the term proportional to $p_\omega(y)p_\gamma(x)$. In general this leads to a sum over $\mu$ and $\rho$ (in the middle expression in \eqref{subl}) of the form
\beq
\mu = (\omega ,\alpha)~~,~~~ \rho = (\gamma ,\alpha)~,
\eeq
so that $m_j(\mu)-m_j(\omega) = m_j(\alpha ) = m_j(\rho) - m_j(\gamma)$ for all $j$, and $\abs{\omega}=\abs{\eta}$. Since we sum over $\mu$ and $\rho$ for a  fixed choice of $\omega$ and $\gamma$, this effectively leads to a sum over $\alpha$. Since there are $ \frac{1}{m_j(\alpha)!}  \frac{m_j(\mu)!}{m_j(\omega)!} \frac{m_j(\rho)!}{m_j(\gamma)!}   $ ways to perform $m_j(\alpha)$ contractions of $p_j$ appearing in $p_\mu$ and $p_\rho$, we have
\beq
\label{pmprgo}
\avg{p_\mu p_\rho } = \dots +  p_\omega(y)p_\gamma(x) \prod_{j\geq 1} \frac{1}{m_j(\alpha)!}  \frac{m_j(\mu)!}{(m_j(\mu)-m_j(\alpha))!} \frac{m_j(\rho)!}{(m_j(\rho)-m_j(\alpha))!} j^{m_j(\alpha)}  +\dots 
\eeq
where we show only the term proportional to $p_\omega(y)p_\gamma(x)$. Using the fact that,
\beq
z_\mu = z_\alpha \prod_{j\geq 1} j^{m_j(\mu) - m_j(\alpha)} \frac{m_j(\mu)!}{ m_j(\alpha)!}~,
\eeq
and again applying the orthogonality of symmetric group characters, we find that the term proportional to $p_\omega(y)p_\gamma(x)$ in the expansion of 
$\avg{s_\lambda(U)  s_\lambda(U^{-1})}$ is given by
\beq
\label{autocorrpp}
 \frac{p_\omega(y)p_\gamma(x)}{z_\omega z_\gamma} \sum_{\{\eta\} , \{ \xi\} } (-1)^{\hg(T_\eta) +\hg(T_\xi)} \delta_{\lambda\setminus \{\eta\}, \lambda \setminus \{ \xi\}}~,
\eeq 
where the sum is over border strips $\eta_1, \eta_2,\dots$ satisfying $\abs{\eta_j} = \omega_j$, as well as border strips $\xi_j$ satisfying $\abs{\xi_j} = \gamma_j$. Further, $T_\eta$ and $T_\xi$ are the (generally disconnected) skew diagrams consisting of the unions of the border strips in $\eta$ and $\xi$, respectively, and $\hg(T)$ is defined in \eqref{htbst}. It is clear that the examples in \eqref{subl} and \eqref{pjpkkcorr} arise as special cases of \eqref{autocorrpp}. The autocorrelation function can then be expanded as a sum $\omega $ and $\gamma$, each contributing a term of the form appearing in \eqref{autocorrpp}. The same reasoning can evidently be applied to more general correlation functions, which are then given by  
\beq
\label{corrpp}
\boxed{%\hspace{.6cm}  
\avg{ s_\lambda (U) s_\nu(U^{-1})} = \sum_{\omega, \gamma} \frac{p_\omega(y)p_\gamma(x)}{z_\omega z_\gamma} \sum_{\{\eta\} , \{ \xi\} } (-1)^{\hg(T_\eta) +\hg(T_\xi)} \delta_{\lambda\setminus \{\eta\}, \nu \setminus \{ \xi\}}~. %\hspace{.6cm}
}
\eeq 
The above expression can be straightforwardly applied by considering the different ways to remove border strips from $\lambda$ and $\nu$ such that the resulting diagrams are identical, as expressed by the presence of $\delta_{\lambda\setminus \{\eta\}, \nu \setminus \{ \xi\}}$. This provides a recursive expression which significantly simplifies various computations, especially when $\lambda, \nu$ are large diagrams for which the computation of the symmetric group characters rapidly grows more difficult. Equation \eqref{corrpp} is particularly useful for applications where one has access to $p_k(x)$ and $p_k(y)$. This includes LRRW models, where $p_k(x)$ and $p_k(y)$  are proportional to the time parameter and the hopping parameters, as we will see in the next section.

%and compare with the more complicated method applied above, which resulted in equations \eqref{mnxmpf}--\eqref{mnxmpl}. In particular, c

We now apply \eqref{corrpp} to $\avg{s_\lambda(U) s_\lambda(U^{-1})}$ for $\lambda =(3,2)$. Consider all distinct ways to remove a border strip of size one, i.e. a single cell, from $~{\tiny \ydiagram{3,2}}$ . As can be seen in figure \ref{y32graph}, this results in the following diagrams.
 %\vspace{.1cm}
%
%
%	\begin{center}
%	\beq
\beq
\label{y32pieri1}
		\hbox{
  \begin{ytableau}
 *(white) & *(white) & *(black)  \\ 
*(white)  & *(white)  
\end{ytableau} 
\hspace{1cm}
  \begin{ytableau}
 *(white) & *(white)    & *(white)  \\ 
*(white)& *(black) 
\end{ytableau} 
}
\eeq
%\eeq
%\end{center} 
%
 \vspace{.1cm}
Removing a single cell twice results in the following diagrams.
 \vspace{.1cm}
	\beq
		\label{y32bs11}
		\hbox{
  \begin{ytableau}
 *(white) & *(white)    & *(white)  \\ 
 *(black)& *(black) 
\end{ytableau} 
\hspace{.7cm}
   {\large  2  } \begin{ytableau}
 *(white) & *(white) & *(black)  \\ 
*(white) & *(black) 
\end{ytableau} 
}
\eeq
%\end{center} 
%\vspace{.2cm}
%\normalsize
%
We get multiplicity two for $~{\tiny \ydiagram{2,1}}$ on the right as one can remove the single cells indicated in black in either order, corresponding to the fact that there are two distinct ways to arrive at $~{\tiny \ydiagram{2,1}}$ from $~{\tiny \ydiagram{3,2}}$ by following $p_1$ in figure \ref{y32graph}. Removing border strips of sizes $2,3,4$ results in the diagrams below.
% read from left to right.
%
\beq
\label{y32bs234}
		\hbox{
	%\beq
  \begin{ytableau}
 *(white) & *(white)    & *(white)  \\ 
 *(black)& *(black) 
\end{ytableau} 
\hspace{1cm}
  \begin{ytableau}
 *(white) & *(black)   & *(black)   \\ 
 *(white)& *(black) 
\end{ytableau} 
\hspace{1cm}
  \begin{ytableau}
 *(white) & *(black)   & *(black)   \\ 
*(black) & *(black) 
\end{ytableau} 
}
\eeq
%\end{center} 
%\vspace{.2cm}
%
Note that the height $\hg$ of the border strips of sizes $3$ and $4$ is given by 1, so that $(-1)^\hg = -1$. It is easy to see that no border strips of size $\geq 5 $ can be removed from $~{\tiny \ydiagram{3,2}}$ , since it is a partition of 5 that is not itself a border strip. From the fact that there are two distinct ways to remove a single cell from $\lambda$, and only a single way to remove border strips of sizes $2,3,4$, we see that the the corresponding contributions are given by
\beq
\label{p1234}
%\avg{s_\lambda(U) s_\lambda(U^{-1})} = 
2 p_1(x) p_1(y) + \frac{p_2(x) p_2(y) }{4} + \frac{p_3(x) p_3(y) }{9}+\frac{p_4(x) p_4(y) }{16}~.% +\dots 
~,
\eeq
%thus recovering equations \eqref{mnxmpf} through \eqref{mnxmp11} via a significantly simpler method.
Consider now the term proportional to $p_1(x)^2 p_1(y)^2$. The two diagrams which arise from consecutively removing two single cells are $~{\tiny \ydiagram{2,1}}$ and $~{\tiny \ydiagram{3}}$ , where the former appears with multiplicity two, as demonstrated above. We thus see that there are four ways to take two copies of $~{\tiny \ydiagram{3,2}}$ , consecutively remove two single cells, and end up with $~{\tiny \ydiagram{2,1}}$ . Conversely, there is only a single way to end up with $~{\tiny \ydiagram{3}}$ via this procedure. Combined with $(z_{(1^2)})^2 = 4$, this demonstrates that we have
\beq
\label{corrp1p1}
\frac{5 p_1(x)^2 p_1(y)^2}{4}
\eeq
appearing in the expansion of $\avg{s_{(3,2)}s_{(3,2)}}$. Lastly, we consider $p_2(x) p_1(y)^2 + p_1(x)^2 p_2(y)$. One may remove either a border strip of size two or two single cells and end up in $~{\tiny \ydiagram{3}}$ , see the leftmost diagrams in \eqref{y32bs11} and \eqref{y32bs234}. Combined with $z_{(1^2)} = 2 = z_{(2)}$, this results in 
\beq
\label{corrp2p1}
\frac{p_2(x) p_1(y)^2 + p_1(x)^2 p_2(y)}{4}~.
\eeq
 Although the effectiveness of \eqref{corrpp} is already clear for $~{\tiny \ydiagram{3,2}}$ , this is still a
rather\\[2pt] small partition. For larger $\lambda$, it becomes progressively harder to compute $\chi^\lambda_{~\alpha}$, increasing the advantage of \eqref{corrpp}. We will work out more complicated examples in section \ref{secpsecf}, where we apply the above results to correlation functions of LRRW models.

We now proceed to compute $\avg{s_{(3,2)}(U) s_{(3,2)} (U^{-1})}$ via the more laborious method briefly outlined at the start of this subsection, where we compute $\chi^{(3,2)}_{~\alpha}$ for all $\alpha$, apply \eqref{schurprodexp}, and perform all possible contractions between the power sum polynomials appearing in \eqref{schurprodexp}. We will see in the following section, below equation \eqref{pnlt}, that one may calculate $\chi^\lambda_{~\alpha}$ by using a relation with fermionic particles hopping on a one-dimensional lattice. One may then apply Wick's theorem to the resulting expression to find an expansion of LRRW correlation funtions in terms of power sums. Although this may provide a convenient method for calculating $\chi^\lambda_{~\alpha}$, we will see here that its application to the computation of $\avg{s_\lambda (U) s_\nu (U^{-1})}$ is much less convenient than simply applying equation \eqref{corrpp}. Using \eqref{chirec} and looking at figure \ref{y32graph}, we see that the non-zero characters $\chi^\lambda_{~\alpha} $ are given by
\begin{alignat}{4}
\label{y32chi}
  \chi^{(3,2)}_{~(1^5)} &  =  5 ~, ~~~ \chi^{(3,2)}_{~(2,1^3)} && =  1~ && , ~~~  \chi^{(3,2)}_{~(2^2,1)} && =   1~, \notag \\
 \chi^{(3,2)}_{~(3,1^2)}  & = 1 ~, ~~~ \chi^{(3,2)}_{~(3,2)} && =  -1~ && , ~~~ \chi^{(3,2)}_{~(4,1)} && =   1~. 
\end{alignat}
%However, we will first apply the former, more complicated method to $\avg{s_\lambda(U)s_\lambda(U^{-1})}$ for $\lambda = (3,2)$, before deriving and applying \eqref{corrpp} for comparison. This provides both an explicit check of equation \eqref{corrpp} as well as a demonstration of its effectiveness for computing $\avg{s_\lambda(U)s_\lambda(U^{-1})}$. More general and complicated examples of $\avg{s_\lambda(U)s_\nu(U^{-1})}$ are treated in section \ref{secpsecf}, where we consider the application to LRRW models. For $\lambda = (3,2)$, b
%
Applying \eqref{schurprodexp} gives
\begin{align}
\label{y32pexp}
s_{(3,2)} = & \frac{p_1^5}{24} + \frac{ p_1^3 p_2}{12} + \frac{ p_1^2 p_3}{6}+\frac{p_2^2p_1}{8} - \frac{ p_2 p_3}{6} + \frac{p_1p_4}{4}\notag \\
= &  \frac{p_{(1^5)}}{24} + \frac{ p_{(2,1^3)} }{12} + \frac{ p_{(3,1^2)} }{6}+\frac{p_{(2^2,1)}}{8} - \frac{ p_{(3,2)} }{6} + \frac{p_{(4,1)}}{4} ~.
\end{align} 
We then apply Wick's theorem to $\avg{s_{(3,2)}(U)s_{(3,2)}(U^{-1}}_c$. We give three examples below, again indicating the compositions $\alpha$ corresponding to the $p_\alpha$ which are contracted.
\begin{align}
%\beq
 \overset{(4)}{\frac{1}{4} p_1(x) p_1(y) } + \overset{(2^2)}{ \frac{1}{8^2} p_1(x)p_1(y) }  & + \overset{(3)}{\frac{3}{~6^2}\left( p_2(x)p_2(y) -   p_2(y)p_1(x)^2 \right)} + %\dots 
 \notag \\
& + \frac{3}{~6^2}\left( p_1(x)^2 p_1(y)^2 - p_2(x)p_1(y)^2  \right) +\dots 
%\eeq
\end{align}
We will use the above method to find the prefactors of $p_\omega(y) p_\gamma(x)$ for some examples of $\omega,\gamma$. Consider first $\gamma = \omega = (4)$, leading to a term proportional to $p_4(x)p_4(y)$. This term is rather simple to find as we only get a contribution from the rightmost term in \eqref{y32pexp}. In particular, we have
\beq
\label{mnxmpf}
\frac{1}{4^2} \avg{p_{(4,1)}(U) p_{(4,1)}(U^{-1}) }_c = \frac{p_4(x)p_4(y) }{4^2} + \frac{p_1(x)p_1(y)}{4} + \frac{1}{4}~,
\eeq
so that $p_4(x)p_4(y)$ appears with a prefactor $\frac{1}{4^2}$ in the expansion of $\avg{s_{(3,2)}(U) s_{(3,2)}(U^{-1})}$. We consider now those terms proportional to $p_3(x)p_3(y)$ arising from \eqref{y32pexp}, where the dots below refer to terms not proportional to $p_3(x)p_3(y)$ and where we omit writing $(U^{\pm 1})$ explicitly henceforth. 
\begin{align}
\label{mnxmpf2}
\frac{1}{6^2} \avg{ p_{(3,1^2)} p_{(3,1^2)}  }  +\frac{1}{6^2} \avg{ p_{(3,2)}  p_{(3,2)}  } & =   \frac{p_3(x)p_3(y)}{36} ( 2 +2)+ \dots \notag \\
 & = \frac{p_3(x)p_3(y)}{9} + \dots
\end{align}
For $p_2(x)p_2 ( y)$, we find
\begin{align}
\label{mnxmpf3}
\frac{\avg{p_{(2,1^3)} p_{(2,1^3)}} }{(12)^2}  +  \frac{\avg{p_{(2^2,1)} p_{(2^2,1)}  }}{8^2}  +\frac{\avg{ p_{(3,2)}  p_{(3,2)} }}{6^2}  & =   p_2(x)p_2(y) \left( \frac{3!}{12}  + \frac{8}{8^2} + \frac{3}{6^2} + \right)+ \dots \notag \\
 & = \frac{p_2(x)p_2(y)}{4} + \dots
\end{align}
We see a pattern emerge as we get $\frac{p_j(x)p_j(y)}{j^2}$ for all $j$ considered thus far. However, this pattern does not continue down to $j=1$. In particular, the following expression
\beq
\frac{\avg{p_{(1^5)}p_{(1^5)} } }{(24)^2}  + \frac{\avg{p_{(2,1^3)} p_{(2,1^3)}}}{(12)^2}  + \frac{\avg{p_{(3,1^2)} p_{(3,1^2)} }}{6^2}  +  \frac{\avg{p_{(2^2,1)} p_{(2^2,1)}  }}{8^2}  +\frac{\avg{ p_{(4,1)}  p_{(4,1)} }}{4^2} ~,
\eeq
%
%\beq
%\frac{1}{(24)^2} \avg{p_{(1^5)}p_{(1^5)} }  + \frac{1}{(12)^2} \avg{p_{(2,1^3)} p_{(2,1^3)}} + \frac{1}{6^2} \avg{p_{(3,1^2)} p_{(3,1^2)} } +  \frac{1}{8^2} \avg{p_{(2^2,1)} p_{(2^2,1)}  } +\frac{1}{4^2} \avg{ p_{(4,1)}  p_{(4,1)} }~,
%\eeq
%
is given by
\beq
\label{mnxmp11}
  p_1(x)p_1(y) \left( \frac{(5!)^2}{(4!)^3}  + \frac{6^2}{(12)^2} + \frac{12}{6^2} + \frac{8}{8^2} +\frac{4}{4^2} \right)+ \dots   =   2 p_1(x)p_1(y) + \dots
\eeq
Combining equations \eqref{mnxmpf}, \eqref{mnxmpf2}, \eqref{mnxmpf3}, and \eqref{mnxmp11} then leads to equation \eqref{p1234}, albeit via a much less convenient method.

Consider now the term proportional to $p_1(x)^2 p_1(y)^2$, which receives contributions from
\beq
\label{mnxmp1111}
\frac{\avg{p_{(1^5)}p_{(1^5)} }}{(24)^2}   + \frac{\avg{p_{(2,1^3)} p_{(2,1^3)}}}{(12)^2}  + \frac{\avg{p_{(3,1^2)} p_{(3,1^2)} }}{6^2}~.  
\eeq
In particular, this contributes
\beq
\label{mn1111res}
p_1(x)^2 p_1(y)^2 \left( \frac{(5!)^2}{(4!)^2 (2!)^2 3!} + \frac{(12)^2}{12} + \frac{3}{6^2} \right)  = \frac{5  p_1(x)^2 p_1(y)^2 }{4}~,
\eeq
thus confirming equation \eqref{corrp1p1}. Lastly, we consider a mixed term, namely, the term proportional to $p_2(x) p_1(y)^2 + p_1(x)^2 p_2(y) $. This is given by 
\begin{align}
\label{mnxmpl}
\frac{ \avg{p_{(1^5)} p_{(2,1^3)} }}{12\cdot 24 }  - \frac{ \avg{ p_{(3,1^2)} p_{(3,2)}}}{6^2}   + \frac{\avg{p_{(2,1^3)} p_{(2^2,1)} }}{12\cdot 8} 
= & p_2(x) p_1(y)^2 \left( \frac{5}{24} - \frac{1}{12} +\frac{1}{8} \right) +\dots \notag \\
 = & \frac{p_2(x) p_1(y)^2}{4}+\dots 
\end{align}
Of course, inverting the order of $p_\mu(U)$ and $p_\rho(U^{-1})$ in $\avg{ p_\mu(U) p_\rho(U^{-1})}$ above gives the same contribution with $x \leftrightarrow y$. The end result is therefore $\frac{ p_2(x) p_1(y)^2 + p_1(x)^2 p_2(y) }{4}$, thereby confirming \eqref{corrp2p1}. It is clear from this simple example that the method applied in the second part of this subsection is much less powerful than equation \eqref{corrpp}, and this becomes more acute when we consider larger partitions than $\lambda= (3,2)$ as the $\chi^\lambda_{~\alpha}$ are then a lot harder to compute.

\section{Applying symmetric polynomial theory to long-range random walks}
\label{secspincorr}
We will now apply some of the results above to LRRW correlation functions by using their relation to weighted $U(N)$ integrals over Schur polynomials in equation \eqref{corrmat} and identities from symmetric polynomial theory, as well as results we derived in section \ref{secwick}. We first consider identities relating to elementary and complete homogeneous symmetric polynomials, before moving on to power sum polynomials and border strips. We will generally take $N\to \infty$ here, although the presentation in sections \ref{secqlph}, \ref{subsubsah}, and \ref{subsubsecbst} is valid for finite $N$ as well.
\subsection{Elementary and complete homogeneous symmetric polynomials}
\label{subsecehcorr}
We repeat here the expression for the weight function in \eqref{spinwfct},
\beq
f(e^{i\theta} ;\tau) = \exp\left(\tau\sum_{k\in\mathds{Z}} a_ke^{i\theta  } \right)  ~,
\eeq
where $a_k$ are the hopping parameters of the hamiltonian in \eqref{tiham}. As noted before, the hamiltonian is a Toeplitz matrix, and $a_k$ is the number on its $k^{\text{th}}$ diagonal. By comparing with  $\eqref{genfuncsymp}$, we see that we can write the weight function as
\beq
f(z) =H(x;z)H(y;z^{-1})~,
\eeq
with the following identification, for $k\geq 1$,
\begin{align}
\label{hpid}
\tau a_k   & = \frac{p_k(x)}{k} ~, \notag \\
  \tau a_{-k} & = \frac{p_k(y)}{k} ~.
%\frac{p_k(x)}{k}  = \tau a_k =\tau a_{-k}
\end{align}
Alternatively, we can write
\beq
f(z) =E(x;z)E(y;z^{-1})~,
\eeq
by identifying, for $k\geq 1$,
\begin{align}
\label{epid}
\tau a_k   & = \frac{(-1)^{k+1} p_k(x)}{k} ~, \notag \\
  \tau a_{-k} & = \frac{(-1)^{k+1} p_k(y)}{k} ~,
%\frac{p_k(x)}{k}  = \tau a_k =\tau a_{-k}
\end{align}
and by transposing the diagrams as in \eqref{largen2}. When $\tau \to 0$, i.e. the CUE limit, we have
\beq
\label{fzero}
F_{\lambda;\mu} (0) = \avg{ s_\lambda(U) s_\mu(U^{-1})}_{\text{CUE}} = %\bket{\lambda}{\mu}_{\text{CUE}}  =
\delta_{\lambda,\mu} ~,
\eeq
which is again simply the orthonormality of Schur polynomials as the irreducible characters of $U(N)$. %condition for the basis $\{\ket{\lambda}\}$ of spin configurations. We see that \eqref{fzero} is satisfied when using \eqref{largen} from the fact that $\avg{ s_\lambda s_\mu}_{\text{CUE}}= \delta_{\lambda \mu}$.
By using the strong Szeg\H{o} limit theorem, we can compute $F_{\emptyset;\emptyset}$ \footnote{ This has been noted before, see e.g \cite{vitiq}.}. Assuming $a_0=0$, i.e. zero on-site energy, we have
\beq
F_{\emptyset;\emptyset}(\tau)= \exp\left( \tau^2 \sum_{k=1}^\infty ka_k^2\right)~.
\eeq
If we have $a_0\neq 0$, we get an additional multiplicative term $e^{-N \tau a_0}$ on the right, where one should remember that we take $N\to \infty$. Considering $a_1 =-1 = a_1$ and $a_k=0$ otherwise, i.e. the XX0-model, and choosing $\tau = it$, we recover the result of \cite{wei} and \cite{vitiq}, see also \cite{klm1},
\beq
F_{\emptyset;\emptyset} (it) = e^{-t^2}~.
\eeq
%Note that the aforementioned works consider the temporal (instead of thermal) autocorrelation, with $it$ instead of $\tau$. 
Remember that equations \eqref{mcdschurs} and \eqref{largen} state that
\beq
F_{\lambda;\mu}(\tau) = F_{\emptyset;\emptyset}(\tau)\avg{s_\lambda(U) s_\mu(U^{-1})}_\tau~,
\eeq
where $\avg{\dots}_\tau$ is given in equation \eqref{schuravg} with weight function given by $f(z;\tau)$ in \eqref{spinwfct}.
We therefore define
\beq
\label{gdef}
G_{\lambda;\mu} (\tau)  \coloneqq \frac{F_{\lambda;\mu}(\tau)}{F_{\emptyset;\emptyset}(\tau)} = \avg{s_\lambda (U) s_\mu(U^{-1})}_\tau~,
\eeq
i.e. we express correlations in terms of their proportionality to $F_{\emptyset;\emptyset}(\tau)$. %When we are considering spin configurations corresponding to some $\lambda,\mu$, 
We will also write this as 
\beq
G_{\lambda;\mu}(\tau) \eqqcolon \bket{\lambda}{\mu}_\tau~.
\eeq
%where, instead of writing $\lambda,\mu$, we will draw the spin configurations corresponding to $\lambda,\mu$.
%
If $\mu=\emptyset$ (or $\lambda=\emptyset$), we will simply write
\beq
G_{\lambda; \emptyset}(\tau) = \bket{\lambda}{\emptyset}_\tau \eqqcolon \avg{\lambda}_\tau~.
\eeq
We also define the following - connected - correlation function,
\beq
G^c_{\lambda;\mu}(\tau) \eqqcolon \bket{\lambda}{\mu}_\tau - \avg{\lambda}_\tau \avg{\mu}^*_\tau~.
\eeq
We consider now some explicit examples. Using \eqref{newtonbell}, we can express 
\beq
\avg{e_n}_\tau=e_n(y)=G_{(1^n); \emptyset} ~,~~ \avg{h_n}_\tau =h_n(y)=G_{(k); \emptyset } ~,
\eeq
and their complex conjugates $G_{\emptyset;(1^n)}$ and $ G_{\emptyset;(n) }$, in terms of $p_k(y)$ with $k\leq n$. For $h_n$ and $e_n$ with $n=4$, the diagrams and corresponding configurations are given below on the left and right, respectively.
 \vspace{.2cm}
\beq
\hspace{-1cm}
\rlap{\tableau{&&&&\emptycell \!\!\!&\emptycell&\emptycell&\hdotscell \\ \emptycell  \\ \emptycell\\ \vdotscell %\\&&\emptycell\!\!\!\\&\emptycell\\\emptycell\\\emptycell \\\vdotscell%
}}
\unitlength=\cellsize
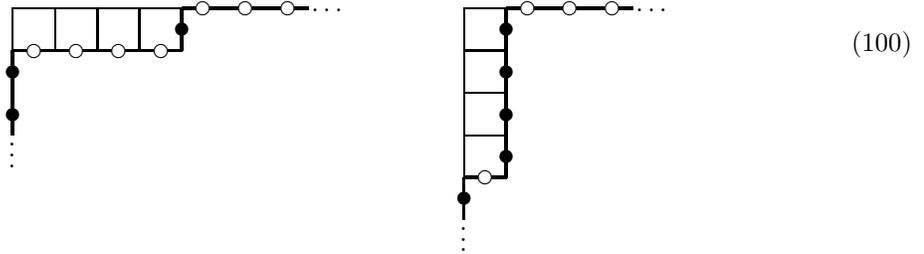
\begin{picture}(0,0)
\thicklines
\drawline(0,-2)(0,0)(4,0)(4,1)(7,1)%(2,0)(4,0)(4,1)(7,1)
\thinlines
%\put(0,-3.5){\circle*{0.3}}
%\put(0,-2.5){\circle*{0.3}}
\put(0,-1.5){\circle*{0.3}}
\put(0,-0.5){\circle*{0.3}}
\put(4,+0.5){\circle*{0.3}}
{\color{white}
\put(6.5,1){\circle*{0.3}}
\put(5.5,1){\circle*{0.3}}
\put(4.5,1){\circle*{0.3}}
\put(3.5,0){\circle*{0.3}}
\put(2.5,0){\circle*{0.3}}
\put(1.5,0){\circle*{0.3}}
\put(0.5,0){\circle*{0.3}}
}
\put(6.5,1){\circle{0.3}}
\put(5.5,1){\circle{0.3}}
\put(4.5,1){\circle{0.3}}
\put(3.5,0){\circle{0.3}}
\put(2.5,0){\circle{0.3}}
\put(1.5,0){\circle{0.3}}
\put(0.5,0){\circle{0.3}}
\end{picture}
\hspace{6cm}
\rlap{\tableau{& \emptycell&\emptycell&\emptycell&\hdotscell% \!\!\!&\emptycell&\emptycell&
\\&\emptycell
\\ &\emptycell\\\!\\\emptycell \\ \vdotscell %\\\emptycell \\ \vdotscell%
}}
\unitlength=\cellsize
\begin{picture}(0,0)
\thicklines
\drawline(0,-4)(0,-3)(1,-3)(1,1)(4,1)
\thinlines
%\put(0,-4.5){\circle*{0.3}}
\put(0,-3.5){\circle*{0.3}}
\put(1,-2.5){\circle*{0.3}}
\put(1,-1.5){\circle*{0.3}}
\put(1,-0.5){\circle*{0.3}}
\put(1,+0.5){\circle*{0.3}}
{\color{white}
\put(1.5,1){\circle*{0.3}}
\put(2.5,1){\circle*{0.3}}
\put(3.5,1){\circle*{0.3}}
\put(.5,-3){\circle*{0.3}}
}
\put(1.5,1){\circle{0.3}}
\put(2.5,1){\circle{0.3}}
\put(3.5,1){\circle{0.3}}
\put(.5,-3){\circle{0.3}}
\end{picture}
\eeq
That is, $h_n$ correspond to taking only the single rightmost particle and moving it $n$ sites to the right, whereas $e_n$ corresponds to taking the $n$ rightmost particles and moving them all a single site to the right. Note that with $x= (x_1,\dots ,x_K)$, we have $e_j(x) =0$ for $j>K$ (and likewise for $y$). This means that, for any $n,m >K$,
\beq
\label{f11e}
F_{(1^m);(1^n) } =\sum_{j=0}^K e_j e_j
\eeq
That is, when we move $n>K$ adjacent particles by a single site, the effect is the same as to move $K$ adjacent particles a single site. 

Not only can we read off $G_{(n);\emptyset}$ and $G_{(1^n) ; \emptyset}$ (and their complex conjugates). From the corollary of the Pieri formula in \eqref{piericorh}, we have that, for any $\lambda$,
\beq
\label{corrhs}
G^c_{\lambda; (n)} = \avg{s_\lambda (U) h_n(U^{-1})}_c =\sum_{j=1}^n h_{n-j} s_{\lambda/(j)}=\sum_{j=1}^n h_{n-j} \sum_{\nu^j} s_{\nu^j} ~,
\eeq
where the rightmost sum is over all $\nu^j$ such that $\lambda/\nu^j$ is a horizontal strip of length $j$. Take, for example, $\lambda= (3,2)$,
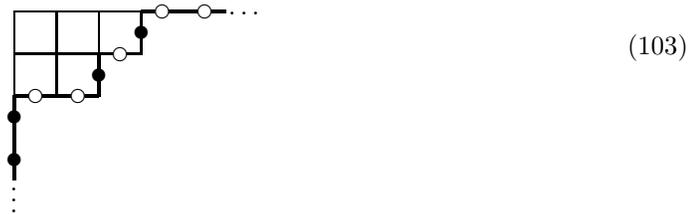
\beq
\label{y32}
\hspace{-1cm}
\rlap{\tableau{&&&\emptycell &\emptycell&\hdotscell\\ && \emptycell \\ \emptycell\\ \emptycell \\\vdotscell % \\\emptycell \\ \\\vdotscell% & % \emptycell %&\emptycell %\!\!\!\\&\emptycell\\ & \emptycell\\ \emptycell \\\vdotscell%
}}
\unitlength=\cellsize
\begin{picture}(0,0)
\thicklines
\drawline(0,-3)(0,-1)(2,-1)(2,0)(3,0)(3,1)(5,1)
\thinlines
%\put(0,-3.5){\circle*{0.3}}
\put(0,-2.5){\circle*{0.3}}
\put(0,-1.5){\circle*{0.3}}
\put(2,-0.5){\circle*{0.3}}
\put(3,+0.5){\circle*{0.3}}
{\color{white}
%\put(6.5,1){\circle*{0.3}}
%\put(5.5,1){\circle*{0.3}}
\put(4.5,1){\circle*{0.3}}
\put(3.5,1){\circle*{0.3}}
\put(2.5,0){\circle*{0.3}}
\put(1.5,-1){\circle*{0.3}}
\put(0.5,-1){\circle*{0.3}}
}
%
%\put(6.5,1){\circle{0.3}}
%\put(5.5,1){\circle{0.3}}
\put(4.5,1){\circle{0.3}}
\put(3.5,1){\circle{0.3}}
\put(2.5,0){\circle{0.3}}
\put(1.5,-1){\circle{0.3}}
\put(0.5,-1){\circle{0.3}}
\end{picture}
\eeq
which corresponds to 
\beq
\label{d32s}
\ket{(3,2)}=\cdots
\vcenterbox{\unitlength=\cellsize\begin{picture}(9,2)
%{\color{gray}\dottedline{0.1}(5,0)(5,2)}%
\linethickness{0.7pt}
\put(0,1){\line(1,0){9}}
\put(1.5,1){\circle*{0.3}}
\put(.5,1){\circle*{0.3}}
\put(4.5,1){\circle*{0.3}}
\put(6.5,1){\circle*{0.3}}
%\put(5.5,1){\circle*{0.3}}
%\put(8.5,1){\circle*{0.3}}
{\color{white}
\put(2.5,1){\circle*{0.3}}
\put(3.5,1){\circle*{0.3}}
\put(5.5,1){\circle*{0.3}}
\put(7.5,1){\circle*{0.3}}
\put(8.5,1){\circle*{0.3}}
%\put(9.5,1){\circle*{0.3}}
%\put(10.5,1){\circle*{0.3}}
}
\put(2.5,1){\circle{0.3}}
\put(3.5,1){\circle{0.3}}
\put(5.5,1){\circle{0.3}}
\put(7.5,1){\circle{0.3}}
\put(8.5,1){\circle{0.3}}
%\put(9.5,1){\circle{0.3}}
%\put(10.5,1){\circle{0.3}}
%\put(5,0){\dashbox{0.04}(0,2){}}
\end{picture}}
\cdots
\eeq
We remind the reader that the dots on the left (right) refer to an infinite string of particles (holes). We take $\lambda = (3,2)$ and $n=2$. For this choice of $\lambda$, the diagrams contributing to the sum over $\nu^1 $ and $\nu^2$ in \eqref{corrhs} are given in equation \eqref{y32pieri1}.  

We assign particles and holes to these diagrams and remind ourselves that
\beq
\label{goes}
G_{(1);\emptyset} (\tau) = 
\langle
\cdots
\vcenterbox{\unitlength=\cellsize\begin{picture}(4,2)
\linethickness{0.7pt}
\put(0,1){\line(1,0){4}}
\put(.5,1){\circle*{0.3}}
%\put(1.5,1){\circle*{0.3}}
\put(2.5,1){\circle*{0.3}}
{\color{white}
\put(1.5,1){\circle*{0.3}}
\put(3.5,1){\circle*{0.3}}
%\put(5.5,1){\circle*{0.3}}
}
\put(1.5,1){\circle{0.3}}
\put(3.5,1){\circle{0.3}}
%\put(5.5,1){\circle{0.3}}
\end{picture}}
\cdots 
\rangle_\tau
\eeq

We then see that equation \eqref{corrhs} is given by the following expression. For each configuration, the dots on the left and right again indicate an infinite sequence of particles and holes, of which we show only the left- and rightmost, respectively.
\begin{align}
G_{(3,2);(2)} (\tau)  = & ~ \langle 
   \cdots
\vcenterbox{\unitlength=\cellsize\begin{picture}(7,2)
%{\color{gray}\dottedline{0.1}(5,0)(5,2)}%
\linethickness{0.7pt}
\put(0,1){\line(1,0){7}}
\put(.5,1){\circle*{0.3}}
\put(3.5,1){\circle*{0.3}}
\put(5.5,1){\circle*{0.3}}
%\put(5.5,1){\circle*{0.3}}
%\put(8.5,1){\circle*{0.3}}
{\color{white}
\put(1.5,1){\circle*{0.3}}
\put(2.5,1){\circle*{0.3}}
\put(4.5,1){\circle*{0.3}}
\put(6.5,1){\circle*{0.3}}
%\put(9.5,1){\circle*{0.3}}
%\put(10.5,1){\circle*{0.3}}
}
\put(1.5,1){\circle{0.3}}
\put(2.5,1){\circle{0.3}}
\put(4.5,1){\circle{0.3}}
\put(6.5,1){\circle{0.3}}
%\put(9.5,1){\circle{0.3}}
%\put(10.5,1){\circle{0.3}}
%\put(5,0){\dashbox{0.04}(0,2){}}
\end{picture}}
\cdots 
\lvert
   \cdots
\vcenterbox{\unitlength=\cellsize\begin{picture}(5,2)
%{\color{gray}\dottedline{0.1}(5,0)(5,2)}%
\linethickness{0.7pt}
\put(0,1){\line(1,0){5}}
\put(.5,1){\circle*{0.3}}
\put(3.5,1){\circle*{0.3}}
%\put(5.5,1){\circle*{0.3}}
%\put(5.5,1){\circle*{0.3}}
%\put(8.5,1){\circle*{0.3}}
{\color{white}
\put(1.5,1){\circle*{0.3}}
\put(2.5,1){\circle*{0.3}}
\put(4.5,1){\circle*{0.3}}
%\put(6.5,1){\circle*{0.3}}
%\put(9.5,1){\circle*{0.3}}
%\put(10.5,1){\circle*{0.3}}
}
\put(1.5,1){\circle{0.3}}
\put(2.5,1){\circle{0.3}}
\put(4.5,1){\circle{0.3}}
%\put(6.5,1){\circle{0.3}}
%\put(9.5,1){\circle{0.3}}
%\put(10.5,1){\circle{0.3}}
%\put(5,0){\dashbox{0.04}(0,2){}}
\end{picture}}
\cdots 
\rangle_\tau \notag \\
 = & \left( \langle  \cdots
\vcenterbox{\unitlength=\cellsize\begin{picture}(6,2)
%{\color{gray}\dottedline{0.1}(5,0)(5,2)}%
\linethickness{0.7pt}
\put(0,1){\line(1,0){6}}
\put(.5,1){\circle*{0.3}}
\put(3.5,1){\circle*{0.3}}
\put(4.5,1){\circle*{0.3}}
%\put(5.5,1){\circle*{0.3}}
%\put(8.5,1){\circle*{0.3}}
{\color{white}
\put(1.5,1){\circle*{0.3}}
\put(2.5,1){\circle*{0.3}}
\put(5.5,1){\circle*{0.3}}
%\put(6.5,1){\circle*{0.3}}
%\put(9.5,1){\circle*{0.3}}
%\put(10.5,1){\circle*{0.3}}
}
\put(1.5,1){\circle{0.3}}
\put(2.5,1){\circle{0.3}}
\put(5.5,1){\circle{0.3}}
%\put(6.5,1){\circle{0.3}}
%\put(9.5,1){\circle{0.3}}
%\put(10.5,1){\circle{0.3}}
%\put(5,0){\dashbox{0.04}(0,2){}}
\end{picture}}
\cdots 
\rangle_\tau
+
 \langle  \cdots
\vcenterbox{\unitlength=\cellsize\begin{picture}(7,2)
%{\color{gray}\dottedline{0.1}(5,0)(5,2)}%
\linethickness{0.7pt}
\put(0,1){\line(1,0){7}}
\put(.5,1){\circle*{0.3}}
\put(2.5,1){\circle*{0.3}}
\put(5.5,1){\circle*{0.3}}
%\put(5.5,1){\circle*{0.3}}
%\put(8.5,1){\circle*{0.3}}
{\color{white}
\put(1.5,1){\circle*{0.3}}
\put(3.5,1){\circle*{0.3}}
\put(4.5,1){\circle*{0.3}}
\put(6.5,1){\circle*{0.3}}
%\put(9.5,1){\circle*{0.3}}
%\put(10.5,1){\circle*{0.3}}
}
\put(1.5,1){\circle{0.3}}
\put(3.5,1){\circle{0.3}}
\put(4.5,1){\circle{0.3}}
\put(6.5,1){\circle{0.3}}
%\put(9.5,1){\circle{0.3}}
%\put(10.5,1){\circle{0.3}}
%\put(5,0){\dashbox{0.04}(0,2){}}
\end{picture}}
\cdots 
\rangle_\tau 
\right) \times G_{(1);\emptyset}(\tau) \notag\\
& + 
 \langle  \cdots
\vcenterbox{\unitlength=\cellsize\begin{picture}(6,2)
\linethickness{0.7pt}
\put(0,1){\line(1,0){6}}
\put(.5,1){\circle*{0.3}}
\put(2.5,1){\circle*{0.3}}
\put(4.5,1){\circle*{0.3}}
{\color{white}
\put(1.5,1){\circle*{0.3}}
\put(3.5,1){\circle*{0.3}}
\put(5.5,1){\circle*{0.3}}
}
\put(1.5,1){\circle{0.3}}
\put(3.5,1){\circle{0.3}}
\put(5.5,1){\circle{0.3}}
\end{picture}}
\cdots 
\rangle_\tau
+
 \langle  \cdots
\vcenterbox{\unitlength=\cellsize\begin{picture}(6,2)
\linethickness{0.7pt}
\put(0,1){\line(1,0){6}}
%\put(.5,1){\circle*{0.3}}
\put(0.5,1){\circle*{0.3}}
\put(4.5,1){\circle*{0.3}}
{\color{white}
\put(1.5,1){\circle*{0.3}}
\put(2.5,1){\circle*{0.3}}
\put(3.5,1){\circle*{0.3}}
\put(5.5,1){\circle*{0.3}}
}
\put(1.5,1){\circle{0.3}}
\put(2.5,1){\circle{0.3}}
\put(3.5,1){\circle{0.3}}
\put(5.5,1){\circle{0.3}}
\end{picture}}
\cdots 
\rangle_\tau
\end{align}
The second line above corresponds to $\nu^1$ in \eqref{corrhs}, whereas the third line corresponds to $\nu^2$. By comparing with equation \eqref{y32}, we see that the sum over $\nu^1$ contains all diagrams related to {\tiny \ydiagram{3,2}} by moving a single particle a single site to the left. The sum over $\nu^2$ consists of all ways to move one or two particles by a total of two sites, with the restriction that we cannot move two adjacent particles. This is the interpretation of the corollary of the Pieri formula in \eqref{piericorh}, and it generalizes to $h_n $ for any $n\in\mathds{Z}^+$. In particular, we can assign a particle-hole interpretation to $s_{\lambda/(n)}$ by noting that it is given by summing over all distinct ways to move $\leq n$ particles a total of $n$ sites to the left without moving any two adjacent particles.
%have the following. %assigning a spin configuration to some configuration $\lambda$ gives the following.
%
%\beq
%\label{corrhsi}
%s_{\lambda/(n)}  = \sum   \eft\{ \begin{tabular}{ll}
%Distinct ways to move $\leq n$ particles a total of %$n$ sites\\
%to the left without moving any two adjacent %particles.\\
%   \end{tabular}
%    \right\}
%\eeq
%begin{align*}
%s_{\lambda/(n)}  = \sum  & \{ \text{Distinct ways to move %$\leq n$ particles $n$ sites to the left} \\ \vspace{-.2cm}
%& \text{without moving any two adjacent spins.}  \}
%\end{align*}
%
We can apply the same reasoning to $\avg{s_\lambda (U) e_n(U^{-1}}$ by using \eqref{piericore}. Similar to \eqref{corrhs}, we have 
 \beq
\label{corres}
G_{\lambda; (1^n)} =\sum_{j=1}^n e_{n-j} s_{\lambda/(1^j)} =\sum_{j=1}^n e_{n-j} \sum_{\nu^j} s_\nu ~,
\eeq
where $\sum_{\nu^j}$ is now a sum over all $\nu$ such that $\lambda/\nu$ is a vertical strip of length $j$. The skew Schur polynomial $s_{\lambda/(1^n)}$ can be interpreted as taking the configuration corresponding to $\lambda$ and summing over all distinct ways to move $ n$ particles a single site to the left.

Consider the case where $x=y$ (e.g. $\tau = \beta \in \mathds{R}$ and $a_k = a_{-k}$). We can then apply the Pieri formula once again, in this case to the products $h_{n-j}s_{\lambda/(j)}$ in \eqref{corrhs}, as
\beq
h_{n-j}s_{\lambda/(j)} = \sum_\nu s_\nu~,
\eeq
where the sum is over all $\nu$ obtained from $\lambda$ by removing a horizontal strip of size $j$ and then adding to the resuting diagram a horizontal strip of size $n-j$. We consider again $G_{(3,2);(2)} = s_{(3,2)/(1)} h_1 +  s_{(3,2)/(2)}$, where $(3,2)/(2) = (2,1) + (3)$. Applying the Pieri formula to $s_{(3,2)/(1)} h_1 = s_{(3,2)/(1)} s_{(1)} $ gives the following diagrams, where the cell that is added again indicated in gray. It is clear that $\lambda = (3,2)$ appears twice, as there are two ways to remove (and then add again) a single cell from $\lambda$.%, given below \eqref{d32s}.
 \vspace{.1cm}
\beq
\hbox{
  \begin{ytableau}
 *(white) & *(white) & *(gray)  \\ 
*(white)  & *(white)  
\end{ytableau} 
\hspace{.4cm}
  \begin{ytableau}
 *(white) & *(white)   \\ 
*(white)  & *(white)  \\
*(gray)
\end{ytableau} 
\hspace{.4cm}
  \begin{ytableau}
 *(white) & *(white) & *(white)  & *(gray)  \\ 
*(white)  
\end{ytableau} 
\hspace{.4cm}
  \begin{ytableau}
 *(white) & *(white) & *(white)   \\ 
*(white)   & *(gray)
\end{ytableau} 
\hspace{.4cm}
  \begin{ytableau}
 *(white) & *(white) & *(white)   \\ 
*(white) \\
  *(gray)
\end{ytableau} 
}
\eeq
%\end{center} 
%\vspace{.2cm}
%
We now briefly consider the single particle correlation functions $F_{j;l}$. The generating function $f(z) = \sum_j F_{j;l}z^{j-l}$ for one-particle correlations is written in \eqref{spinwfct}. Writing again $f(z) =H(x;z)H(x;z^{-1})$ for some $x$, we have
\beq
f(z) = \prod_j(1-x_jz)^{-1}(1-y_j z^{-1})^{-1} =\sum_{j,k=0}^\infty h_j(x) h_k(y) z^{j-k} \eqqcolon \sum_{m=0}^\infty  d_m(z^m+z^{-m})~, 
\eeq 
where $h_j$ are complete homogeneous symmetric polynomials. Then,
\beq
\label{dmh}
d_m=\sum_{j=m}^\infty h_j(y)h_{j-m}(x) = \avg{h_M(U) h_{M-m}(U^{-1})} = G_{j;l}(\tau)~~,~~~ j-l=m  ~. 
\eeq
where $M$ is the largest number $k$ such that $h_k\neq 0$, which is typically infinite. Remember that the $G_{j;l}(\tau)$ are the single-particle wavefunctions at site $j$ and time $\tau$ for a particle released from site $l$. The last two equalities in \eqref{dmh} relate $F_{j;l}(\tau)$ to single-particle wavefunctions at site $M$ and time $\tau$ for a particle released from site $M-m$. Since $M$ is infinite for most choices of hopping parameters $a_k\sim p_k$, this correspond to releasing a particle at infinity and finding its wavefunction $m$ lattice sites away from where it was released at some Euclidean time $\tau$. The fact that $h_{M-m}(U^{-1}) = G_{j;l}(\tau)$ for $M\to \infty$ agrees with the intuition that a single particle at infinity does not feel the presence of the other remaining particles, which are infinitely far away from it.

Lastly, we note that one can apply the Jacobi-Trudi identity \eqref{jtid} to $h_n$ and $e_n$ to find any skew Schur polynomials on the right hand side of \eqref{largen}. This, in effect, provides a way to compute any correlation function. Although the resulting expressions generally grow quickly as one considers large partitions, the application of this method itself is rather simple as it only requires the computation of a determinant. In \ref{subsecpsbs}, we will outline various other, closely interrelated, methods to calculate any $\avg{s_\lambda  s_\mu }$, based on power sum polynomials rather than elementary and complete homogeneous symmetric polynomials. 
%
%
%
%
%
%
%
%The involution between $e_n$ and $h_n$ and q
 \subsubsection{Quasi-local particle-hole duality}
\label{secqlph}
Remember that one can replace the generating function $f_1(z) =  H(x;z) H(y;z^{-1})$ by $f_2(z) =  E(x;z) E(y;z^{-1})$ at the cost of transposing the partitions appearing in the correlation functions. This leads to a duality between models related by $a_n \to (-1)^{n+1}a_n$, as we demonstrate here. Though mathematically trivial, the physical interpretation of this duality is quite surprising. As mentioned at the start of this section, this duality holds for finite as well as infinite $N$. From \eqref{genfuncsymp}, it follows that switching between $f_1(z) $ and $f_2(z)$ corresponds to taking
\beq
p_k(x) \to (-1)^{k+1} p_k(x)~.
\eeq
From equations \eqref{hpid} and \eqref{epid}, it is clear that replacing as $p_k(x) \to (-1)^{k+1} p_k(x)$ corresponds to $a_k \to (-1)^{k+1} a_k$. In particular, we consider two hamiltonians for the same $a_n$,
\begin{align}
\label{invham}
\hat{H}_1 & =- \sum_{m=0}^{\infty}\sum_{n}a_n \left(\sigma^-_m \sigma_{m+n}^+ + \sigma_m^-\sigma_{m-n}^+  \right)~, \notag \\
\hat{H}_2 & =- \sum_{m=0}^{\infty}\sum_{n} (-1)^{n+1} a_n \left(\sigma^-_m \sigma_{m+n}^+ + \sigma_m^-\sigma_{m-n}^+  \right)~.
\end{align}
Correlation functions for $\hat{H}_1$ correspond to weight function $f_1(z) $, whereas those for $\hat{H}_2$ correspond to $f_2(z) $. Let us compare correlation functions for $\hat{H}_1$ and $\hat{H}_2$, which we will write as $
F^{(1)}_{\lambda;\mu} (\tau) $ and $
F^{(2)}_{\lambda;\mu} (\tau) $, respectively. From Szeg\H{o}'s theorem, we know that $F_{\emptyset;\emptyset}$ only depends on $p_k(x) p_k(y) $, so taking $p_k \to (-1)^{k+1}p_k$, has no effect on $F_{\emptyset;\emptyset}$. Therefore,
\beq
F^{(1)}_{\emptyset;\emptyset} (\tau) = F^{(2)}_{\emptyset;\emptyset} (\tau)  ~,
\eeq
that is, the return probability for $N$ adjacent particles for $\hat{H}_1$ is identical to that of $\hat{H}_2$ for any $\tau$. Moreover, by comparing \eqref{largen} and \eqref{largen2}, we see that, for general $\lambda$ and $\mu$,
\beq
\label{qlph}
\boxed{\hspace{.2cm}
F^{(1)}_{\lambda;\mu} (\tau)  = F^{(2)}_{\lambda^t;\mu^t} (\tau)  ~. \hspace{.15cm}}
\eeq
This establishes the following duality between correlations functions of $\hat{H}_1$ and $\hat{H}_2$. It is well known that transposition of diagrams induces a particle-hole and parity transformation on the corresponding particle-hole configuration. This simply follows from the fact that transposition exchanges left and right, and that it maps vertical edges to horizontal ones (and vice versa). However, we are not implementing particle-hole symmetry in \eqref{qlph} but instead we establish an equality between different correlation functions corresponding to different models. Let us consider the configuration associated to $\lambda = (5,4,2,1^3)$ and $\lambda^t=(6,3,2^2,1)$, given on the left and right below, respectively.
 \vspace{.4cm}
 \beq
\label{ydinvtr}
\mkern-18mu\mkern-18mu\mkern-18mu\mkern-18mu\mkern-18mu\mkern-18mu
\rlap{\tableau{&&&&&\emptycell &\hdotscell\\ &&&& \emptycell \\& \emptycell\\ & \emptycell \\&\emptycell\\&\emptycell\\ \emptycell\\ \vdotscell 
}}
\unitlength=\cellsize
\begin{picture}(0,0)
\thicklines
\drawline(0,-6)(0,-5)(1,-5)(1,-2)(2,-2)(2,-1)(4,-1)(4,0)(5,0)(5,1)(6,1)
\thinlines
\drawline(.6,-5.6)(-.5,-4.5)
\drawline(5.6,0.4)(4.5,1.5)
\put(0,-5.5){\circle*{0.3}}
\put(1,-4.5){\circle*{0.3}}
\put(1,-3.5){\circle*{0.3}}
\put(1,-2.5){\circle*{0.3}}
\put(2,-1.5){\circle*{0.3}}
\put(4,-.5){\circle*{0.3}}
\put(5,.5){\circle*{0.3}}
{\color{white}
\put(0.5,-5){\circle*{0.3}}
\put(1.5,-2){\circle*{0.3}}
\put(1.5,-2){\circle*{0.3}}
\put(2.5,-1){\circle*{0.3}}
\put(3.5,-1){\circle*{0.3}}
\put(4.5,0){\circle*{0.3}}
\put(5.5,1){\circle*{0.3}}
}
\put(0.5,-5){\circle{0.3}}
\put(1.5,-2){\circle{0.3}}
\put(1.5,-2){\circle{0.3}}
\put(2.5,-1){\circle{0.3}}
\put(3.5,-1){\circle{0.3}}
\put(4.5,0){\circle{0.3}}
\put(5.5,1){\circle{0.3}}
\end{picture}
\qquad\qquad\qquad\qquad\qquad\qquad\qquad
\rlap{\tableau{&&&&&&\emptycell &\hdotscell\\ &&& \emptycell \\&& \emptycell\\ && \emptycell \\&\emptycell\\ \emptycell\\ \emptycell \vdotscell % \\\emptycell \\ \\\vdotscell% & % \emptycell %&\emptycell %\!\!\!\\&\emptycell\\ & \emptycell\\ \emptycell \\\vdotscell%
}}
\unitlength=\cellsize
\begin{picture}(0,0)
\thicklines
\drawline(0,-5)(0,-4)(1,-4)(1,-3)(2,-3)(2,-1)(3,-1)(3,0)(6,0)(6,1)(7,1)
\thinlines
\drawline(.6,-4.6)(-.5,-3.5)
\drawline(6.6,0.4)(5.5,1.5)
\put(0,-4.5){\circle*{0.3}}
\put(1,-3.5){\circle*{0.3}}
\put(2,-2.5){\circle*{0.3}}
\put(2,-1.5){\circle*{0.3}}
\put(3,-.5){\circle*{0.3}}
\put(6,.5){\circle*{0.3}}
{\color{white}
\put(0.5,-4){\circle*{0.3}}
\put(1.5,-3){\circle*{0.3}}
\put(2.5,-1){\circle*{0.3}}
\put(3.5,0){\circle*{0.3}}
\put(4.5,0){\circle*{0.3}}
\put(5.5,0){\circle*{0.3}}
\put(6.5,1){\circle*{0.3}}
%\put(7.5,1){\circle*{0.3}}
}
\put(0.5,-4){\circle{0.3}}
\put(1.5,-3){\circle{0.3}}
\put(2.5,-1){\circle{0.3}}
\put(3.5,0){\circle{0.3}}
\put(4.5,0){\circle{0.3}}
\put(5.5,0){\circle{0.3}}
\put(6.5,1){\circle{0.3}}
%\put(7.5,1){\circle{0.3}}
\end{picture}
\eeq
These correspond to the following configurations, 
\begin{align}
\label{transpdiag2}
    \ket{\lambda} = &  \cdots
\vcenterbox{\unitlength=\cellsize\begin{picture}(13,2)
\linethickness{0.7pt}
\drawline(1,.3)(1,1.7)
\drawline(12,.3)(12,1.7)
\put(0,1){\line(1,0){13}}
\put(.5,1){\circle*{0.3}}
\put(2.5,1){\circle*{0.3}}
\put(3.5,1){\circle*{0.3}}
\put(4.5,1){\circle*{0.3}}
\put(6.5,1){\circle*{0.3}}
\put(9.5,1){\circle*{0.3}}
\put(11.5,1){\circle*{0.3}}
{\color{white}
\put(1.5,1){\circle*{0.3}}
\put(5.5,1){\circle*{0.3}}
\put(7.5,1){\circle*{0.3}}
\put(8.5,1){\circle*{0.3}}
\put(10.5,1){\circle*{0.3}}
\put(12.5,1){\circle*{0.3}}
}
\put(1.5,1){\circle{0.3}}
\put(3.5,1){\circle{0.3}}
\put(5.5,1){\circle{0.3}}
\put(7.5,1){\circle{0.3}}
\put(8.5,1){\circle{0.3}}
\put(10.5,1){\circle{0.3}}
\put(12.5,1){\circle{0.3}}
\end{picture}}
\cdots  ~, \notag\\
        \ket{\lambda^t} = &   \cdots
\vcenterbox{\unitlength=\cellsize\begin{picture}(13,2)
\linethickness{0.7pt}
\drawline(1,.3)(1,1.7)
\drawline(12,.3)(12,1.7)
\put(0,1){\line(1,0){13}}
\put(.5,1){\circle*{0.3}}
\put(2.5,1){\circle*{0.3}}
\put(4.5,1){\circle*{0.3}}
\put(5.5,1){\circle*{0.3}}
\put(7.5,1){\circle*{0.3}}
\put(11.5,1){\circle*{0.3}}
{\color{white}
\put(1.5,1){\circle*{0.3}}
\put(3.5,1){\circle*{0.3}}
\put(6.5,1){\circle*{0.3}}
\put(8.5,1){\circle*{0.3}}
\put(9.5,1){\circle*{0.3}}
\put(10.5,1){\circle*{0.3}}
\put(12.5,1){\circle*{0.3}}
}
\put(1.5,1){\circle{0.3}}
\put(3.5,1){\circle{0.3}}
\put(6.5,1){\circle{0.3}}
\put(8.5,1){\circle{0.3}}
\put(9.5,1){\circle{0.3}}
\put(10.5,1){\circle{0.3}}
\put(12.5,1){\circle{0.3}}
\end{picture}}
\cdots  ~.
\end{align}
where the vertical lines on the left and right correspond to the diagonal lines at the lower left and upper right corners of $\lambda$ and $\lambda^t$ in \eqref{ydinvtr}, respectively. We see that the regions in between the vertical lines for $\ket{\lambda} $ and $\ket{\lambda^t}$ are related by performing a parity transformation and exchanging between particles and holes. Only a finite interval $\ell(\lambda) \leq r \leq \lambda_1$ is affected non-trivially by this combination of particle-hole and parity transformations. We thus establish a bijection between the correlation functions corresponding to $\hat{H}_1$ and $\hat{H}_2$, given by the transposition of diagrams, which we refer to as \emph{quasi-local particle-hole duality}. It follows trivially from the above treatment that if $
a_{2k} =0~~,~~~\forall~ k \in \mathds{Z}^+~$, then $H(x;z)  = E(x;z)$, as can be seen immediately from \eqref{genfuncsymp}. It follows that when all even hopping parameters $a_{2k}$ are zero, we have $F^{(1)}_{\lambda;\mu} = F^{(2)}_{\lambda;\mu}  $, so we can suppress the superscript. In this case, it follow from \eqref{qlph} that
\beq
F_{\lambda;\mu} (\tau)  =F_{\lambda^t;\mu^t} (\tau)  ~. 
 \eeq
 That is, any system with $a_{2k}=0$ satisfies quasi-local particle-hole duality with itself.

\subsection{Power sum symmetric polynomials and border strips}
\label{subsecpsbs}
  We consider now the particle-hole configurations corresponding to power sum polynomials $p_k$, which are a very natural basis for the application to LRRW models due to their proportionality to $\tau a_k$. Consider specifically the case where $\tau = it$. Then, from \eqref{hpid}, we have
		\beq
		\label{pkpkcc}
		p_k(y) = -it k a_k^*  = (-p_k(x))^* = e^{i \phi} p_k(x)~~,~~~\phi \coloneqq \pi-2\arg(p_k(x))~.
		\eeq
		Since $p_k(x)$ only differs from $p_k(y)$ by a phase, we can apply the usual multiplication rules to objects of the form $p_\rho(x) p_\mu(y)$, which is an advantage over the usage of $h_j$, $e_j$. 
  
  Remember that power sum polynomials are expanded as an alternating sum over hook-shaped Schur polynomials, see equation \eqref{powersumschur}. Note that 
\beq
\avg{\tra U^n}_\tau =  p_n =  \sum_{r=0}^{n-1}(-1)^r G_{(n-r,1^r);\emptyset}~,
\eeq
can be read off from the hamiltonian by the identification in either \eqref{hpid} or \eqref{epid}. As an example, the configuration corresponding to a hook shape $\lambda= (4,1^2)$ is as follows.
\vspace{.1cm}
\beq
\label{yhookxmp}
\hspace{-2cm}
\rlap{\tableau{&&&&\emptycell &\emptycell&\hdotscell\\ &\emptycell \\ &\emptycell\\ \emptycell \\\vdotscell % \\\emptycell \\ \\\vdotscell% & % \emptycell %&\emptycell %\!\!\!\\&\emptycell\\ & \emptycell\\ \emptycell \\\vdotscell%
}}
\unitlength=\cellsize
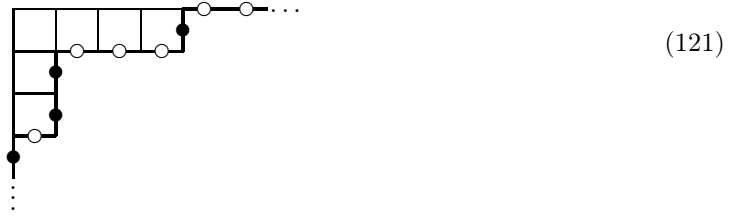
\begin{picture}(0,0)
\thicklines
\drawline(0,-3)(0,-2)(1,-2)(1,0)(2,0)(4,0)(4,1)(6,1)
\thinlines
%\put(0,-3.5){\circle*{0.3}}
\put(0,-2.5){\circle*{0.3}}
\put(1,-1.5){\circle*{0.3}}
\put(1,-0.5){\circle*{0.3}}
\put(4,+0.5){\circle*{0.3}}
{\color{white}
%\put(6.5,1){\circle*{0.3}}
\put(5.5,1){\circle*{0.3}}
\put(4.5,1){\circle*{0.3}}
\put(3.5,0){\circle*{0.3}}
\put(2.5,0){\circle*{0.3}}
\put(1.5,0){\circle*{0.3}}
\put(0.5,-2){\circle*{0.3}}
}
%
%\put(6.5,1){\circle{0.3}}
\put(5.5,1){\circle{0.3}}
\put(4.5,1){\circle{0.3}}
\put(3.5,0){\circle{0.3}}
\put(2.5,0){\circle{0.3}}
\put(1.5,0){\circle{0.3}}
\put(0.5,-2){\circle{0.3}}
\end{picture}
\eeq
%
% moving it $a$ sites to the right, then taking the next $b$ particles and moving them a single site to the right. 
That is, a hook-shaped diagram corresponds to taking the particle at site $-b$ and moving it $a+b$ sites to the right. More generally, consider 
\beq
%\frac{1}{\sqrt{n}}
\bket{p_n}{s_\lambda} = %\frac{1}{\sqrt{n}}
\sum_{r=0}^{n-1}(-1)^r  G^c_{\lambda ;(n-r,1^r) }= %\frac{1}{\sqrt{n}} 
\sum_{r=0}^{n-1}(-1)^r  \bket{(n-1,1^r)}{\lambda} ~.
\eeq
%Note that $\bket{p_n}{s_\lambda}$ is not normalized. % where the factor $ \frac{1}{\sqrt{n}}$ arises from normalization. 
Using equation \eqref{pnskew1}, this is given by
\beq
\label{corrps}
%\frac{1}{\sqrt{n}}
\sum_{r=0}^{n-1}(-1)^r s_{\lambda/(n-r,1^r)}=\sum_\eta (-1)^{\hg(\eta)}s_{\lambda \setminus \eta}~,  %\frac{1}{\sqrt{n}} 
%\sum_\nu (-1)^{\hg(\lambda/\nu)}s_\nu~,
\eeq
where the sum is over all $\eta$ which are border strips of size $n$, see e.g. the example in \eqref{bsremxmp} for $n=4$. Note that there is only a single Schur polynomial appearing on the right hand side, as opposed to \eqref{corrhs} and \eqref{corres}, which contain factors of $e_{n-j}$ and $h_{n-j}$, respectively. We consider a few specific examples of \eqref{corrps}, already noted in \cite{tftis}, before treating it in generality. In particular, consider $ \avg{s_\lambda p_n}_c$ with $(n-r,1^r) \nsubseteq \lambda ~\forall~ r \in \{ 0,\dots, n-1\}$, such that $\lambda_1+\lambda_1^t-1 < n$. This gives
	\beq 
	\label{tralambda}
	\avg{ \tra U^{-n} \tra_\lambda U}_c =0~.
	\eeq
Writing $\lambda $ in Frobenius notation as $\lambda = (a_1,\dots ,a_k| b_1,\dots,b_k)$ with $a_j$ satisfying $a_1> \dots >a_k$ and similar for $b_j$. In this case, $a_1+b_1+1$ equals the hook-length of the top left cell of $\lambda$. Equation \eqref{tralambda} then states that $\avg{ \tra U^{-n} \tra_\lambda U}_c =0 $ if $ a_1+b_1 +1<n$. In terms of particle-hole configurations, \eqref{tralambda} states that $\sum_{r=0}^{n-1}(-1)^r  G^c_{\lambda ;(n-r,1^r) } = 0$ if $n$ is greater than the distance (in units of lattice spacing) between the leftmost hole and rightmost particle in the configuration corresponding to $\lambda$ . 
	
Further, write $m = a_1+b_1+1-n $ and consider $\lambda=(a|b)=(a_1,\dots,a_k|b_1,\dots b_k)$ such that $m \leq a_1-a_2-1 $ and $m \leq b_1 -b_2-1 $, respectively. % or, equivalently, $m\leq \lambda_1 -\lambda_2$ and $m\leq \lambda_1^t - \lambda_2^t$, respectively.
Take $\mu = (a_2,\dots,a_k|b_2,\dots,b_k) $, obtained by removing the first row and column from $\lambda$. %If $a_2+b_2+1 -n < 0 $ 
Any $(n-r,1^r) \subseteq \lambda$ then satisfies
	\beq
	\lambda/(n-r,1^r) = (a_1+1,1^{b_1}) /(n-r,1^r) \otimes \mu = (a_1+1-n+r)\otimes (1^{b_1-r}) \otimes \mu~.
	\eeq
Consider the example of $(6,4,3,1^2)/(5,1^3)$, which is given by the diagram below.
%
%\begin{center}
\vspace{.1cm}
\beq
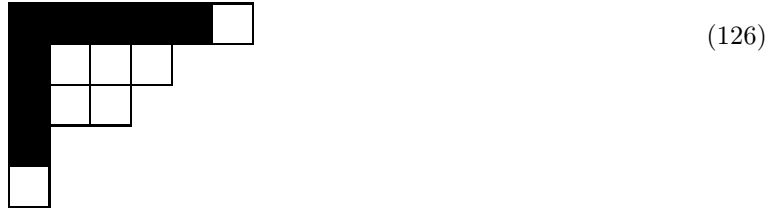
\hbox{
\begin{ytableau}
 *(black) &  *(black) &  *(black) &  *(black) &  *(black) &  *(white)  \\ 
 *(black) &   *(white) & *(white) &   *(white)\\
 *(black) &   *(white) & *(white) \\
 *(black)  \\
  *(white)  
\end{ytableau} 
}
\eeq
%\end{center} 
%	\vspace{.3cm}
	
	We then apply \eqref{ehzero} to find
	\beq
	\label{traidlong}
	\avg{s_\lambda (U) \tra U^{-n} }_c =    \sum_{r=n-a_1-1}^{b_1} (-1)^r s_{\lambda/(n-r,1^r)}  =\pm  s_\mu \sum_{k=0}^{m} (-1)^k  h_{m-k}e_k =0~.
	\eeq
Assigning particle-hole configurations to the diagrams above, we see that the correlation function corresponding to $	\avg{s_\lambda (U) \tra U^{-n} }_c$ vanishes when the distance between the leftmost hole and rightmost particle, as well as the distance between the rightmost particle and the second-to-rightmost particle (and vice versa for holes), are sufficiently small compared to $n$.

We now consider \eqref{corrps} more generally. In this expression, $\nu$ is related to $\lambda$ by the removal of a border strip, i.e. a connected skew diagram not containing a subdiagram that is a 2 by 2 block. In terms of particle-hole configurations, a 2 by 2 block corresponds to moving two adjacent particles by two sites to the right, see below.
 \beq
\label{y22}
\hspace{-.7cm}
\rlap{\tableau{&&\emptycell &\emptycell&\hdotscell\\ && \emptycell \\ \emptycell\\ \emptycell \\\vdotscell % \\\emptycell \\ \\\vdotscell% & % \emptycell %&\emptycell %\!\!\!\\&\emptycell\\ & \emptycell\\ \emptycell \\\vdotscell%
}}
\unitlength=\cellsize
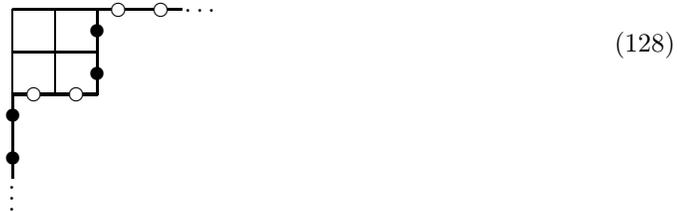
\begin{picture}(0,0)
\thicklines
\drawline(0,-3)(0,-1)(2,-1)(2,1)(4,1)
\thinlines
\put(0,-2.5){\circle*{0.3}}
\put(0,-1.5){\circle*{0.3}}
\put(2,-0.5){\circle*{0.3}}
\put(2,0.5){\circle*{0.3}}
{\color{white}
\put(3.5,1){\circle*{0.3}}
\put(2.5,1){\circle*{0.3}}
\put(1.5,-1){\circle*{0.3}}
\put(0.5,-1){\circle*{0.3}}
}
\put(3.5,1){\circle{0.3}}
\put(2.5,1){\circle{0.3}}
\put(1.5,-1){\circle{0.3}}
\put(0.5,-1){\circle{0.3}}
\end{picture}
\eeq
In equation \eqref{corrps}, the fact that a border strip has no 2 by 2 subdiagram therefore states that (the configuration corresponding to) $\nu$ is related to $\lambda$ by moving (a) particle(s) left by $n$ sites without moving two or more adjacent particles by two or more sites. The number of rows that $\lambda/\nu$ occupies (which is $\hg(\lambda/\nu)+1$) equals the number of particles that are involved in this process, which follows immediately from the fact that vertical edges on the boundary of a diagram correspond to particles. From the fact that the border strip is connected, it follows that it occupies only consecutive rows. For example, the skew diagram on the left is a border strip, whereas the one on the right clearly is not.
\vspace{.1cm}
% (naturally, this is a necessary but not sufficient condition for a border strip)
%
\beq
\ydiagram{3+3,3+1,1+3,2} \hspace{2cm}
\ydiagram{4+2,0,1+3,2}
\eeq
\vspace{.1cm}
\vspace{-.4cm}

This means that only consecutive (but not necessarily adjacent) particles are affected. However, connectedness is a stronger condition, and in the case of border strips this leads to the following observation. For a skew diagram $\lambda/\nu$, we call the \emph{outer rim} the (horizontal and vertical) edges on the bottom right of the diagram of $\lambda/\nu$, as in e.g. \cite{white}. Conversely, the \emph{inner rim} consists of the edges on the top left of $\lambda/\nu$. For a border strip containing $n$ cells, we number the edges of the inner and outer rims by $j=\{1,\dots,n+1 \}$. We will refer to the $j^\text{th}$ edge on the outer rim as the $j^\text{th}$ outer edge, and likewise for the inner rim. We consider an explicit example, where $\lambda = (8,6^2,4,1)$ and $\nu = (5^2,3,2,1)$, leading to a border strip of size $n=9$ below. On the right hand side, we number the vertical edges of the inner and outer rims of $\lambda/\nu$.
\vspace{.3cm}
%
% \begin{center}
 %\ytableausetup{notabloids}
 \beq
 \begin{ytableau}
 *(black) &  *(black) &  *(black) &  *(black) &  *(black) &  *(white) &  *(white) &  *(white) \\
  *(black) &  *(black) &  *(black) &  *(black) &  *(black) &  *(white) \\
  *(black) &  *(black) &  *(black) &  *(white)  &  *(white)  &  *(white) \\
   *(black) &  *(black) &  *(white)  &  *(white)   \\
      *(black)
\end{ytableau} 
%\hso\hso\hso {\Large    =} 
\hspace{.3cm}
\begin{ytableau}
\none & \none &  \none  & \none[ \,\,\, \scriptstyle 7] & {} &{} &{} & \none[ \!\!\! \scriptstyle 10] \\
\none & \none &  \none  & \none[ \,\,\, \scriptstyle 6] & {} & \none[ \!\!\! \scriptstyle 7]  \\
\none & \none[ \,\,\, \scriptstyle 3] & {} & {}& {}& \none[ \!\!\! \scriptstyle 6]  \\ \none[ \,\,\, \scriptstyle 1] & {} & {} & \none[ \!\!\! \scriptstyle 3] 
\end{ytableau}
\eeq
\vspace{.1cm}
 %\end{center}
\vspace{-.4cm}

One can see in the above example that the first edge on the inner rim and the ledge edge on the outer rim are both vertical, which is clearly true for any (skew) diagram. Further, we see that all other vertical edges occupy the same positions on the inner and outer rims, namely $\{3,6,7\}$. If e.g. the third edge on the inner rim were horizontal, the resulting skew diagram would be disconnected and therefore not a valid border strip, shown below.
%
%\begin{center}
\beq
\begin{ytableau}
\none & \none &  \none   & {} &{} &{}  \!\!\!  \\
\none & \none &  \none   & {}   \\
\none  & \none  &  {}& {}  \\  
{} & {} 
\end{ytableau}
\eeq
 %\end{center}
%\vspace{.3cm}
%
\vspace{-.4cm}

It is clear that this holds generally for border strips. In particular, the inner and outer edges of any border strip are identical for $j=2,3,\dots,n$, and the outer (inner) edge for $j=1$ is horizontal (vertical), whereas the outer (inner) edge for $j=n+1$ is vertical (horizontal). 

 The configurations corresponding to $\lambda$ and $\nu$ are related by taking the configuration on the outer rim of $\lambda/\nu$ (as a subset of the outer rim of $\lambda$) and replacing it by the configuration corresponding to the inner rim of $\lambda/\nu$. What we effectively get is the following. We take a particle at some site $k$ and move it to site $k-n$, and we get a factor $(-1)^{\hg(\lambda/\nu)}$. Here, $\hg(\lambda/\nu)$ equals the number of particles the affected particle jumps over, that is, the number of particles occupying sites $\{ k-n+1,\dots,k-1\}$. We thus see that $(-1)^{\hg(\lambda/\nu)}$ simply implements \emph{fermionic statistics}. Below, we show the border strip $\lambda/\nu$ with empty cells as a subset of $\lambda$, where we indicate particles and holes. The inner edge $j$ and outer edge $j$ for $j=2,\dots ,n $ are connected by diagonal lines.%, it is clear that they are identical. Further, the outer (inner) rim starts with a horizontal (vertical) edge, and ends with a vertical (horizontal) edge. 
 \vspace{.1cm}
\beq
\!\!\!\!\!\!\!\!\!\!\!\!\!\!\!\!\!\!\!\!\!\!\!\!\!\!\!\!\!\!\!\!\!\!\!\!\!\!\!\!\!\!\!\!\!\!\!\!\!\!\!\!\!\!\!\!\!\!\!\!\!\!\!\!
{\rlap{\tableau{&&&&&\emptycell\emptycell\emptycell\emptycell&\emptycell&\emptycell&\emptycell&\hdotscell\\&&&&&\emptycell\\&&&\emptycell\\&&\emptycell\\&\emptycell\\\emptycell \\\vdotscell%
}}
\unitlength=\cellsize
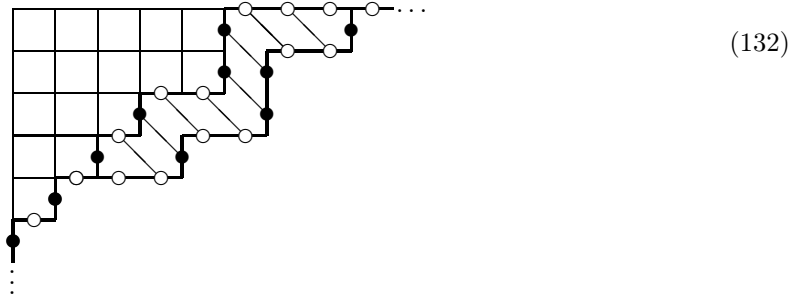
\begin{picture}(0,0)
\thicklines
\drawline(0,-5)(0,-4)(1,-4)(1,-3)(4,-3)(4,-2)(6,-2)(6,0)(8,0)(8,1)(9,1)
\drawline(0,-5)(0,-4)(1,-4)(1,-3)(2,-3)(2,-2)(3,-2)(3,-1)(5,-1)(5,1)(9,1)
\thinlines
\drawline(3.5,-3)(2.5,-2)
\drawline(4,-2.5)(3,-1.5)
\drawline(4.5,-2)(3.5,-1)
\drawline(5.5,-2)(4.5,-1)
\drawline(6,-1.5)(5,-.5)
\drawline(6,-.5)(5,.5)
\drawline(6.5,0)(5.5,1)
\drawline(7.5,0)(6.5,1)
\drawline(6.5,0)(5.5,1)
%
%
%\put(7.5,0){\vector(-1,1){.88}}
%\put(6.5,0){\vector(-1,1){.88}}
%\put(6,-.5){\vector(-1,1){.88}}
%\put(6,-1.5){\vector(-1,1){.88}}
%\put(5.5,-2){\vector(-1,1){.88}}
%
%
%
\put(0,-4.5){\circle*{0.3}}
\put(1,-3.5){\circle*{0.3}}
\put(4,-2.5){\circle*{0.3}}
\put(6,-1.5){\circle*{0.3}}
\put(6,-0.5){\circle*{0.3}}
\put(8,0.5){\circle*{0.3}}
{\color{white}
\put(.5,-4){\circle*{0.3}}
\put(1.5,-3){\circle*{0.3}}
\put(2.5,-3){\circle*{0.3}}
\put(3.5,-3){\circle*{0.3}}
\put(4.5,-2){\circle*{0.3}}
\put(5.5,-2){\circle*{0.3}}
\put(6.5,0){\circle*{0.3}}
\put(7.5,0){\circle*{0.3}}
\put(8.5,1){\circle*{0.3}}
}
\put(.5,-4){\circle{0.3}}
\put(1.5,-3){\circle{0.3}}
\put(2.5,-3){\circle{0.3}}
\put(3.5,-3){\circle{0.3}}
\put(4.5,-2){\circle{0.3}}
\put(5.5,-2){\circle{0.3}}
\put(6.5,0){\circle{0.3}}
\put(7.5,0){\circle{0.3}}
\put(8.5,1){\circle{0.3}}
\put(0,-4.5){\circle*{0.3}}
\put(1,-3.5){\circle*{0.3}}
\put(2,-2.5){\circle*{0.3}}
\put(3,-1.5){\circle*{0.3}}
\put(5,-0.5){\circle*{0.3}}
\put(5,0.5){\circle*{0.3}}
{\color{white}
\put(1.5,-3){\circle*{0.3}}
\put(2.5,-2){\circle*{0.3}}
\put(3.5,-1){\circle*{0.3}}
\put(4.5,-1){\circle*{0.3}}
\put(5.5,1){\circle*{0.3}}
\put(6.5,1){\circle*{0.3}}
\put(7.5,1){\circle*{0.3}}
}
\put(1.5,-3){\circle{0.3}}
\put(2.5,-2){\circle{0.3}}
\put(3.5,-1){\circle{0.3}}
\put(4.5,-1){\circle{0.3}}
\put(5.5,1){\circle{0.3}}
\put(6.5,1){\circle{0.3}}
\put(7.5,1){\circle{0.3}}
\end{picture}
}
\eeq
The configurations corresponding to $\lambda$ and $\nu$ are given by
\begin{align}
\label{transpdiag}
    \ket{\lambda} = &  \cdots
\vcenterbox{\unitlength=\cellsize\begin{picture}(15,2)
\linethickness{0.7pt}
%\drawline(1,.3)(1,1.7)
%\drawline(12,.3)(12,1.7)
\put(0,1){\line(1,0){15}}
\put(.5,1){\circle*{0.3}}
\put(2.5,1){\circle*{0.3}}
\put(6.5,1){\circle*{0.3}}
\put(9.5,1){\circle*{0.3}}
\put(10.5,1){\circle*{0.3}}
\put(13.5,1){\circle*{0.3}}
%\put(11.5,1){\circle*{0.3}}
{\color{white}
\put(1.5,1){\circle*{0.3}}
\put(3.5,1){\circle*{0.3}}
\put(4.5,1){\circle*{0.3}}
\put(5.5,1){\circle*{0.3}}
\put(7.5,1){\circle*{0.3}}
\put(8.5,1){\circle*{0.3}}
\put(11.5,1){\circle*{0.3}}
\put(12.5,1){\circle*{0.3}}
\put(14.5,1){\circle*{0.3}}
}
\put(1.5,1){\circle{0.3}}
\put(3.5,1){\circle{0.3}}
\put(4.5,1){\circle{0.3}}
\put(5.5,1){\circle{0.3}}
\put(7.5,1){\circle{0.3}}
\put(8.5,1){\circle{0.3}}
\put(11.5,1){\circle{0.3}}
\put(12.5,1){\circle{0.3}}
\put(14.5,1){\circle{0.3}}
\end{picture}}
\cdots  ~, \notag\\
        \ket{\nu} = &  \cdots
\vcenterbox{\unitlength=\cellsize\begin{picture}(15,2)
\linethickness{0.7pt}
%\drawline(1,.3)(1,1.7)
%\drawline(12,.3)(12,1.7)
\put(0,1){\line(1,0){15}}
\put(.5,1){\circle*{0.3}}
\put(2.5,1){\circle*{0.3}}
\put(4.5,1){\circle*{0.3}}
\put(6.5,1){\circle*{0.3}}
\put(9.5,1){\circle*{0.3}}
\put(10.5,1){\circle*{0.3}}
\put(13.5,1){\circle*{0.3}}
%\put(11.5,1){\circle*{0.3}}
{\color{white}
\put(1.5,1){\circle*{0.3}}
\put(3.5,1){\circle*{0.3}}
\put(5.5,1){\circle*{0.3}}
\put(7.5,1){\circle*{0.3}}
\put(8.5,1){\circle*{0.3}}
\put(11.5,1){\circle*{0.3}}
\put(12.5,1){\circle*{0.3}}
\put(13.5,1){\circle*{0.3}}
\put(14.5,1){\circle*{0.3}}
}
\put(1.5,1){\circle{0.3}}
\put(3.5,1){\circle{0.3}}
\put(4.5,1){\circle{0.3}}
\put(5.5,1){\circle{0.3}}
\put(7.5,1){\circle{0.3}}
\put(8.5,1){\circle{0.3}}
\put(11.5,1){\circle{0.3}}
\put(12.5,1){\circle{0.3}}
\put(13.5,1){\circle{0.3}}
\put(14.5,1){\circle{0.3}}
\end{picture}}
\cdots  ~.
\end{align}
It is clear that they are identical except for a single particle which has moved $n=9$ sites to the left, whereby it jumps over 3 other particles, leading to a factor $(-1)^3$. Summarizing the above, we have  
\beq
\label{pnlt}
\sum_{r=0}^{n-1}(-1)^r G^c_{\lambda ;(n-r,1^r) } = \sum (-1)^{P} s_{\lambda \setminus \eta }(y) \left\{ \begin{tabular}{ll}
Distinct ways to move a particle in $\lambda$\\ to the left by $n$ sites, thereby hopping \\
 over $P \leq n-1$ other particles.\\
   \end{tabular}
    \right\} ~,~
\eeq
%
 %\begin{align*}
 % \frac{1}{\sqrt{n}}\sum_{r=0}^{n-1}(-1)^r G^c_{\lambda ;(n-r,1^r) } = \sum (-1)^{P}  &  \{ \text{Distinct ways to move a single particle by $n$ sites,  } \\
  % & \text{thereby hopping over $P \leq n-1$ other particles.} \}
   %\end{align*}
   %
   where $\lambda\setminus \eta $ again represents a diagram obtained from $\lambda$ by the removal of a border strip, in this case of size $n$. We can apply the reasoning presented above to the calculation of $\chi^{\lambda/\mu}_{~\alpha}$, by considering $ p_\alpha s_\mu$ % =\sum_{\lambda} \chi^{\lambda/\mu}_{~\alpha} s_\lambda$ 
   in terms of fermionic particles hopping on a 1D lattice. In particular, starting from some partition $\mu$ and fixing a choice of $\alpha$, we consider all way to consecutively take a particle in the configuration corresponding to $\mu$ and move it $\alpha_j$ sites to the right, summing over $j\geq 1$, and adding a multiplicative factor $-1$ for each other particle which it hops over. We then add the resulting numbers $(\pm 1)$ for all cases where the end result of this process is the configuration $\lambda$. The outcome of this computation is precisely $\chi^{\lambda/\mu}_{~\alpha} $. This might provide a convenient method for computing $\chi^{\lambda/\mu}_{~\alpha}$, as it involves moving particles around on a line instead of a border strip tiling problem. Although these two problems evidently identical, the former might be simpler to implement practically.
    \subsubsection{The action of the hamiltonian in terms of Young diagrams}
    \label{subsubsah}
   When we consider the action of the hamiltonian in \eqref{tiham} in terms of diagrams, the following picture arises. The presentation given here does not depend on taking $N\to \infty$ and holds for finite $N$ as well. We first consider the XX0-model, in which case the action of $H^n$ on some state with $k$ particles can be described in terms of $k$ non-intersecting (vicious) random walkers that are allowed to take a single site to the left or right at each time step, for $n$ time steps \cite{bogoxx}. For the XX0-model, considering $H\ket{\lambda}$ in terms of diagrams then corresponds to summing over all ways to add one cell to and to remove one cell from $\lambda$, as this gives vicious random walkers that can move a single site to the right or left. For a general long-range hamiltonian such as in \eqref{tiham}, the action of $H$ is given by vicious random walkers which can move $n$ sites right or left, weighted by $a_{\pm n}$ and summed over $n$. We saw above that removing a border strip of size $n$ corresponds to taking a single particle and moving it by $n$ sites to the left, so it appears that
   \beq
   \label{hamact}
   H \ket{\lambda} = \sum_{n} a_n  \left(\sum_{\eta } \ket{\lambda \setminus \eta} + \sum_{\nu}  \ket{\nu} \right)~,
   \eeq
   where $\eta$ is a border strip of size $n$ and $\nu$ is related to $\lambda$ by the addition of a border strip of size $n$, and we sum over all such $\eta$ and $\nu$. Note that we do not get the minus sign that we get when considering $\avg{p_n(U) s_\lambda(U^{-1}}_c$ as in \eqref{corrps}. For $n=1$, when we consider all ways to start with the empty partition and look at all ways to successively add single cells, we simply get Young's lattice. The action of the XX0-model on some state $\ket{\lambda}$ then corresponds to moving from $\lambda$ along all edges in Young's lattice. For a general hamiltonian as in \eqref{tiham}, the picture is clearly more involved. Take, for example, the case where $a_4\neq 0$ and $a_n=0$ for $n\neq 4$. If we have  $\lambda=(5,3,1)$, given below,
 \vspace{.1cm}
 \beq
\label{y531}
\hspace{-2cm}
\rlap{\tableau{&&&&&\emptycell&\hdotscell\\ &&& \emptycell \\&  \emptycell\\ \emptycell \\ \vdotscell  % \\\emptycell \\ \\\vdotscell% & % \emptycell %&\emptycell %\!\!\!\\&\emptycell\\ & \emptycell\\ \emptycell \\\vdotscell%
}}
\unitlength=\cellsize
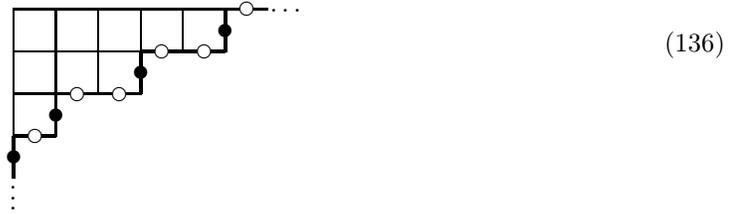
\begin{picture}(0,0)
\thicklines
\drawline(0,-3)(0,-2)(1,-2)(1,-1)(3,-1)(3,0)(5,0)(5,1)(6,1)
\thinlines
%\put(0,-3.5){\circle*{0.3}}
\put(0,-2.5){\circle*{0.3}}
\put(1,-1.5){\circle*{0.3}}
\put(3,-0.5){\circle*{0.3}}
\put(5,+0.5){\circle*{0.3}}
{\color{white}
%\put(6.5,1){\circle*{0.3}}
\put(5.5,1){\circle*{0.3}}
\put(4.5,0){\circle*{0.3}}
\put(3.5,0){\circle*{0.3}}
\put(2.5,-1){\circle*{0.3}}
\put(1.5,-1){\circle*{0.3}}
\put(0.5,-2){\circle*{0.3}}
}
%
%\put(6.5,1){\circle{0.3}}
\put(5.5,1){\circle{0.3}}
\put(4.5,0){\circle{0.3}}
\put(3.5,0){\circle{0.3}}
\put(2.5,-1){\circle{0.3}}
\put(1.5,-1){\circle{0.3}}
\put(0.5,-2){\circle{0.3}} 
\end{picture}
\eeq
 which corresponds to 
 \beq
    \ket{\lambda} = &  \cdots
\vcenterbox{\unitlength=\cellsize\begin{picture}(10,2)
\linethickness{0.7pt}
%\drawline(1,.3)(1,1.7)
%\drawline(12,.3)(12,1.7)
\put(0,1){\line(1,0){10}}
\put(.5,1){\circle*{0.3}}
\put(2.5,1){\circle*{0.3}}
\put(5.5,1){\circle*{0.3}}
\put(8.5,1){\circle*{0.3}}
{\color{white}
\put(1.5,1){\circle*{0.3}}
\put(3.5,1){\circle*{0.3}}
\put(4.5,1){\circle*{0.3}}
\put(6.5,1){\circle*{0.3}}
\put(7.5,1){\circle*{0.3}}
\put(9.5,1){\circle*{0.3}}

}
\put(1.5,1){\circle{0.3}}
\put(3.5,1){\circle{0.3}}
\put(4.5,1){\circle{0.3}}
\put(6.5,1){\circle{0.3}}
\put(7.5,1){\circle{0.3}}
\put(9.5,1){\circle{0.3}}
\end{picture}}
\cdots  ~.
\eeq
It is easy to see that there are two ways to move a single particle four sites to the left, and there are six ways to move a particle four sites to the right. Correspondingly, the action of the hamiltonian $H\ket{\lambda}$ produces the following eight diagrams.\\
%\vspace{.1cm}
%	
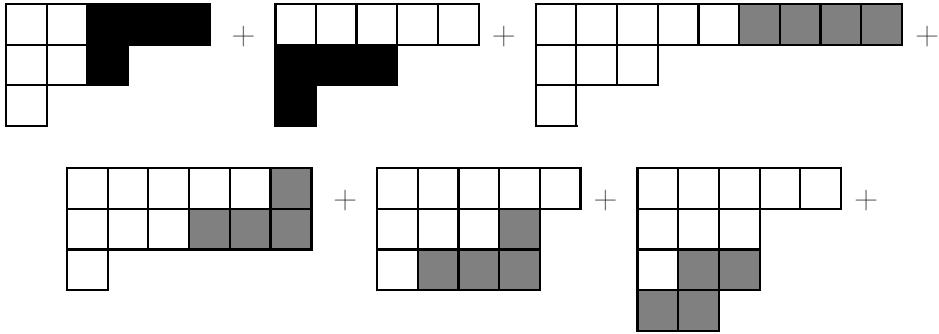
	\begin{center}
  \begin{ytableau}
 *(white) & *(white) & *(black)  & *(black)  & *(black)\\ 
*(white)  & *(white)     & *(black) \\
*(white) 
\end{ytableau} 
	\hspace{.05cm} {\large   + }  	
    \begin{ytableau}
 *(white) & *(white) & *(white) & *(white) & *(white) \\ 
 *(black)  & *(black)   & *(black) \\
  *(black) 
\end{ytableau} 
\hspace{.05cm}{\large  + }    
    \begin{ytableau}
 *(white) & *(white) & *(white) & *(white) & *(white) & *(gray) & *(gray) &  *(gray) & *(gray)  \\ 
 *(white) & *(white) & *(white) \\
 *(white) 
\end{ytableau} 
\hspace{.05cm}{\large  + }    
\end{center} 
\vspace{.3cm}
	\begin{center}
  \begin{ytableau}
 *(white) & *(white) & *(white) & *(white) & *(white) & *(gray)  \\ 
 *(white) & *(white) & *(white) & *(gray)  & *(gray)  & *(gray)   \\
 *(white) 
\end{ytableau} 
	\hspace{.05cm} {\large   + }  	
    \begin{ytableau}
 *(white) & *(white) & *(white) & *(white) & *(white)  \\ 
 *(white) & *(white) & *(white) & *(gray)    \\
 *(white) & *(gray)  & *(gray)  & *(gray) 
\end{ytableau} 
\hspace{.05cm}{\large  + }    
    \begin{ytableau}
 *(white) & *(white) & *(white) & *(white) & *(white)  \\ 
 *(white) & *(white) & *(white)  \\
 *(white) & *(gray)  & *(gray) \\
 *(gray)  & *(gray) 
\end{ytableau} 
\hspace{.05cm}{\large  + }   
\end{center} 
\vspace{.3cm}
%
%
%	\begin{center}
\beq
\hbox{
  \begin{ytableau}
 *(white) & *(white) & *(white) & *(white) & *(white)  \\ 
 *(white) & *(white) & *(white)  \\
 *(white) & *(gray) \\
 *(gray)  & *(gray)  \\
 *(gray) 
\end{ytableau} 
	\hspace{.05cm}  %{\textstyle + } % 
	{\large   + }  	
  \begin{ytableau}
 *(white) & *(white) & *(white) & *(white) & *(white)  \\ 
 *(white) & *(white) & *(white)  \\
 *(white) \\
 *(gray)  \\
 *(gray)  \\
 *(gray)  \\
 *(gray) 
\end{ytableau} 
}
\eeq
%\end{center} 
%\vspace{.3cm}
%
Consider a generalization of the Young's lattice in the form of a graph where each edge connects two diagrams that are related by addition or removal of some size $n$ border strip, for all $n$. The action $H^n \ket{\lambda}$ of a general long-range hamiltonian as in \eqref{tiham} corresponds to an $n$-site random walk on this graph, where edges corresponding to a border strip of size $n$ carry weight $a_n$. This provides a general description of the action of a hamiltonian in \eqref{tiham} in terms of (addition and removal of) border strips. We know from equation \eqref{numbncorpart} how many diagrams there are with some $n$-core $\mu$ and $n$-weight $w$, but due to the fact that the graph described above has varying connectivity, the $n$-site random walk is more likely to end up in some diagrams than others.
\subsubsection{Border strip tableaux and fermionic models}
		 \label{subsubsecbst}
   As in section \ref{subsubsah}, the present treatment up to equation \eqref{pnschur1} does not require $N\to \infty$. We saw that adding or removing a border strip $\eta$ and multiplying with $(-1)^{\hg(\eta)}$ implements fermionic statistics, and that this can be applied to the calculation of $\chi^{\lambda/\mu}_{~\alpha}$. Inverting this line of reasoning, we will now derive two relations for fermionic models using the properties of $\chi^{\lambda/\mu}_{~\alpha}$. The first of these involves the fact, noted at equation \eqref{prodschurexp} and below, that the order in which one adds border strips in the construction of a border strip tableau $\text{BST}(\lambda/\mu,\alpha)$ is irrelevant to its outcome. That is, $\chi^{\lambda/\mu}_{~\alpha}$ does not depend on the order of the entries of $\alpha$. The same is true for the removal of border strips, as can be seen in equation \eqref{chirec} and the example in \eqref{chi32xmp} and figure \ref{y32p21}. From the relation between removing or adding border strips in the construction of $\chi^{\lambda/\mu}_{~\alpha}$ and fermions hopping on a line, we can make the following observation. Consider a one-dimensional fermion configuration corresponding to some diagram $\mu$, and consider all ways of moving not necessarily distinct fermions to the right by $\alpha_1,\alpha_2,\dots$ sites, where $\alpha_j$ are unordered non-negative integers. We then see that \emph{the order of the step sizes $\alpha_j$ by which we move fermions has no effect on the outcome of this process}. That is, starting from a fermionic configuration and consecutively moving fermions to the right by various step sizes, the outcome depends only on the distribution of the step sizes and not the order in which the steps are taken. It is clear from applying \eqref{chirec} and the removal of border strips in the computation of $\chi^{\lambda/\mu}_{~\alpha}$ that the same statement holds when we consider fermions that can only hop to the left instead of the right. 

Consider a simple example which is partly given in figure \ref{y32p21} and in its entirety in figure \ref{y32graph}. Namely, consider again $\lambda = (3,2) = {\tiny \ydiagram{3,2}}~$ and now remove twice a single cell and once a border strip of size 2 from $\lambda$.\\[2pt] We can see this gives ${\tiny ~ \ydiagram{1}}$ , regardless of the order of the border strip sizes.
\begin{align}
\label{y32p211}
 {\small   \ydiagram{3,2} } & ~ \overset{p_2}{\longrightarrow}~  {\small   \ydiagram{3} } ~ \overset{p_1}{\longrightarrow} ~ {\small   \ydiagram{2} } ~\overset{p_1}{\longrightarrow}  ~{\small   \ydiagram{1} } \notag\\[5pt]
  {\small   \ydiagram{3,2} } & ~  \overset{p_1}{\longrightarrow}  ~{\small   \ydiagram{3,1} } + {\small   \ydiagram{2,2} }  ~\overset{p_2}{\longrightarrow} ~ {\small   \ydiagram{2} } + ( 1-1) {\small \times ~ } {\small   \ydiagram{1,1} }~ \overset{p_1}{\longrightarrow} ~ {\small   \ydiagram{1} } \notag\\[5pt]
    {\small   \ydiagram{3,2} } & ~  \overset{p_1}{\longrightarrow}  ~{\small   \ydiagram{3,1} } + {\small   \ydiagram{2,2} }  ~\overset{p_1}{\longrightarrow} ~ 2 {\small \times ~ } {\small   \ydiagram{2,1} } +  {\small   \ydiagram{3} }~ \overset{p_2}{\longrightarrow} ~ {\small   \ydiagram{1} } 
\end{align}
It is interesting to see that the fact that we arrive at ${\tiny ~ \ydiagram{1}~}$ follows either from the cancellation between two different ways to arrive at ${\tiny ~ \ydiagram{1,1}}$ , in the second line of \eqref{y32p211}, or the fact that one cannot remove a size 2 border strip from ${\tiny ~ \ydiagram{2,1}}$ , in the third line of \eqref{y32p211}. Assigning particle-hole configurations to the diagrams above, given in \eqref{y32} for ${\tiny ~ \ydiagram{3,2}}$ , and treating the particles as fermions gives an example of the irrelevance of the order of step sizes.

Our second result for fermionic systems arises from equations \eqref{clt2}, \eqref{clt3}. These state that $\chi^{\lambda/\mu}_{~(n^k)}$ is cancellation-free, where $\mu$ is the $n$-core of $\lambda$ \cite{jkbook}, \cite{white}. This leads to the following observation. Take a fermionic hamiltonian $\hat{H}_f$, where fermions are only allowed to hop $n$ sites, and denote a fermionic state corresponding to the $n$-core $\mu$ of some $\lambda$ as $\lVert \mu \rrangle$. The result of $\hat{H}_f \lVert \mu \rrangle $ can then be expressed in terms of symmetric functions as%is then given by %the following diagrams
\beq
%\hat{H}_f \lVert \lambda \rrangle = 
s_\mu p_n + \sum_{r=0}^{n-1} (-1)^r  s_{\mu/(n-r,1^r)} =  \sum_{\nu}  (-1)^{\hg (\nu/\mu ) } s_{\nu} + \sum_{\nu_n}  (-1)^{\hg (\eta ) }  s_{\mu \setminus \eta } ~, 
\eeq
where, as before, $\nu$ and $\mu \setminus \eta$ are related to $\mu$ by the addition and removal of a border strip of size $n$, respectively. Compare with equation \eqref{hamact}, especially the factors $(-1)^\hg$. We can keep iterating this step by adding and removing border strips of size $n$ to and from the resulting diagrams. From \eqref{clt2} and the comments below, it then follows that all diagrams appearing in the expansion of $\left(\hat{H}_f\right)^k \lVert \mu \rrangle $ have the same sign \emph{for any $k$}. Further, we saw that adding or removing border strips $\mu/\nu$ and multiplying by $(-1)^{\hg(\mu/\nu)}$ corresponds to letting fermions hop over a distance of $\abs{\mu/\nu}$ lattice sites. It follows that, for a fermionic model where particles can only hop $n$ sites, \emph{all distinct ways to go from some configuration $\mu$ to another configuration $\lambda$ appear with the same sign, with the sign depending only only the choice of $\mu$ and $\lambda$}. In other words, all different ways to go from any configuration $\mu$ to any other configuration $\lambda$ involves fermions hopping over either an even or an odd number of other fermions. Physically, this means that there is no interference between various ways to arrive at some fermionic state. Fermionic states will spread through Hilbert space rapidly upon time evolution, as they are not restricted by destructive interference.  
				
		The same reasoning can be applied to various expectation values of the LRRW models with hamiltonian \eqref{tiham}. In particular, we will consider
		%
	%	\beq
		$\avg{  (\tra U^n)^k s_\lambda(U^{-1}) }_c$ ~.
	%	\eeq
		%
		Repeatedly applying \eqref{pnskew1}, which we have used at various points above, we get
		\beq
		\label{pnschur1}
		\avg{  (\tra U^n)^k s_\lambda(U^{-1}) }_c  = \sum_{j=1}^k \sum_{\nu_{n,j}} (-1)^{\hg(T_j)} s_{\nu_{n,j}} (x) p_n(y)^{k-j}~. 
		\eeq 
		where we use the notation $\nu_{n,j} $ for diagrams related to $\lambda$ by the consecutive removal of $j$ border strips of size $n$. For example, consider again $\lambda =  (6,5,2^2,1)$ and $(n^k) = (4^4)$. Then, the diagrams $\nu_{4,2}$ are given by removing the yellow and blue regions from the diagrams in \eqref{y65bs4}, for $\nu_{4,3}$ the orange regions are also removed, and after removing the red regions we end up with $\nu_{4,4} = \emptyset$. Note that $\chi^{\lambda}_{~(n^k)} =0$ for most $\lambda$, one sufficient but far from necessary condition for this being that $nk \neq \abs{\lambda}$. In general, consecutively removing border strips of some size $n$ leads eventually to the aforementioned $n$-core of $\lambda$ \cite{stanley}. From \eqref{pkpkcc}, we have %it follows that there appears in equation \eqref{pnschur1} the product
		\beq
		s_{\nu_{n,j}}(x)p_n^{k-j}(x) = \sum_\mu d^\mu_{n,j,k} s_\mu(x) ~,
		\eeq
		for some coefficients $d^\mu_{n,j,k} $. From \eqref{schurpowerprod}, we know that $s_\lambda p_n^k $ is expanded as a sum over all diagrams related to $\lambda$ by subsequently adding a $k$ border strips of size $n$. On the other hand, $\nu_{n,j}$ is related to $\lambda$ by removal of $j$ border strips of size $j$. Therefore,
		\beq
		\label{pnschur2}
		\avg{  (\tra U^n)^k s_\lambda(U^{-1}) }_c  =  \sum_{j=1}^k \sum_{\tilde{\nu}_{n,j}^k} e^{i\phi(k-j)} (-1)^{\hg(T_1) + \hg(T_2)} s_{\tilde{\nu}_{n,j}^{k-j}} (x)~.
		\eeq
		In the above expression, the sum is over all $\tilde{\nu}_{n,j}^{k-j}$ constructed by first removing $j$ border strips of size $n$ from $\lambda$ (which results in $\nu_{n,j}$) and then adding $k-j$ border strips of size $n$, including multiplicities. Further, $T_1$ is given by the BST constructed from of the union of the $j$ border strips that are removed from $\lambda$, and $T_2$ is the tableau that is the union of the $k-j$ border strips that are added to $\nu_{n,j}$ to construct $\tilde{\nu}_{n,j}^{k-j}$. For $\lambda =  (6,5,2^2,1)$, we consider
		\beq
		\label{trausl}
		\avg{  (\tra U^4)^2 s_\lambda(U^{-1}) }_c ~.
		\eeq
		As mentioned above, contracting both copies of $\tra U^4$ with $s_\lambda$ gives rise to the diagrams in \eqref{y65bs4} after removing the yellow and blue border strips. The $\tilde{\nu}_{4,1}^{1} $ are given by the ways to remove from and then add to $\lambda$ a border strip of size four. We see from \eqref{pnschur2} that $\lambda$ appears in the expansion of \eqref{trausl} with a multiplicity two, as there are two distinct border strips of size four that one can remove from $\lambda$, namely, the yellow and blue border strips on the leftmost diagram in \eqref{y65bs4}. To find the remaining diagrams appearing in \eqref{trausl}, one should add border strips of size four to the diagrams obtained after removing the yellow and blue border strips on the leftmost diagram in \eqref{y65bs4}. In general, we get the following picture.
		\beq
		\label{pkcf}
		  \avg{  (\tra U^n)^k s_\lambda(U^{-1}) }_c  =  \pm \sum_{j=1}^k e^{i\phi(k-j)}  s_{\tilde{\nu}_{n,j}^{k-j}} (x) \left\{ \begin{tabular}{lll}
		  Distinct ways to consecutively\\  move $j$ particles in $\lambda$ to  
the left\\ by $n$ sites and then move $k-j$ \\ particles to the right by $n$ sites . %\\
%, thereby hopping over $P$ other particles.\\
   \end{tabular}
    \right\} ~~
\eeq
%	\begin{align*}
%	\avg{  (\tra U^n)^k s_\lambda(U^{-1}) }_c  =  \sum_{j=1}^k e^{i\phi(k-j)}  \sum (-1)^P  \{ & \text{All ways to consecutively move $j$ particles $n$ sites }  \notag\\
%& \text{to the left and then move $k-j$ particles $n$ steps to} \notag\\ 
%& \text{the right, thereby hopping over $P$ other particles.} %\} 
%		\end{align*}
		Note that the particles that are moved by $n$ sites are not necessarily distinct. Further, the right hand side appears with a positive or negative sign depending only on the final configuration. This again follows from the fact that $\chi^{\lambda/\mu}_{~\alpha}$ is cancellation-free. % We thus see that the correlation function corresponding to $\avg{  (\tra U^n)^k s_\lambda(U^{-1}) }_c $ counts the number of ways to go from some configuration $\lambda$ to any other via the procedure spelled out on the right hand side of equation \eqref{pkcf}.
 \subsubsection{Schur polynomial expansions of correlation functions}
 \label{specf}
 We consider now the expansions for general correlation functions which were derived in section \ref{secschurbs}. We first consider the application of \eqref{spsavg} here, before moving on to \eqref{corrpp} in section \ref{secpsecf}. Equation \eqref{spsavg} expresses $\avg{s_\lambda(U) s_\nu(U^{-1})}$ as a sum over diagrams obtained form $\lambda $ and $\nu$ by removing border strips of size $\alpha_1$, $\alpha_2 ,\dots$ for a composition $\alpha$, summed over $\alpha$ up to permutations of the entries $\alpha_j$. In particular, it establishes a relation between $\avg{s_\lambda s_\nu}_c $ and the correlation functions $\avg{s_{\lambda\setminus \{\eta\} }}= \bket{\emptyset}{s_{\lambda\setminus \{\eta\} }}$, involving only a single non-trivial configuration $\lambda\setminus \{\eta\}$ (and likewise for $\nu\setminus \{\eta\}$). Fixing some $\alpha$, we sum over all ways to start from $\lambda$ and $\nu$ and move particles $\alpha_j$ sites to the left with fermionic statistics (in the form of $(-1)^{\hg (T_\eta)}$). Consider the autocorrelation for $\nu=\lambda  = (8,6^2,4,1)$, which we show below.
   \vspace{.1cm}
 \beq
 \ydiagram{8,6,6,4,1}
 \eeq
 \vspace{-.3cm}

Equation \eqref{spsavg} gives an expression in terms of diagrams obtained by removing border strips from $\lambda$. Below, we show from left to right all border strips of size $1,2,3$ that can be removed from $\lambda=\nu$. 
\vspace{.2cm}
 \beq
 
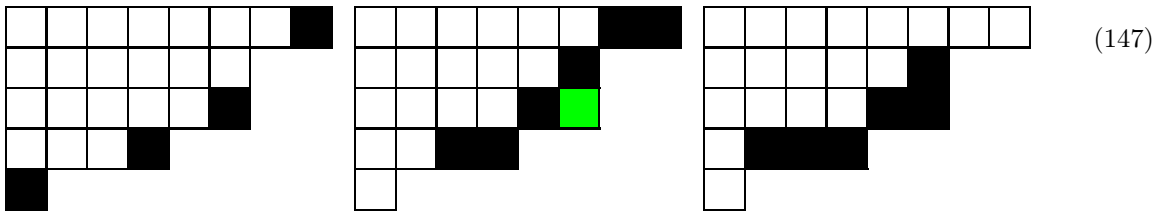
\label{y86bs}
 \begin{ytableau}
 *(white) &  *(white) &  *(white) & *(white) &  *(white) &  *(white)&  *(white) &  *(black)  \\
  *(white) &  *(white) &  *(white) &  *(white) &  *(white) &  *(white)  \\
  *(white) &  *(white) &  *(white) &  *(white) &  *(white) & *(black)  \\
 *(white) &  *(white) &  *(white) & *(black)   \\ *(black) 
\end{ytableau} 
\hso\hso\hso
 \begin{ytableau}
 *(white) &  *(white) &  *(white) & *(white) &  *(white) &  *(white)&  *(black) &  *(black)  \\
  *(white) &  *(white) &  *(white) &  *(white) &  *(white) &   *(black)  \\
  *(white) &  *(white) &  *(white) &  *(white) &  *(black) & *(green)  \\
 *(white) &  *(white)  & *(black)  & *(black)   \\
  *(white)
\end{ytableau} 
\hso\hso\hso
 \begin{ytableau}
 *(white) &  *(white) &  *(white) & *(white) &  *(white) &  *(white)&  *(white) &  *(white)  \\
  *(white) &  *(white) &  *(white) &  *(white) &  *(white)& *(black)  \\
  *(white) &  *(white) &  *(white) &  *(white)& *(black)   & *(black)   \\
 *(white) & *(black) & *(black) & *(black)   \\
 *(white)
\end{ytableau} 
 \eeq
 \vspace{-.3cm}
 
 In the middle diagram above, the green cell is shared between both a horizontal and a vertical strip of size two. We can thus remove four border strips of both size 1 and 2, and two of size 3. It is easy to see that there is a single border strip of size 4 that can be removed from $\lambda$, three border strips for both sizes 5 and 6, and so on, up to a single border strip of size $\lambda_1 + \ell(\lambda) -1 = 12$. Applying \eqref{spsavg} then gives
 \begin{align}
\avg{s_\lambda (U) s_\lambda(U^{-1})}_c = & \left( s_{(7,6^2,4,1)} +  s_{(8,6,5,4,1)}+s_{(8,6^2,3,1)} + s_{(8,6^2,4)}  \right)^2+ \notag \\
 & + \frac{1}{2} \left( s_{(6^3,4,1)} - s_{(8,5^2,4,1)} + s_{(8,6,4^2,1)} \right)^2 + \frac{1}{3} \left( - s_{(8,5,4,1^2 )} + s_{(8,6^2,1^2)} \right)^2+ \dots 
 \end{align}
 In the above expression, the first, second, and third terms correspond to the removal of border strips of size one, two, and three, respectively. Further, we again write $\left( s_\mu  + s_\rho +\dots \right)^2 $ instead of\\ $\left( s_\mu(x)  + s_\rho(x) +\dots \right)\left( s_\mu(y)  + s_\rho(y) +\dots \right)$.

 Consider now $\avg{s_\lambda (U) s_\nu (U^{-1})}$ with $\lambda$ as above and $\nu = (8^3,4,1)$, the latter of which is shown below. We have here $\lambda \subset \nu$, and we indicate $\nu/\lambda$ in gray.
  \vspace{.2cm}
 \beq
  
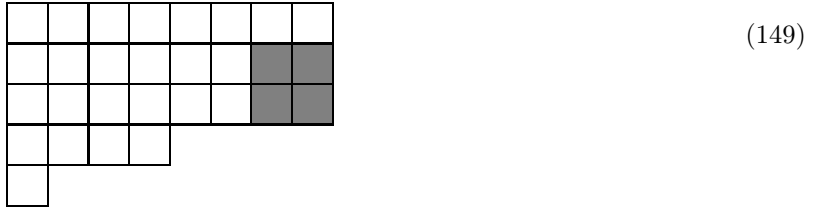
\label{ynuxmp}
 \begin{ytableau}
 *(white) &  *(white) &  *(white) & *(white) &  *(white) &  *(white)&  *(white) &  *(white)  \\
  *(white) &  *(white) &  *(white) &  *(white) &  *(white) &  *(white) &  *(gray) &  *(gray) \\
  *(white) &  *(white) &  *(white) &  *(white) &  *(white) &  *(white) &  *(gray) &  *(gray) \\
 *(white) &  *(white) &  *(white) &  *(white)  \\ *(white)
\end{ytableau} 
 \eeq
 
\vspace{-.2cm}
 Consider again all ways to remove border strips from $\nu$, and apply \eqref{spsavg}. This then leads to the following terms appearing in the expansion of $\avg{s_\lambda(U) s_\nu(U^{-1})}$,
\begin{align}
\label{slsnxmp}
 & \left( s_{(7,6^2,4,1)} +  s_{(8,6,5,4,1)}+s_{(8,6^2,3,1)} + s_{(8,6^2,4)}  \right)\left( s_{(8^2,7,4,1)} +s_{(8^3,3,1)} + s_{(8^3,4)} \right) + \notag \\
& +  \frac{1}{2} \left( s_{(6^3,4,1)} - s_{(8,5^2,4,1)} + s_{(8,6,4^2,1)} \right)\left( s_{(8,7^2,4,1)} + s_{(8^2,6,4,1)} + s_{(8^3,2,1)} \right) +  \dots 
\end{align}
The diagrams appearing on the left (right) in the top and bottom lines of \eqref{slsnxmp} are found by removing border strips of sizes 1 and 2 from $\lambda$ ($\nu$), respectively. This expansion can easily be continued by considering more or larger border strips. From the relation between removal or addition of border strips as in \eqref{spsavg} and fermionic particles hopping on a line, we can interpret \eqref{spsavg} in the following manner.% terms of spin configurations gives the following.
 \beq
 \label{spsint}
 \avg{s_\lambda s_\nu} = \sum_\alpha \frac{1}{z_\alpha} \left( (-1)^{P} s_{\lambda\setminus \{\eta\} }(y) \left\{ \begin{tabular}{lll}
		  Distinct ways to take particles in $\lambda$ \\  and move them $\alpha_1, \alpha_2,\dots$ sites to the\\
		  left, thereby hopping over $P$ particles
   \end{tabular}
    \right\} \right)\times \left( { y \to x \atop \lambda \to \nu } \right)~~
 \eeq 
 We remind the reader that $\{ \eta \} = \{ \eta_1 ,\eta_2,\dots\}$ are border strips of sizes $\abs{\eta_j}  = \alpha_j$. Note that the particles mentioned above are not required to be distinct. 
 \subsubsection{Power sum expansions of correlation functions}
 \label{secpsecf}
  From the fact that, for $\tau = it$, $p_k(x)$ and $p_k(y)$ are related by a phase, it is useful to express LRRW correlation functions as expansions in terms of $p_k(x)$ and $p_k(y)$ in order to reveal these phases. To do so, one may use the relation between Schur and power sum polynomials in equation \eqref{schurprodexp} and apply the Murnaghan-Nakayama rule, as done for the relatively simple example of $\avg{s_\lambda(U) s_\lambda(U^{-1})}$ for $\lambda = (3,2)$ in section \ref{secepsp}. However, as noted there, even this relatively simple example is already somewhat non-trivial, as it requires the computation of $\chi^\lambda_{~\alpha}$ for all $\alpha$. For these purposes, it is more convenient to employ \eqref{corrpp}, which provides an expansion of $\avg{s_\lambda(U) s_\nu(U^{-1})}$ in terms of $p_k(x)$ and $p_k(y)$. The prefactors appearing in this expansion depend on the number of ways to remove border strips from $\lambda$ and $\nu$ in such a way that the resulting diagram is the same for both $\lambda$ and $\nu$. Taking again $\nu=\lambda  = (8,6^2,4,1)$, for which the diagrams resulting from removal of border strips were treated in section \ref{specf} above. Applying \eqref{corrpp} then gives
 %
 %\beq
 \begin{align}
 \label{scpe}
 \avg{s_\lambda(U)s_\lambda(U^{-1})}_c = &  ~ 4 p_1(x)p_1(y)+ p_2(x)p_2(y) + \frac{2}{9} p_3(x)p_3(y) +\frac{1}{16} p_4(x)p_4(y)+ \notag \\ 
  & + 3 \left( \frac{1}{25} p_5(x) p_5(y) + \frac{1}{36} p_6(x) p_6(y) \right) + \notag \\
  & + \frac{1}{2} \left(p^2_1(x)p_2(y)  + p_2(x) p_1^2(y)  \right)+ \notag \\    & +
  \frac{1}{6} \left(p_3(x)p_1(y)p_2(y) + p_3(y)p_1(x)p_2(x)\right)  + \dots
 \end{align}
% \eeq
In the above expression, the first two lines on the right hand side are given by \eqref{subl}, which is a special case of \eqref{corrpp}. As we saw previously, the terms $\sim \frac{p_j(x)p_j(y)}{j^2}$ in the first two lines of \eqref{scpe} arise as follows. The denominator is given by the inverse of $z_{(j)}z_{(j)} = j^2$, and the numerator equals the number of distinct ways to remove a border strip of size $j$ from $\lambda =(3,2)$. The third and fourth line give mixed power sums obtained from not contracting a single $p_j(U^\pm)$ and two copies of $p_k(U^\mp)$, $p_m(U^\mp)$. From the diagrams in \eqref{y86bs}, one can see that there are four ways to remove two border strips of unit size and arrive at a diagram that can alternatively be obtained by removing a single border strip of size 2. This follows from the fact that there are four ways to remove a border strip of size two, which can alternatively be achieved by removing the two cells of such a border strip successively. However, due to the factors $(-1)^{\hg(T)}$ in \eqref{corrpp}, one of those four contributes with a negative sign, leading to a prefactor of $\frac{2}{z_{(1^2)}z_{(2)}} = \frac{1}{2}$ multiplying $\sim p_1^2 p_2 $ in the third line of \eqref{scpe}. There is a similar cancellation occurring from the term proportional to $p_3p_1p_2$. This expansion can then be continued by removing further border strips. %It is clear that finding $\chi^\lambda_{~\alpha}$ for $\lambda  =(8,6^2,4,1) $ 
Consider again $\avg{s_\lambda (U) s_\nu(U^{-1})}$ with $\lambda=(8,6^2,4,1) $ and $\nu = (8^3,4,1)$, as at the end of section \ref{specf}. applying \eqref{corrpp} leads to
\beq
\avg{s_\lambda (U) s_\nu(U^{-1})} = -\frac{2}{3}p_1(x)p_3(x) + \frac{1}{4} p_2^2(x) + \frac{1}{12}p_1(x)^4 + \frac{1}{2} p_2^2(x) p_1(x)p_1(y) +\dots
\eeq
This can be seen by considering diagram corresponding to $\nu$ in \eqref{ynuxmp}, and covering the gray 2 by 2 square with border strips of sizes from 1 to 3. We get no contribution proportional to $p_2(x)p_1(x)^2$ since the two ways to cover a 2 by 2 diagram with a single border strip of size 2 and 2 border strips of size 1 appear with opposite sign. This expansion can be continued by removing more or larger border strips from $\lambda$ and $\nu$, which can be conveniently done by using the relation to fermionic configurations. 

The above examples show that \eqref{corrpp} can be conveniently applied to $\avg{s_\lambda (U) s_\nu(U^{-1})}$ for larger $\lambda$, $\nu$ as well, for which $\chi^\lambda_{~\alpha}$ and $\chi^\nu_{~\alpha}$ would be very hard to compute. Indeed, as hinted at above, it is particularly useful for the following three reasons.
\begin{enumerate}
    \item Equation \eqref{corrpp} provides a controlled expansion of general correlation functions $\avg{s_\lambda(U) s_\nu(U^{-1})}$ in terms of power sums. These power sums can be directly read off from the hamiltonian, see \eqref{hpid} and \eqref{epid}.
    \item This expansion can be straightforwardly applied, including to correlation functions involving large diagrams $\lambda$, $\nu$. Using the comments below \eqref{y22}, the removal of border strips is related to fermionic particles hopping on a line, which could further simplify the application of this method.
    \item The power sums $p_k(x)$ and $p_k(y)$ are proportional to $\tau$, so that \eqref{corrpp} provides an expansion in terms of powers of $\tau$. Depending on the application and the range of $\tau$ one would like to consider, this expansion can be truncated at any desirable order that provides sufficient precision. For $\tau = it$, this $p_k(x)$ and $p_k(y)$ are related by a complex phase, leading to various simplifications that are hard to reveal otherwise.
\end{enumerate}

From the treatment above, it follows that the expression for $\avg{ s_\lambda (U) s_\nu(U^{-1})} $ in equation \eqref{corrpp} has the following particle-hole interpretation.
%
%\begin{align}
\beq
\label{pspint}
\avg{ s_\lambda (U) s_\nu(U^{-1})} =  \sum_{\omega, \gamma}   \frac{p_\omega(y)p_\gamma(x)}{z_\omega z_\gamma} (-1)^{P} %\times \notag \\
%& \times 
\left\{ \begin{tabular}{llll}
		  Distinct ways to move particles in $\lambda$ and $\nu$ \\  to the left by $\gamma_1, \gamma_2,\dots$  and $\omega_1, \omega_2,\dots$ sites,\\
		   respectively, hopping over $P$ other particles \\
		  and ending up in the same configuration.
   \end{tabular}
    \right\}  
~~~~ 
\eeq
%\end{align}
%
 %
 %
 %
 %
  \subsubsection{Correlations for power sums and applications to experimental benchmarking\label{sec:Benchmarking}}
%
%
%set $\tau = -it$ and 
We give here some suggestions for the benchmarking of experimental setups using correlation functions involving power sum polynoials. In experimental contexts such as trapped ion systems, one could measure the correlation functions corresponding to $\avg{p_n(U^{\pm 1} )}=\avg{\tra U^{\pm n}} = p_n $ experimentally. Remember that $p_n$ can be read off from the hamiltonian as in \eqref{hpid} or \eqref{epid} due to their direct proportionality to $a_{\pm n}$, and is given by a superposition of configurations corresponding to hook-shaped diagrams as in \eqref{yhookxmp}. Therefore, one could measure $\avg{p_n(U^{\pm 1} )}$ and compare them with the intended values of $a_{\pm n}$ to benchmark experimental setups. Further, one can use equation \eqref{trakn1},
\beq
\avg{\abs{\tra U^n}^2} = n + p_n(x)p_n(y) ~,
\eeq
 and consider $\tra U^n$ as a superposition of states (when properly normalized), as before. Then, the correlation function corresponding $\avg{\abs{\tra U^n}^2}$ is proportional to $F_{\emptyset;\emptyset}$ with proportionality given by $n+p_n(x)p_n(y) $, up to normalization. When we have certain hopping parameters $a_n \neq 0$, the only way to get independence from time is to have no dependence on the power sums $p_n \sim \tau n a_n$, as they contain a factor of $\tau$. The $\tau$-independence of the connected part of the correlation function corresponding to \eqref{trakn1} therefore sets it apart from other correlation functions, and might offer an effective way to benchmark experimental setups. Remember that the $p_\lambda$ form a basis for all symmetric polynomials.  In case $\ell(\lambda) \geq 2$ (or $\ell(\nu) \geq 2$) so that $p_\lambda = p_{\lambda_1} p_{\lambda_2} \dots$, Wick's theorem tells us that $\avg{p_\lambda p_\mu}_c$ will contain terms where not all $p_{\lambda_j}$ and $p_{\mu_k}$ are contracted. These give contributions containing $p_k(x) \sim \tau$ (and/or $p_j( y)\sim \tau $). Therefore, $\frac{1}{\sqrt{n}}\avg{\abs{\tra U^n}^2}_c$  for any integer $n$ is the only non-zero connected correlation function that does not depend on $\tau$. Lastly, one could use $\avg{\tra U^n \tra U^{-k}}_c = n\delta_{n,k}$ (after proper normalization) to see if the system is truly translationally invariant, as its derivation is predicated on the assumption of translational invariance. %Note that $\avg{\tra U^n \tra U^{-k}}_c = n\delta_{n,k}$, which means that the connected autocorrelation corresponding to $p_n$ is equal to that of the empty diagram. In particular,
%\beq
%\label{pnlosch}
%n^{-1}  \avg{\tra U^n \tra U^{-n}}_c = n^{-1} \sum_{r,s=0}^{n-1}(-1)^{r+s}  G^c_{(n-s,1^s) ;(n-r,1^r) }%= n^{-1}  \sum_{r=0}^{n-1}(-1)^r  \bket{(n-1,1^r)}{\lambda}
% =\bket{\emptyset}{\emptyset}~.
%\eeq
% This result could also be used for benchmarking experimental setups such as trapped ions. 
%
%
%
\section{Conclusions}
This work focused on weighted $U(N)$ integrals over symmetric polynomials and their relation with  correlation functions of LRRW models. The weighting is given by a weight function $f(z)$, which is required to satisfy Szeg\H{o}'s strong limit theorem, corresponding to hopping parameters $a_k$ whose modulus decays faster than $k^{-1}$ for $k\to \infty$. Writing the weight function in terms of generating functions of elementary or complete homogeneous symmetric polynomials as $f(z) = E(x;z)E(y;z^{-1})$ or $f(z) = H(x;z)H(y;z^{-1})$, respectively, general correlation functions can be expressed in terms of Schur polynomials with variables $x$ and $y$. %The Fourier coefficients of $\log f(e^{i\theta)})$ can be identified with $\pm k^{-1} p_k(x) 
In particular, the power sum polynomials are related to the hopping parameters as $ p_k(x) =\pm \tau k a_k $ and $ p_k(y) = \pm \tau k a_{-k}$ for $k\geq 1$. By using our earlier result \cite{tftis} that 
$\avg{\tra U^n \tra U^{-k}}_c = n \delta_{n,k}$ and applying Wick's theorem, we derive various identities. In particular, we compute $\avg{p_\mu(U) p_\rho(U^{-1})}$, generalizing a %
result due to Diaconis and Shahshahani, who computed said object in the CUE, corresponding to $f=1$. Further, we derive two expressions for correlation functions $\avg{s_\lambda(U) s_\nu (U^{-1})}$ for general $\lambda$ and $\nu$, %
both of which are obtained by removing border strips from $\lambda$ and $\nu$. In particular, the first of these expressions is an expansion in terms of diagrams related to $\lambda$ and $\nu$ by the removal of border strips. The second is an expansion in terms of $p_\gamma(x)$ and $p_\omega(y)$, where the expansion coefficients are determined by the number of ways to remove border strips from $\lambda$ and $\nu$ and arrive at the same diagram, up to a sign determined by the height of the (generally disjoint) border strip tableau that is removed.

We applied these results to correlation functions of LRRW models, where we consider configurations starting with an infinite number of particles (down spins) and ending %
with and infinite number of holes (up spins). Before applying our own results, we considered various standard relations in the theory of symmetric %
polynomials, such as the Pieri formula, which lead to a simple ways to compute and interpret various correlation functions. The most striking of these results is the fact that the involution between elementary and complete %
homogeneous symmetric polynomials, corresponding to the transposition of diagrams, leads to a duality between models related by switching the sign of the even hopping parameters, $a_k \to (-1)^{k+1}a_k$. In particular, we found that the correlation functions of these two models equal each other when we perform a particle-hole and parity transformation, which acts non-trivially on the %
finite interval in between the infinite strings of particles and holes. We therefore refer to this duality as quasi-local particle-hole duality. Further, we noted that the Jacobi-Trudi %
identity allows one to calculate any $\avg{s_\lambda(U) s_\nu (U^{-1})}$ directly. Although the number of terms appearing in this expansion grows quickly with the size of $\lambda$, $\nu$, %
it is simple to implement as it only involves the calculation of a determinant of a matrix with known entries.

We then considered the basis of power sum polynomials $p_\gamma(x)$ and $p_\omega(y)$ and the closely related border strips, which is particularly useful in the application to LRRW models due to their aforementioned relation to hopping parameters $a_k$ and generalized time $\tau$. As mentioned above, we derived expressions for LRRW correlation functions involving addition or removal of a border strip $T$ of size $n$ and height $\hg(T)$, leading to a sign $(-1)^{\hg(T)}$. We found that this corresponds to moving a fermionic particle a distance $n$ to the right or left, respectively, thereby hopping %
over $\hg (T)$ other particles. This provides an interpretation of various expansions of correlation functions in terms of particles hopping with fermionic exchange statistics, %
even though the LRRW model consists of hard-core bosons rather than fermions. For example, the procedure of the second expansion of $\avg{s_\lambda  s_\nu}$, in \eqref{corrpp}, which involves the removal of border strips, can be interpreted as replacing the (hard-core boson) particles in $\lambda$ and $\nu$ by fermions and considering all distinct ways to move them to the left such that the final configurations are identical. Due to the aforementioned relation between power sums and hopping parameters, this expression provides an expansion in powers of $\tau$ where the numbers appearing in the expansion coefficients can be read off from the hamiltonian.

The mathematical results derived in section \ref{secwick} may be generally applied to the case where the strong Szeg\H{o} limit theorem applies. The expansion of general correlation functions in terms of power sum polynomials is particularly useful in such applications where these can be accessed directly, such as for LRRW models, where they are provided as input. In section \ref{secspincorr}, we saw that the addition or removal of border strips with sign $(-1)^{\hg}$ can be interpreted as fermionic particles hopping to the right or left, respectively. One may apply this `physical' interpretation of this process to the power sum expansion of $\avg{s_\lambda s_\nu}$, by considering all ways to move fermionic particles in $\lambda $ and $\nu$ to the left to arrive at the same configuration. Further, the same fermionic interpretation may be applied to the Murnaghan-Nakayama rule, which could provide a more convenient practical method for the calculation of general $\chi^{\lambda/\mu}_{~\alpha}$. Inverting this reasoning, we applied properties of these character to long-range fermionic models. First of all, the fact that $\chi^{\lambda/\mu}_{~\alpha}$ does not depend on the order of the entries of $\alpha$, means that taking a fermionic configuration and considering all ways to move particles by $\alpha_1,\alpha_2,\dots$ sites to the right gives the same outcome regardless of the ordering of the step sizes $\alpha_j$. Further, the fact that the symmetric group character are cancellation-free for $\alpha = (n^k)$ lead to the conclusion that all ways to go from any configuration to any other configuration by taking only $n$ sites involves either an even or an odd number of particles being hopped over, depending only on $n$ and the choice of initial and final configurations. %This appears to be a novel result as well (at least it was not found elsewhere by the authors) which may have independent mathematical applications or generalizations.

As mentioned in the introduction, LRRW models such as we consider here have seen increasing activity in both experimental and theoretical contexts due to the experimental accessibility of such systems and the surprising phenomena they exhibit. We believe our work can be applied along some of these lines of research, including the consideration of localization by addition of (diagonal) disorder \cite{Deng_2018}, \cite{Nosov_2019}. Adding such disorder will generally break translational invariance, and the expressions in this work will no longer apply. One may also consider translationally invariant disorder such as random hopping parameters, as in \cite{Bogomolny_2020}, \cite{Bogomolny_2021}, in which case the results derived here would still apply as long as the hopping parameters satisfy the aforementioned asymptotic fall-off conditions. % of experimental setups. 
All results in section \ref{secspincorr} can, in principle, be checked experimentally, where the expressions in this work would be expected to hold with reasonable accuracy up to the time that finite size effects start to occur. Besides checking our results in experimental setups, correlation functions involving power sum polynomials may be used for experimental benchmarking.
\section{Acknowledgements}
We would like to thank Ivan Khaymovich, Vladimir Kravtsov, Ji\v{r}\'{i} Min\'{a}\v{r}, and Arghavan Safavi-Naini for useful discussions and suggestions.
This work is part of the DeltaITP
consortium, a program of the Netherlands Organization for Scientific Research (NWO) funded
by the Dutch Ministry of Education, Culture and Science (OCW). 
\begin{appendices}
\section{Symmetric polynomials and Young diagrams}
\label{secsymmpol}
%
%
%We review here some aspects of symmetric polynomials and Young diagrams that will be useful to us later. %We fist consider elementary and complete homogeneous symmetric polynomials and Young diagrams consisting only of a single column or row, before continuing to power sum polynomials, hook-shaped Young diagrams, and border strips.
%
%
%
%\subsection{Elementary and complete homogeneous symmetric polynomials}
%\label{subsecelhom}
%
We review here some aspects of symmetric polynomials and Young diagrams, starting with some simple examples. Take a set of variables $x =( x_1,x_2,\dots )$. The \emph{elementary symmetric polynomials} are then defined as
	\beq
	\label{epol}
	e_k (x) = \sum_{i_1<\dots<i_k} x_{i_1}\dots x_{i_k}~.
	\eeq
	Some examples include
	\begin{align}
	e_0& = 1 ~, \notag \\
	e_1(x_1)& = x_1 ~, \notag\\
	e_1(x_1, x_2 )& = x_1 +x_2~, \notag\\
	e_2(x_1, x_2 )& = x_1 x_2~.
	\end{align}
	Closely related are the \emph{complete homogeneous symmetric polynomials}, defined as
	\beq
	\label{hpol}
	h_k (x) = \sum_{i_1\leq\dots\leq i_k} x_{i_1}\dots x_{i_k} , 
	\eeq
	which contain all monomials of degree $k$. Note the difference in the summation bounds between \eqref{epol} and \eqref{hpol}. Some examples of $h_k$ include
	\begin{align}
	h_0 & = 1 ~,  \notag\\
	h_1(x_1)& = x_1 ~, \notag\\
	h_1(x_1, x_2 )& = x_1 +x_2~, \notag\\
	h_2(x_1, x_2 )& = x_1 x_2 +  x_1^2+x_2^2~.
	\end{align}
	Another type of symmetric polynomial is the \emph{power sum symmetric polynomial}, also simply referred to as \emph{power sum polynomial},
	\beq
	\label{powersum}
	p_k(x)= \sum_j x_j^k =x_1^k+x_2^k+\dots 
	\eeq
The generating functions of $e_k$ and $h_k$ and their relation with power sums are as follows [see e.g \cite{mcd}],
	\begin{align}
	\label{genfuncsymp}
	E(x;z) &= \sum_{k=0}^\infty e_k(x) z^k = \prod_{k=1}^\infty (1+x_k z) = \exp \left[ \sum_{k=1}^\infty \frac{(-1)^{k+1}}{k} p_k(x) z^k   \right]  ~,\notag \\
	H(x;z) & = \sum_{k=0}^\infty h_k(x) z^k = \prod_{k=1}^\infty \frac{1}{1-x_k z} = \exp \left[ \sum_{k=1}^\infty \frac{1}{k} p_k(x) z^k   \right]  ~.
	\end{align}
	From the above expressions, it is clear that $H(x;z) E(x;-z) = 1$. Checking every order of $z$ then gives, for all $n\geq 1$ and any choice of $x$ [e.g. (2.6') from \cite{mcd}],
	\beq
	\label{ehzero}
	\sum_{r=0}^{n}(-1)^r h_{n-r} (x) e_{r} (x) =0~.
	\eeq
	Partitions play an important role in the study of symmetric polynomials. A partition of $n\in \mathds{Z}^+$ is a sequence of non-negative integers $\lambda = ( \lambda_1 , \lambda_2,\dots,\lambda_{\ell(\lambda)} )$, which we will order as $\lambda_1 \geq \lambda_2 \geq \dots $, satisfying $\sum_j \lambda_j=n$. The \emph{size} (or \emph{weight}) of a partition is given by the sum of its terms $\abs{\lambda}= \sum_j\lambda_j$ and its \emph{length} $\ell(\lambda) $ is the largest value of $j$ such that $\lambda_j \neq 0$. Closely related to partitions of $n$ are \emph{compositions} of $n$, consisting also of a sequence of non-negative integers which sum to $n$, but where a different ordering in these integers defines a different composition. A \emph{weak composition} of $n$ is a composition which may include zeroes as its entries, that is, a set of non-negative integers which sum up to $n$. A partition of $n$ corresponds to a Young (or Ferrers) diagram containing $n$ cells, or `boxes'. We will use these terms interchangeably. As an example, the diagram corresponding to a partition of 12 given by $\lambda = (6,4,2,1)$ is given below, where $\lambda_j$ equals the number of cells in the $j^\text{th}$ row.
	%
%
%\begin{center}
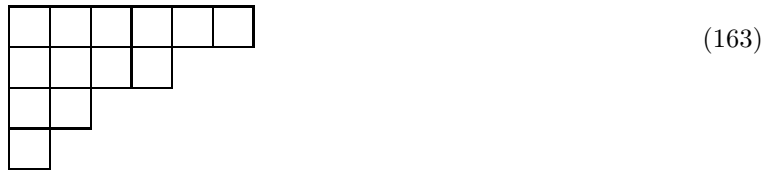
\beq
\hbox{
\ydiagram{6,4,2,1}
}
\eeq
%\end{center}
%
We will denote a diagram consisting of $b$ rows of $a$ cells by $(a^b)$. %E.g. $(4^3)$ is given by the following diagram.\\
%\begin{center}
%\ydiagram{4,4,4}
%\end{center}
For a partition $\lambda$, we will write 
 \beq
 \label{generalizedsympol}
 e_\lambda = \prod_{j\geq 1} e_{\lambda_j}~,~~  h_\lambda = \prod_{j\geq 1} h_{\lambda_j}~,~~  p_\lambda = \prod_{j\geq 1} p_{\lambda_j}~.
 \eeq
 Further, we write
\beq
\label{zmf}
z_\lambda=\prod_{j\geq 1}j^{m_j}m_j!~,~~  m_j(\lambda)=\text{Card} \{k:\lambda_k=j\}~,
\eeq
i.e.  $m_j(\lambda)$ is the number of rows in $\lambda$ of length $j$. We also write $\varepsilon_\lambda=(-1)^{\abs{\lambda} -\ell(\lambda)}$. Newton's identities then read
\begin{align}
h_n & = \sum_{\abs{\lambda}=n} z_\lambda^{-1} p_\lambda~, \notag \\
e_n & = \sum_{\abs{\lambda}=n} \varepsilon_\lambda z_\lambda^{-1} p_\lambda~.
\end{align}
In terms of the complete exponential Bell polynomial $B_n$ \footnote{Complete Bell polynomials $B_{n}(x_{1}\ldots x_{n}) $ can be defined by their generating function, $\sum_{n=0}^{\infty}B_{n}(x_{1}\ldots x_{n})t^{n}/n!=\exp(\sum_{j=1}^{\infty}x_{j}t^{j}/j!)$}, we have
\begin{align}
\label{newtonbell}
h_n & = \frac{1}{n!} B_n\left( p_1,p_2,2! p_3,\dots, (n-1)!p_n\right)~,\notag \\
e_n & = \frac{(-1)^n}{n!} B_n\left( -p_1,-p_2,-2! p_3,\dots, -(n-1)!p_n\right)~.
\end{align} 
Another type of symmetric polynomial is the \emph{Schur polynomial}. Schur polynomials play an important role as characters of irreducible representations of general linear groups and subgroups thereof. Schur polynomials are associated to a partition $\lambda$ and a set of variables $x=(x_1,x_2,\dots)$ in the following way. For a choice of $\lambda$, a \emph{semistandard Young tableau} (SSYT) is given by positive integers $T_{i,j}$ satisfying $1 \leq i \leq \ell(\lambda)$ and $1\leq j \leq \lambda_i$. These integers are required to increase weakly along every row and increase strongly along every column, i.e. $T_{i,j} \geq T_{i,j+1}$ and $T_{i,j}  > T_{i+1,j}$ for all $i,j$. Label by $\alpha_i$ the number of times that the number $i$ appears in the SSYT. We then define
	\begin{equation}
	x^T = x_1^{\alpha_1} x_2^{\alpha_2 } \dots~.
	\end{equation}
	The \emph{Schur polynomial} $s_\lambda(x) $ is given by \cite{stanley}.
	\begin{equation}
	\label{sp}
	s_\lambda(x) =\sum_T x^T~,
	\end{equation}
	where the sum runs over all SSYT's corresponding to $\lambda$ i.e. all possible ways to inscribe the diagram corresponding to $\lambda$ with positive integers that increase weakly along rows and strictly along columns. If $\lambda_j=0  $ for all $j$, then $\lambda $ is the empty partition, which we denote by $\lambda = \emptyset$. The Schur polynomial of the empty partition is set to unity, i.e.
	\beq
	s_\emptyset (x) = 1~, %~, ~~~\text{ any } x=(x_1,x_2,\dots)~.
	\eeq
	which is independent of the choice of variables $x$. We give an example of an SSYT corresponding to a Young diagram $\lambda=(3,2)$ and with non-zero variables $x_1,x_2,x_3$. 	
%	\begin{center}
	\vspace{.1cm}
	\beq
%	\hbox{
		%\hspace{1cm}
		\begin{ytableau}
			1 & 1 & 3  \\
			2 & 3\\
		\end{ytableau}
	%	}
		\eeq
%	\end{center}
%	\vspace{.2cm}

	From \eqref{ssp} one can see that the contribution of this SSYT is given by $x_1^2x_2 x_3^2$. Summing over all monomials corresponding to all SSYT's then gives the Schur polynomial $s_{(3,2)}(x_1,x_2,x_3)$. We emphasize that the result is generally a symmetric polynomial, as this may not be obvious from the definition. Consider the Schur polynomials corresponding to a row or a column of $n$ cells, shown below for $n=4$.
		\vspace{.1cm}
	%
	%	\begin{center}
		\beq
%	\hbox{
	   \ydiagram{4} \hspace{.8cm} \ydiagram{1,1,1,1}
	  % }
	   \eeq
	%   \end{center}
%	\vspace{.2cm}
	
	One can see that
	\beq
	\label{schureh}
	s_{(1^n)}=e_n	~~,~~~ s_{(n)}=h_n ~,
	\eeq
	for any choice of $x$. That is, the Schur polynomial of a column or row of $n$ cells is given by the degree $n$ elementary or complete homogeneous symmetric polynomial, respectively. Equation \eqref{schureh} simply follows from the requirement for SSYT's that integers increase weakly along rows and strongly along columns, compare with \eqref{epol} and \eqref{hpol}. It follows that we can exchange between $e_n$ and $h_n$
	by transposing diagrams, that is, by reflecting across the main diagonal of the diagram, as this exchanges rows and columns. For a diagram $\lambda$, its transpose is denoted as $\lambda^t$. Since transposition is a reflection, it is an involution, i.e. $  (\lambda^t)^t=\lambda$. It is clear that $(n)^t = (1^n)$ i.e. this involution maps rows to columns and vice versa. As a less trivial example, take $\lambda = (5,4,2,1)$, shown below on the left, and $\lambda^t = (4,3,2^2,1)$, shown on the right.
	%
	%\begin{center}
	\beq
%	\hbox{
	   \ydiagram{5,4,2,1} \hspace{.8cm} \ydiagram{4,3,2,2,1}
	  % }
	   \eeq
%	\end{center}
%	\vspace{.2cm}
%
	Power sum polynomials can also be expressed in terms of Schur polynomials, in this case in the form of a sum,
	\beq
	\label{powersumschur}
	p_n = \sum_{r=0}^{n-1} (-1)^r s_{(n-r,1^r)}~.
	\eeq
	Here, $(a,1^b)$ is a hook-shaped diagram consisting of a row with $a$ cells  followed by $b$ rows with a single cell. For example, $\lambda =(5,1^3)$ is given by the following diagram.
	\vspace{.1cm}
%
%\begin{center}
	\beq
	\hbox{
    \ydiagram{5,1,1,1}
    }
    \eeq
%\end{center}
%	\vspace{.2cm}
	%
Although equation \eqref{powersumschur} may not be immediately obvious, it can easily be seen to arise in simple examples. If $x=(x_1,x_2)$ and $n=1$, we have the following SSYT's.
\vspace{.1cm}
%
%	\begin{center}
	\beq
	\hbox{
		\begin{ytableau}
			1 
		\end{ytableau} 
		\hspace{.05cm}{\large   + }
			\begin{ytableau}
			2
		\end{ytableau} 
		}
		\eeq
%	\end{center}
	Which gives $p_1(x) = s_{(1)}(x_1,x_2) = x_1+x_2$. For $n=2$, the following SSYT's contribute,
	\vspace{.2cm}
	%
	%	\begin{center}
		\beq
	\hbox{
		\begin{ytableau}
			1 & 1 
		\end{ytableau} 
		\hso {\large   + }
			\begin{ytableau}
		1 & 	2
		\end{ytableau}
		\hso {\large   + }
			\begin{ytableau}
		2 & 	2
		\end{ytableau}
		\hspace{.1cm}{\Large $\mathrm{-}$ }\hspace{-.1cm}
			\begin{ytableau}
		1 \\ 	2
		\end{ytableau}
		}
		\eeq
%	\end{center}
%	\vspace{.1cm}
	Note that the rightmost SSYT corresponding to $(1^2)$ contributes with a minus sign, which results in $p_2(x) = s_{(2)}(x_1,x_2) - s_{(1^2)}(x_1,x_2) = x_1^2  +x_2^2$. One may convince oneself that this generalizes to higher $n$ and general choice of $x$.
	
	Schur polynomials have a natural generalization to so-called \emph{skew Schur polynomials}, which are associated to skew diagrams. Skew diagrams are constructed from two non-skew diagrams $\lambda$ and $\mu$ such that $\mu \subseteq \lambda$, which means that $\mu_i \leq \lambda_i, ~\forall ~ i$. The skew diagram denoted by $\lambda/\mu$ is then the complement of $\mu$ in the diagram corresponding to $\lambda$. For $\lambda = (4,3,2)$ and $\mu = (2,1)$, the skew diagram $\lambda/\mu$ is given by the following, where we indicate in black those cells which are removed from $\lambda$.	
	\vspace{.2cm}
	%\begin{center}
		\beq
  \label{bseqn}
	\hbox{
\ydiagram{4,3,2}  { \Bigg   /  }  \ydiagram{2,1} \vspace{.2cm}  { \huge   \vspace{.4cm} =  \vspace{.1cm}}  \begin{ytableau}
*(black) & *(black) &  *(white) & *(white)   \\ 
*(black) &  *(white)&  *(white)  \\
 *(white) &  *(white)
\end{ytableau} 
{ \huge   \vspace{.4cm} =  \hspace{-.2cm}} 
\ydiagram{2+2,1+2,2}
}
\eeq
%\end{center}
%\vspace{-.2cm}

The skew diagram on the right hand side is a \emph{border strip}, which is a connected skew diagram not containing a 2 by 2 subdiagram. This is an important class of skew diagrams which we will encounter again later. For a general skew diagram, define a skew semistandard Young tableau corresponding to $\lambda/\mu$ as above, namely, as an array of positive integers $T_{ij}$ satisfying $1\leq i \leq \ell(\lambda)$ and $\mu_i\leq j \leq \lambda_i$ which increase weakly along rows and strictly along columns. We then define the \emph{skew Schur polynomial} corresponding to $\lambda/\mu$ as 
	\begin{equation}
	\label{ssp}
	s_{\lambda/\mu}=\sum_T x^T,
	\end{equation}
	where the sum again runs over all SSYT's corresponding to $\lambda/\mu$. Note that if $\mu = \emptyset$, we have $s_{\lambda/\mu} = s_\lambda$, and if $\mu = \lambda $, $s_{\lambda/\lambda} = s_\emptyset =1$. Let us consider $\lambda =(3,2)$ and $\mu= (1)$. Below, we give a skew SSYT corresponding to the skew partition $\lambda/\mu$, which would contribute $x_1^2 x_2  x_3$ to the skew Schur polynomial.
	%
	%	\begin{center}
	\beq
	\hbox{
		\begin{ytableau}
			\none & 1 & 3  \\
			1 & 2\\
		\end{ytableau}
		}
		\eeq
%	\end{center}
%	\vspace{.2cm}
	From the strong increase of integers along the rows of a (skew) SSYT, it follows that,
	\beq
	\label{mcdfe}
	s_{\lambda/\mu}(x_1,\dots,x_n) = 0 ~\text{   unless   }~  0 \leq \lambda_i^t-\mu_i^t \leq n  ~ \text{ for all }~ i \geq 1~.
	\eeq
	Note that an example of \eqref{mcdfe} is given by the fact that $e_k(x_1,\dots,x_N)= s_{(1^k)}(x_1,\dots,x_N) = 0$ for $ k>N$.
	
	We have the following expressions for skew Schur polynomials,
\beq
s_{\lambda / \mu} =\sum_\nu c^\lambda_{\mu \nu} s_\nu~,
\eeq
and products of non-skew Schur polynomials,
\beq
\label{schurproduct}
s_\lambda s_\mu = \sum_\nu c_{\lambda\mu}^\nu s_\nu~.
\eeq
The expansion coefficients $c^\lambda_{\mu \nu}$ are known as \emph{Littlewood-Richardson coefficients}, which are given by the number of Littlewood-Richardson tableaux of shape $\nu/\lambda $ and weight $\mu$. A Littlewood-Richardson tableau is an SSYT such that, when we read its entries from right to left and top to bottom, any positive integer $j$ appears at least as many times as $j+1$. Note from \eqref{schurproduct} that $c^\lambda_{ \mu \nu} =c^\lambda_{ \nu \mu}  $. For example, of the SSYT's pictured below, the one on the left is a Littlewood-Richardson tableau while the one on the right is not.
 \vspace{.1cm}
	\beq
	\hbox{
%	\begin{center}
		\begin{ytableau}
			\none & 	\none & 1 & 1  \\
			\none & 	2 & 2\\
			1 & 3
		\end{ytableau}  \hspace{1cm}  
				\begin{ytableau}
			\none & 	\none & 1 & 2  \\
			\none & 	2 & 2\\
			1 & 3
		\end{ytableau}
		}
		\eeq
%	\end{center}
\vspace{-.3cm}

We apply the Littlewood-Richardson rule to the special case where one of the diagrams consists of a single row. The result is known as the Pieri formula \cite{mcd},
%A special case of the Littlewood-Richardson rule is the Pieri formula \cite{mcd}
\beq
\label{hpieri}
s_\lambda h_n = \sum_\nu s_\nu~,
\eeq
where the sum is over all $\nu$ such that $\nu/\lambda$ is a horizontal strip, i.e. a skew diagram with at most one cell in each column. Applying the involution which transposes all diagrams, and therefore exchanges $h_n=s_{(n)}$ with $e_n=s_{(1^n)}$, we have
\beq
\label{epieri}
s_\lambda e_n =\sum_\nu s_\nu~,
\eeq
where the sum is now over all $\nu$ such that $\nu/\lambda$ is a vertical strip, i.e. a skew diagram with at most one cell in each row. The Pieri formula states that $c^\nu_{\lambda \mu}$ for $\mu=(n)$ is equal to 1 when $\nu/\lambda$ is a horizontal strip, and zero otherwise. Applying this to $s_{\lambda /(n)}$, we have
\beq
\label{piericorh}
s_{\lambda/(n)} = \sum_\nu s_\nu~,
\eeq
where the sum is now over all $\nu$ such that $\lambda/\nu$ is a horizontal strip. From \eqref{epieri}, we also have
\beq
\label{piericore}
s_{\lambda/(1^n)} = \sum_\nu s_\nu~,
\eeq
where $\lambda/\nu$ is a vertical strip. The Pieri formula in \eqref{piericorh} is illustrated below for $  \lambda = (2,1)$ and $n=2$, with the cells that are added onto $\lambda$ are indicated in gray. It is clear that $\nu/\lambda$ is a horizontal strip for all diagrams $\nu$ on the right hand side, as there are no two gray cells in any column. For equation \eqref{epieri}, one should simply transpose the diagrams. 
 \vspace{.1cm}
%
%	\begin{center}
	\beq
	\hbox{
\ydiagram{2,1} \hspace{.05cm} \ydiagram{2} \hspace{.05cm} {\Large    = }  \begin{ytableau}
 *(white) & *(white) & *(gray)  & *(gray) \\ 
*(white) 
\end{ytableau} 
\hspace{.05cm}{\large   + }    
\begin{ytableau}
 *(white) & *(white) & *(gray)  \\ 
*(white) & *(gray)  
\end{ytableau} 
\hspace{.05cm}{\large   + }  
\begin{ytableau}
 *(white) & *(white) & *(gray)   \\ 
*(white)  \\
*(gray)
\end{ytableau} 
\hspace{.05cm} {\large   + }  
\begin{ytableau}
 *(white) & *(white)   \\ 
*(white) & *(gray) \\
*(gray) 
\end{ytableau} 
}
\eeq
%\end{center}
%\vspace{.2cm}
%
We can use the Pieri rule to demonstrate $\sum_{r=0}^{n}(-1)^r h_{n-r} e_{r} =0$, equation \eqref{ehzero}. In particular, for $a,b\geq 1$, we have 
\beq
h_a e_b = s_{(a,1^b)} + s_{(a+1,1^{b-1})} ~.
\eeq
The example for $a=4$ and $b=3$ gives the following.
 \vspace{.1cm}
	\beq
	\hbox{
%\begin{center}
\ydiagram{4} \hspace{.2cm}%{\Large$ \times$}  
\ydiagram{1,1,1} \hso \hso\hso {\Large    = } \hso \begin{ytableau}
 *(white) & *(white) &  *(white) & *(white) \\ 
*(gray)  \\
*(gray) \\
*(gray)
\end{ytableau}   \hspace{.05cm} {\large   + } \begin{ytableau}
 *(white) & *(white) &  *(white) & *(white) & *(gray) \\ 
*(gray)  \\
*(gray) 
\end{ytableau} 
}
\eeq
%\end{center}
%{\huge   $ \times$}
%
Plugging this in gives
\beq
\label{ehp}
\sum_{r=0}^{n}(-1)^r h_{n-r} e_{r} = h_n +(-1)^n e_n +  \sum_{r=1}^{n-1} (-1)^r \left( s_{(n-r,1^r)} +s_{(n-r+1,1^{r-1})}\right) =0~.
\eeq
For $1\leq r \leq n-1$, the summand on the left for any $r$ is cancelled with terms coming from $r-1$ and $r+1$. For $r=0$ and $r=n$, we get the contributions $h_n$ and $(-1)^n e_n$ which cancel with term from $r=1$ and $r=n-1$, respectively. For example, for $n=3$, applying the Pieri formula gives the diagrams below.

 \vspace{.3cm}

\begin{center}
\hspace{-3.3cm} $ \sum_{r=0}^3 (-1)^r h_{3-r}e_r$ {\large \hso = \hso  }  % \hspace{2.1cm} 
 \ydiagram{3} \hso  {\large  $\mathrm{-}$ }\ydiagram{1} \hso \hso\ydiagram{2} \hso {\large +}  \ydiagram{1,1}\hso \hso \ydiagram{1} \hso {\large  $\mathrm{-}$}    \ydiagram{1,1,1}\hso 
 \vspace{-.3cm}
  
  	\beq
	\hbox{  
 \hspace{3.4cm}    {\large \hso = \hso  }   \ydiagram{3} \hso  {\large  $\mathrm{-}$ }\ydiagram{3}\hso  {\large  $\mathrm{-}$} \ydiagram{2,1} \hso {\large +}  \ydiagram{2,1}\hso {\large +}  \ydiagram{1,1,1} \hso {\large  $\mathrm{-}$}    \ydiagram{1,1,1}\hso {\large \hso = \hso 0 }   
 }
 \eeq
\end{center}
Further, we have [e.g. theorem 7.17.1, \cite{stanley}]
\beq
\label{schurpowerprod}
s_\lambda p_n = \sum_\nu (-1)^{\hg(\nu/\lambda)} s_\nu~,
\eeq
where $\nu/\lambda$ is a border strip of size $n$, i.e. a border strip containing $n$ cells. The height $\hg(\nu/\lambda)$ equals the numbers of rows that the border strip occupies minus one. We will also denote these border strips by (e.g.) $\eta$ with $\abs{\eta } = n$ and write $\nu \setminus \eta$ for the partition obtained from $\nu$ after removing the border strip $\eta$. For $\lambda=(3,2)$ and $n=3$, equation \eqref{schurpowerprod} is as follows. 
 \vspace{.2cm}
%
%	\begin{center}
	\beq
	\hbox{
  \begin{ytableau}
 *(white) & *(white) & *(white) & *(gray)  & *(gray)  & *(gray)\\ 
*(white)  & *(white)
\end{ytableau} 
\hspace{.05cm}{\large  $\mathrm{-}$ }    
  \begin{ytableau}
 *(white) & *(white) & *(white) & *(gray)   \\ 
*(white)  & *(white) & *(gray)  & *(gray)
\end{ytableau} 
\hspace{.05cm}{\large   $\mathrm{-}$ }  
  \begin{ytableau}
 *(white) & *(white) & *(white)  \\ 
*(white)  & *(white)  \\
*(gray)  & *(gray)\\
*(gray)  
\end{ytableau} 
\hspace{.05cm} {\large   + }  
  \begin{ytableau}
 *(white) & *(white) & *(white)  \\ 
*(white)  & *(white)  \\
*(gray)  \\
*(gray)  \\
*(gray)
\end{ytableau} 
}
\eeq
%\end{center} 
%\vspace{.2cm}
%
%
Similar to the Pieri formula, equation \eqref{schurpowerprod} can be inverted to give the following expression for skew Schur polynomials,
\beq
\label{pnskew1}
\sum_{r=0}^{n-1}(-1)^r s_{\lambda/(n-r,1^r)}= \sum_\nu (-1)^{\hg(\lambda/\nu)}s_\nu~,
\eeq
where the sum is now over all $\nu$ such that $\lambda/\nu$ is a border strip of size $n$. 
	
Schur polynomials can be expressed in determinantal form. First of all, the (antisymmetric) Vandermonde determinant can be expressed as
	\beq
	\label{vdmdet}
%{	\scriptstyle
	\det\left(x_j^{(N-k)}\right)_{j,k=1}^N = \prod_{1\leq j<k \leq N}(x_j-x_k)  ~.
%	}
	\eeq
	We then have 
	\beq
	\label{schurxi}
 s_\lambda(x_j) = \frac{\det\left(x_j^{N-k+\lambda_k}\right)_{j,k=1}^N}{\det\left(x_j^{N-k}\right)_{j,k=1}^N}~.
	\eeq
	% It was Cauchy who first introduced this expression and demonstrated that it is a symmetric polynomial.
(Skew) Schur polynomials can be expressed in terms of elementary symmetric polynomials or complete homogeneous symmetric polynomials via the following determinantal expressions, known as the Jacobi-Trudi identities,
	\begin{align}
	\label{jtid}
s_{(\mu/\lambda)}  =	& \det(h_{\mu_j -\lambda_k -j+k }  )_{j,k=1}^{\ell(\mu)}  = \det(e_{\mu^t_j -\lambda^t  _k -j+k }  )_{j,k=1}^{\mu_1}~, \notag\\ %  = D^{\lambda,\mu}_N(H(x;z)) ~,\notag\\
	s_{(\mu/\lambda)^t}  = 	& \det(e_{\mu_j -\lambda_k -j+k }  )_{j,k=1}^{\ell(\mu)}  = \det(h_{\mu^t_j -\lambda^t  _k -j+k }  )_{j,k=1}^{\mu_1} ~. % = D^{\lambda,\mu}_N(E(x;z))~.
	\end{align}
	Again, we see that the expressions in terms of $h_j$ and $e_j$ are related by transposition of the skew diagram, $(\mu/\lambda)\to (\mu/\lambda)^t$.
	
	Schur polynomials can also be expanded in terms of power sum polynomials,
\beq
\label{schurprodexp}
s_{\lambda} = \sum_{\alpha} \frac{\chi_{~\alpha}^{\lambda}}{z_\alpha} p_\alpha  ~,
\eeq
where the sum is over all partitions $\alpha$, $z_\alpha$ is defined in \eqref{zmf}, and where $\chi_{~\alpha}^{\lambda}$ is the character of the symmetric group $S_n$ with $  n = \abs{ \lambda} $ of an irreducible representation $\lambda$ associated to a permutation of cycle type $\alpha$, see e.g. \cite{mcd}, \cite{stanley}. In fact, $\alpha$ can generally be a weak composition, and $\chi_{~\alpha}^{\lambda}$ does not depend on the order of the entries of $\alpha$. However, in \eqref{schurprodexp}, we only sum over a single $\alpha$ corresponding to each cycle type, which is equivalent to summing over partitions. Equation \eqref{schurprodexp} generalizes to the case of a skew partition $\lambda/\mu$ instead of $\lambda$. The inverse of \eqref{schurprodexp} is given by
\beq
\label{prodschurexp}
p_\alpha = \sum_{\lambda}\chi_{~\alpha}^{\lambda}s_\lambda ~.
\eeq
It is clear that $p_\alpha $ does not depend on the order of the entries of $\alpha$, it then follows from the above expression that the same is true for $\chi^\lambda_{~\alpha}$. To construct the latter objects, we first define a \emph{border-strip tableau} (BST) of shape $\lambda $ and type $\alpha$  as follows. We take a diagram $\lambda$ and inscribe it with positive integers such that
 %z
 \begin{enumerate}
     \item The integers are weakly increasing along both rows and columns
   %  \item The integer $j$ appears $\mu_j$ times
     \item The cells of $\lambda$ that are inscribed by $j$ form a border strip of size $\alpha_j$
 \end{enumerate}
 The resulting object is called a \emph{border strip tableau}, which we denote as $ T \in \text{BST}(\lambda , \alpha)$. We show an example below for $\lambda = (7,5^2,3,1)$ and $\alpha =(4^2 ,5,3,5)$ (remember that $\alpha $ is a composition and its entries are not generally in non-decreasing order) where cells belonging to a single border strip share the same color.
 	\beq
%	\hbox{ 
 %\begin{center}
     \begin{ytableau}
 *(red) 1 & *(red)  1 & *(red)  1 &  *(orange)  2 & *(orange)  2 & *(orange)  2 & *(orange)  2  \\
 *(red) 1 &  *(yellow)  3 &  *(yellow) 3 &  *(yellow) 3 &   *(blue)   5  \\
 *(yellow) 3 &  *(yellow) 3 &    *(blue) 5 &   *(blue) 5 &    *(blue) 5 \\
 *(purple)  4 & *(purple)  4 &  *(blue)  5 \\
 *(purple) 4
\end{ytableau}
%}
\eeq
% \end{center}
%\vspace{.2cm}

Denoting the border strips of length $\alpha_j$ that appear in $T$ as $B_j$, the height of $T$ is defined as
 \beq
 \label{htbst}
 \hg(T)= \sum_{j=1}^{\ell(\alpha)} \hg( B_j) ~.
 \eeq
 For example, for the above BST for  $\lambda = (7,5^2,3,1)$ and $\alpha =(4^2 ,5,3,5)$, we have
 \beq
 \hg(T) = 1+ 0+1+1+2=5~.
 \eeq
We then have
 \beq
 \label{mnrule}
 \chi^\lambda_{~\alpha}  = \sum_{T \in \text{BST}(\lambda , \alpha)} (-1)^{\hg(T)} ~.
 \eeq
 This is known as the Murnaghan-Nakayama rule, see e.g. \cite{mcd} or \cite{stanley}. The Murnaghan-Nakayama rule generalizes to skew diagrams $\lambda/\mu$. 
 
 Consider a simple example we have encountered before. From \eqref{prodschurexp}, we have
 \beq
 p_n = \sum_{\lambda} \chi^\lambda_{~(n)} s_\lambda  = \sum_{r=0}^{n-1} (-1)^r s_{(n-r,1^r)}~.
 \eeq
 This arises simply from the fact that any Young diagram consisting of a single border strip is a hook shape, as this is the only type of non-skew diagram that has no two by two subdiagram. This gives a sum over all hook shapes containing $n$ cells, $(n-r,1^r)$, where the sign appears from the fact that $\hg((n-r,1^r) ) = r$. We thus see how equation \eqref{powersumschur} arises as a special case of \eqref{prodschurexp}. To calculate $ \chi^\lambda_{~\alpha}$, one can use the following recursive formula [e.g.(2.4.4) \cite{jkbook}]
\beq
\label{chirec}
\chi^\lambda_{~\alpha} = \sum_\rho (-1)^{\hg(\rho)} \chi^{\lambda\setminus \rho}_{~\alpha - \alpha_1}~,
\eeq
where the sum runs over all border strips $\rho$ of $\lambda$ containing $\alpha_1$ cells, $\lambda\setminus \rho$ is the results of removing $\rho $ from $\lambda$, and  $\alpha - \alpha_j  =  (\alpha_1 , \alpha_2,\dots, \alpha_{j-1} ,\alpha_{j+1}, \dots)$. We emphasize again that $\chi^\lambda_{~\alpha}$ does not depend on the order of the entries of $\alpha$. This implies that we can consecutively apply equation \eqref{chirec} by removing border strips of different sizes in different orders, and end up with the same result. This may at first sight be surprising, as removing border strips in different orders generally leads to a different set of diagrams. Let us consider a simple example, where we remove border strips of sizes 1 and 2 from $\lambda = (3,2) = {\tiny \ydiagram{3,2}}~$ in the two\\[1pt] different orders. This is indicated in the figure below, where two diagrams connected by an arrow as
\vspace{-.1cm}
\beq
\begin{tikzpicture}
\begin{scope}%[every node/.style={circle,thick,draw}]
    \node (A) at (0,0) {$\mathlarger{\mathlarger{\lambda}}$};
    \node (B) at (2.2,0) {$\mathlarger{\mathlarger{\mu}}$};
\end{scope}
\begin{scope}[>={Stealth[black]},
              every node/.style={fill=white,circle},
              every edge/.style={draw=black,thick}]
    \path [->] (A) edge node {$p_j$} (B);
\end{scope}
\end{tikzpicture} 
\eeq

\vspace{-.4cm}

indicates that partitions $\lambda$ and $\mu$ are related by the removal of a border strip of size $j$. 

%\vspace{-.5cm}
%
%
%\begin{center}
%\begin{align}
\begin{figure}[ht]
 \centering
 \begin{tikzpicture}
\begin{scope}%[every node/.style={circle,thick,draw}]
    \node (A) at (0,0) {$\ydiagram{3,2}$};
    \node (B) at (4,.83) {$\ydiagram{3,1}$};
    \node (C) at (3.8,-1.15) {$\ydiagram{2,2}$};  
    \node (D) at (7.8,1.5) {$\ydiagram{3}$};
    \node (F) at (11.5,.45) {$\ydiagram{2}$};
    \node (G) at (8.3,-.2) {$\ydiagram{1,1}$};  
   % \node (I) at (13,0) {$\mathlarger{\mathlarger{\mathlarger{\mathlarger{\emptyset}}}}$};      
\end{scope}
\begin{scope}[>={Stealth[black]},
              every node/.style={fill=white,circle},
              every edge/.style={draw=black,very thick}]
    \path [->] (A) edge node {$p_1$} (B);
    \path [->] (A) edge node {$p_1$} (C);
    \path [->] (B) edge node {$p_1$} (D);   
     \path [->] (B) edge%[bend left=20] 
     node {$p_2$} (G);
    \path [->] (D) edge node {$p_1$} (F);      
    \path [->] (A) edge[bend left=30] node {$p_2$} (D);
    \path [->] (C) edge[bend right=30]
    node {$p_2$} (F); 
\end{scope}
\begin{scope}[>={Stealth[black]},
              every node/.style={fill=white,circle},
              every edge/.style={draw=black,ultra thick,dashed}]
        \path [->] (C) edge%[bend right=3]
    node {$p_2$} (G); 
    \end{scope}
\end{tikzpicture}
\caption{The Young diagram for $\lambda = (3,2)$ and the removal of border strips of sizes 1 (single cell) and 2 (domino), indicated by $p_1$ and $p_2$, respectively. The dashed line connecting $(2,2)$ %${\tiny \ydiagram{2,2}}~$
and %${\tiny \ydiagram{1,1}}~$ 
$(1,1)$ indicates the only case where a border strip of odd height is removed. \label{y32p21}}
\end{figure}
It is clear from figure \ref{y32p21} that going first along $p_2$ and then along $p_1$ only leads to the result ${\tiny \ydiagram{2}}~$, where going first along $p_1$ and then $p_2$ leads to $~{\tiny \ydiagram{2}}~$ as well as $~{\tiny \ydiagram{1,1}}~$, the latter in two different ways. As indicated in the figure, the dashed line connecting ${~\tiny \ydiagram{2,2}}~$ and ${~\tiny \ydiagram{1,1}}~$ involves the removal of a border strip of height\\[2pt] equal to 1, whereas other border strips that are removed in figure \ref{y32p21} have height equal to 0. Consider a composition $\alpha$ containing at least one row of length one and length two. Applying \eqref{chirec} then gives 
\begin{align}
\label{chi32xmp}
\chi^{(3,2)}_{~\alpha} = & \sum_{\mu,\rho} (-1)^{\hg(\rho) + \hg(\mu)}  \chi^{((3,2)\setminus\mu)\setminus \rho }_{~\alpha-(2,1)} =  \chi^{(2) }_{~\alpha-(2,1)} + (1-1)  \chi^{(1,1) }_{~\alpha-(2,1)}  \notag \\
= &  \sum_{\mu,\rho} (-1)^{\hg(\rho) + \hg(\mu)}  \chi^{((3,2)\setminus\rho)\setminus \mu }_{~\alpha-(2,1)}  = \chi^{(2) }_{~\alpha-(2,1)}
\end{align}
where $\mu$ and $\rho$ are BS of sizes 1 and 2, respectively, and where $\alpha-(2,1)$ indicates that we remove a row of length 1 and a row of length 2 from $\alpha$. Note the different order of removal of $\mu$ and $\rho$ in the top and bottom rows. We see that the sign given by $(-1)^{\hg(\rho)}$ ensures that removing border strips of different sizes in different orders, as it leads to the cancellation between various ways to arrive at certain diagrams that are unattainable via a different order of removal. We treat the removal of border strips from $\lambda = {\tiny \ydiagram{3,2}}~$ more extensively in section \ref{secschurbs}, see figure \ref{y32graph}.

Let us consider the object $(p_n)^k$. Using \eqref{prodschurexp}, we have
 \beq
 \label{trucexp}
 (p_n)^k = \sum_{\lambda}   \chi^\lambda_{~(n^k)} s_\lambda~.
 \eeq
 From \eqref{mnrule} (or \eqref{schurpowerprod} with $\lambda = \emptyset$), the $\chi^\lambda_{(n^k)}$ appearing in \eqref{trucexp} are of the following form 
 \beq
 \label{clt}
\chi^\lambda_{~(n^k)} = \sum_{T \in \text{BST}(\lambda , (n^k))} (-1)^{\hg(T)} ~,
 \eeq
 where the sum is over all \emph{border strip tableaux} of shape $\lambda$ and type $\alpha = (n^k)$. For $\alpha =(n^k)$, it has been shown that the expansion in \eqref{clt} is cancellation-free [corollary 10, \cite{white}]. That is, for any fixed choice of $\lambda$, all BST's appear with the same sign, so that
  \beq
 \label{clt2}
\chi^\lambda_{~(n^k)}  = \pm  \sum_{T \in \text{BST}(\lambda , (n^k))}  1 ~.
 \eeq
 In fact, there is a more general result [theorem 2.7.27 in \cite{jkbook}], which states that
 \beq
  \label{clt3}
 \chi^{\lambda / \mu }_{~(n^k)} = \pm \sum_{T \in \text{BST}(\lambda/\mu , (n^k))}  1 ~.
 \eeq
 In the above expression, $\mu$ is the so-called $n$-core of $\lambda$, which is the diagram that remains after removing the maximum possible of border strips of size $n$. We denote by $w(\lambda)$ the $n$-weight of $\lambda$, which is the number of border strips of size $n$ which one has to remove to obtain the $n$-core of $\lambda$. The number of partitions with $n$-core given by $\mu$ is then given by [theorem 2.7.17 in \cite{jkbook}]
 \beq
 \label{numbncorpart}
 b(w) = \sum_{\{w_j\}} \prod_{j=1}^n p(w_j) ~,
 \eeq
 where the sum is over all sets of $n$ non-negative integers $w_j$ satisfying $\sum_{j=1}^n w_j = w (\lambda)$, and where $p(w_j$ is the number of partitions of $w_j$. 
 
 We illustrate \eqref{clt3} for $\lambda$ with empty $n$-core (i.e. \eqref{clt2}), in particular, for $\lambda = (6,5,2^2,1) $ and $ \alpha = (n^k ) = (4^4)$. This gives the following border strip tableaux, where border strips again share the same color.
  \vspace{.2cm}
 	\beq
 	\label{y65bs4}
%	\hbox{
% \begin{center}
     \begin{ytableau}
 *(red)  & *(red)   & *(red)  &   *(red)    &  *(blue)  &  *(blue) \\
 *(orange)  &*(orange)  & *(orange)  &  *(blue)   &  *(blue) \\
   *(orange)   & *(yellow)  \\
  *(yellow)  &  *(yellow)  \\
   *(yellow)  
\end{ytableau}
\hso \hso
     \begin{ytableau}
 *(red)  & *(red)  & *(orange)  & *(orange)    &  *(blue)  &  *(blue)  \\
   *(red)    &   *(orange)  &*(orange)  &  *(blue) &  *(blue)  \\
    *(red) & *(yellow)  \\
  *(yellow)  &  *(yellow)  \\
   *(yellow)  
\end{ytableau}
\hso \hso
     \begin{ytableau}
 *(red)  & *(red)  & *(orange)  & *(orange)  & *(orange)  & *(orange)   \\
   *(red)     &  *(blue)  &  *(blue) &  *(blue) &  *(blue)  \\
    *(red) & *(yellow)  \\
  *(yellow)  &  *(yellow)  \\
   *(yellow)  
\end{ytableau}
\hso\hso
     \begin{ytableau}
 *(red)  & *(red)  &  *(red)  & *(red)  &  *(blue) &  *(blue)  \\
     *(orange)  & *(yellow) & *(yellow) &  *(blue) &  *(blue)  \\
     *(orange)   & *(yellow)  \\
  *(orange)    &  *(yellow)  \\
   *(orange) 
\end{ytableau}
\eeq
 \vspace{-.4cm}
% \end{center}
% \vspace{.2cm}

 We see that the heights of the tableaux are given, from left to right, by 4, 6, 4, 6, so that $(-1)^{\hg(T)} =1$.  Note that the two leftmost BST's appear with multiplicity two, as one can remove the yellow and blue border strips in either order. This means that $\chi^{(6,5,2^2,1)}_{~(4^4)} = 6$. We see, then, that all BST's contribute with the same sign as they are all of even height. Equation \eqref{clt2} states that the BST's \emph{always} appear with the same sign for any choice of $\lambda $ and $(n^k)$. One can see that this generalizes to skew partitions $\lambda/\rho$ by considering $\lambda $ and $\rho$ for which $\text{BST} (\lambda,(n^k))$ and $BST (\rho,(n^k))$ are non-empty, and using the fact that $\hg(T_{\lambda / \rho}) = \hg(T_{\lambda }) - \hg(T_{ \rho})$, where $T_{\lambda / \rho} $ is a border strip tableau of type $(n^k)$ for some $k$. 
\section{Correlation functions of long-range random walks}
\label{seccorrev}
We summarize here the derivation which relates correlation functions of one-dimensional LRRW models to weighted integrals over $U(N)$ with insertion of Schur polynomials, following \cite{bogoxx}, \cite{bogpronko}, and \cite{pereztierz}. We first consider the XX0-model, that is, the XX-model with zero magnetic field,
\beq
\label{bogham}
\hat{H}=\sum_{m,n}\Delta_{n,m}\sigma^-_n \sigma^+_m~~,~~~\Delta_{m,n} =\delta_{n+1,m}+\delta_{n-1,m}~.
\eeq
We start with state with holes at all sites
\beq
\ket{\Uparrow} = \otimes_n \ket{\uparrow}_n = \otimes_n \begin{pmatrix}1 \\0 \end{pmatrix}_n~,
\eeq
which satisfies 
\beq
\hat{H}\ket{\Uparrow}=0~.
\eeq
Define the correlation function
\beq
F_{j;l}(\tau ) =\obket{\Uparrow}{\sigma_j^+ e^{-\tau \hat{H}} \sigma_l^-}{\Uparrow}~.
\eeq
For now, we consider $\tau$ to be a general complex number, but we will mostly be interested in the case where $\tau = it$  where $t$ is a real-valued time parameter. Using 
%As long as $\Delta_{n,n}=0$,
\beq
\left[\sigma_n^+ ,\sigma_m^-\right] = \sigma_n^z \delta_{m,n}~~,~~~ \left[\sigma_n^z ,\sigma_m^{\pm}  \right] = \pm 2 \sigma_n^{\pm} \delta_{m,n}~,
\eeq
we have
\beq
\left[ \hat{H} ,\sigma_k^- \right] =\sum_m \Delta_{m,k}\sigma^-_m \sigma^z_k~,
\eeq
which we apply to find
\beq
\frac{d}{d\tau } F_{j;l}(\tau ) = - \obket{\Uparrow}{\sigma_j^+ e^{-\tau \hat{H}} \hat{H} \sigma_l^-}{\Uparrow} = \sum_m \Delta_{l,m} F_{j;m}(\tau )~.
\eeq
In particular, for the XX0-model, where $\Delta_{m,n} =\delta_{n+1,m}+\delta_{n-1,m}$, 
\beq
\frac{d}{d\tau }F_{j;l}(\tau )= F_{j;l+1}(\tau ) +F_{j;l-1}(\tau )~.
\eeq
We will generally consider the case where $\Delta_{m,n}= \Delta_{m-n}=a_{m-n}$, i.e. the hopping amplitude depends only on the (positive or negative) distance between lattice sites $m$ and $n$. In this case, $ \Delta_{m,n}$ is a Toeplitz matrix, i.e. it is constant along its diagonals. Taking $L+1$ lattice sites, as in \cite{bogoxx}, the hamiltonian is generally of the form, % [generalize to $a_{-k} = a_k^\dagger$, add magnetic field],
\beq
\label{tiham}
\hat{H}=- \sum_{m=0}^{L}\sum_{n=1}^{(L-1)/2} \left( a_n %\left(
\sigma^-_m \sigma_{m+n}^+ + a_{-n}  %\left(
\sigma^-_m \sigma_{m-n}^+ \right) +\frac{h}{2} \sum_{m=0}^L \left(\sigma_m^z - \mathds{1}\right)  ~, % \sigma_m^-\sigma_{m-n}^+  \right)~.
\eeq
where we demand that $a_{-k } = a_k^*$, the complex conjugate of $a_k$. The hopping parameters may acquire a non-zero imaginary component e.g. as the result of gauge flux attachment. We then have
\beq
\left[ \hat{H} ,\sigma_k^- \right] = - \sum_n \left( a_{n } \sigma^-_{k-n}\sigma_n^z + a_{-n}\sigma^-_{k+n}\sigma^z_k \right) - h\sigma_k^-   ~,
\eeq
so that
\beq
\frac{d}{d\tau} F_{j;l}(\tau) =\sum_n \left(  a_n F_{j-n;l} +  a_{-n} F_{j+n;l}\right) + h F_{j;l} ~.
\eeq
It is clear that $F_{j;l}$ only depends on $\abs{j-l}$. The generating function $f(z ;\tau) = \sum_{j\in \mathds{Z}} F_{j;l}z^{j-l}$ is given by \cite{bogoxx}, \cite{pereztierz}
\beq
\label{spinwfct}
f(z ;\tau) =  \exp\left(\tau\sum_{k\in\mathds{Z}} a_kz^k \right)~,
\eeq
where we set $a_0 = h$. Consider now the multi-particle correlation function, with $N\leq L$,
\beq
\label{multicorr}
F_{j_1,\dots,j_N;l_1,\dots,l_N}(\tau ) = \obket{\Uparrow}{\sigma_{j_1}^+ \dots \sigma_{j_N}^+ e^{-\tau \hat{H}} \sigma_{l_1}^-  \dots  \sigma_{l_N}^- }{\Uparrow}~.
\eeq
We impose periodic boundary conditions for simplicity. However, we will be taking the thermodynamic limit and considering only configurations which contain an infinite sequence of adjacent holes, with the particles either confined to a finite interval in their initial configuration or starting with an infinite sequence of adjacent particles. As such, imposing periodic boundary conditions will have no effect on our final result. We then have,
\begin{align}
\label{fcorrdif}
\frac{d}{d \tau } F_{j_1,\dots,j_N;l_1,\dots,l_N}(\tau ) = & \sum_{k,m}  \left(  a_{k} F_{j_1,\dots ,j_N;l_1,\dots , l_m+k ,\dots,l_N}(\tau ) + a_{k} F_{j_1,\dots   j_N;l_1,\dots, l_m-k  ,\dots,l_N}(\tau )  \right) + \notag \\
& + Nh F_{j_1,\dots,j_N;l_1,\dots,l_N}(\tau )~.
\end{align}
Remember that the summations over $k$ and $m$ are over different ranges. The solution to equation \eqref{fcorrdif} is of determinantal form \cite{bogoxx}, \cite{bogpronko}, \cite{pereztierz},
\beq
 F_{j_1,\dots,j_N;l_1,\dots,l_N}(\tau) = \det \left( F_{j_r;l_s}(\tau)\right)_{r,s=1}^N~.
\eeq
From the initial condition $F_{j;l}(0)=\delta_{j,l}$, it follows that, 
\beq
F_{j;l}(\tau) = \frac{1}{L+1} \sum_{s=0}^L \exp\left(\tau % \left( 
\sum_{k}^{(L-1)/2} a_k e^{ik\theta_s} %+ a_{-k} e^{-ik\theta_s}  \right) 
\right) e^{i(j-l)\theta_s}~~, ~~~ \theta_s =\frac{2\pi}{L+1}\left( s-L/2 \right) ~.
\eeq
Using the determinantal expression for Schur functions in equation \eqref{schurxi}, we have
\begin{align}
\label{corrdet}
F_{j_1,\dots,j_N;l_1,\dots,l_N}(\tau) = &  \frac{1}{(L+1)^N} \sum_{\{ s_j \} } \exp\left( \tau \sum_{j=1}^N %\left( 
\sum_{k}^{(L-1)/2} a_k e^{ik\theta_{s_j}} %+ a_{-k} e^{-ik\theta_{s_j}}  \right)  
\right) \prod_{1\leq j < k \leq N}  \abs{e^{i\theta_{s_j}}-e^{i\theta_{s_k}}}^2 ~  \times \notag\\
&  ~~~~~~ \times 
 s_\lambda(e^{i\theta_{s_1}},\dots,e^{i\theta_{s_N}}) s_\mu (e^{-i\theta_{s_1}},\dots,e^{-i\theta_{s_N}}) ~,
\end{align}
with $ \lambda_r = j_r-N+r$ and $ \mu_s =l_s-N +s$. We shift $j_r$ and $l_r$ by $N$, such that
\beq
\label{lmjr}
\lambda_r = j_r+r~~,~~~\mu_s =l_s +s~.
\eeq
This is merely a convenient relabelling of our lattice sites. We now take $L\to \infty$. Demanding that the hopping parameters $a_k$ decay at least as $a_k \sim k^{-1-\epsilon}$ for some $\epsilon > 0 $, we then have \cite{bogoxx}, \cite{bogpronko}, \cite{pereztierz},
\begin{align}
\label{mpcorr}
F_{j_1,\dots,j_N;l_1,\dots,l_N}(\tau) = & ~ \mathcal{N} e^{ Nh \tau} \int_{-\pi}^\pi d\theta_1 \dots \int_{-\pi}^\pi d\theta_N   \prod_{1\leq r < s\leq N} \abs{e^{i\theta_r}-e^{i\theta_s}}^2 \prod_{k=1}f(e^{i\theta_k} ;\tau) \times \notag\\
 & \times s_\lambda(e^{i\theta_1},\dots,e^{i\theta_N}) s_\mu (e^{-i\theta_1},\dots,e^{-i\theta_N}) ~,
\end{align}
where the weight function $f(z ;\tau) $ is given in \eqref{spinwfct}. We include in equation \eqref{mpcorr} a normalization factor $\mathcal{N}$ which is determined by demanding 
\beq
F_{j_1,\dots,j_N;l_1,\dots,l_N}(0) = \prod_{k=1}^M \delta_{j_k , l_k}~.
\eeq
Note that \eqref{mpcorr} is the expression for an integral over $U(N)$ weighted by some weight function $f$ with insertion of Schur polynomials $s_\lambda$ and $s_\mu$. In particular, writing $s_\lambda(U) =  s_\lambda(e^{i\theta_j})$, were $e^{i\theta_j}$ are the eigenvalues of $U$, we have
\begin{align}
\label{corrmat}
F_{j_1,\dots,j_N;l_1,\dots,l_N}(\tau) = \int_{U(N)} dU \det (f(U)) s_\lambda(U) s_{\mu}(U^{-1}) ~.
\end{align}
When we take the limit $\tau \to 0$, we have $f=1$, which recovers the circular unitary ensemble (CUE) i.e. the integral over the Haar measure on $U(N)$. In this case $s_{\lambda/\mu}(x)=0$ for any $\lambda/\mu \neq \emptyset$, which greatly simplifies many calculations. If $\lambda = \emptyset  = \mu$, we have $j_r=-r=l_r$, and we are simply considering the return probability for $N$ adjacent particles \cite{vitiq}, \cite{wei}, see also \cite{klm1}. In general, we will write 
\beq
F_{\lambda;\mu} (\tau)= F_{j_1,\dots,j_N;l_1,\dots,l_N}(\tau)~,
\eeq 
where equation \eqref{lmjr} expresses the relation between $\lambda$, $\mu$ and $\{ j_r \}$, $\{ l_r \}$, respectively. We will write the state corresponding to an empty diagram as $\ket{\emptyset}$. This can be illustrated as follows, where a particle is indicated by a black dot and hole by a white dot and the vertical line separates lattice sites 0 and 1.
\beq
\ket{\emptyset}=\cdots
\vcenterbox{\unitlength=\cellsize\begin{picture}(10,2)
{\dottedline{0.1}(5,0)(5,2)}
\linethickness{0.7pt}
\put(0,1){\line(1,0){10}}
\multiput(0.5,1)(1,0){5}{\circle*{0.3}}
{\color{white}\multiput(5.5,1)(1,0){5}{\circle*{0.3}}}
\multiput(5.5,1)(1,0){5}{\circle{0.3}}
\put(5,0){\dashbox{0.04}(0,2){}}
\end{picture}}
\cdots 
\eeq
For non-empty $\lambda, \mu$, the object $F_{\lambda;\mu}$ corresponds to a correlation function where certain particles are shifted by a finite number of sites. The well-known association between Young diagrams and 1D configurations of spins (or fermions, or any other binary variable) is as follows. We place a diagram in the corner where infinitely long horizontal and vertical lines meet. We number the edges of these horizontal and vertical lines, as well as external edges (i.e. those on the lower right) of the diagram, such that the main diagonal passes between sites 0 and 1. We then associate a particle to all vertical edges and a hole to all horizontal edges. This association is illustrated\footnote{We made grateful use of the illustrations in the excellent review \cite{zinnjustin6v} by Zinn-Justin, which we adapt here for our purposes.} below for $\lambda=(5,3,1^2)$, where we add a dotted diagonal line which separates lattice sites 0 and 1.
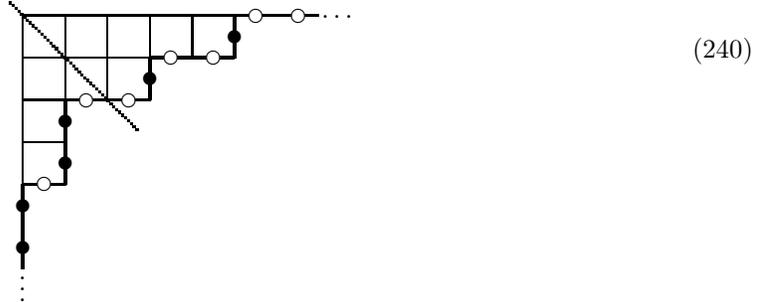
\beq
\hspace{-2.5cm}\rlap{\tableau{&&&&&\emptycell \!\!\!&\emptycell&\hdotscell\\&&&\emptycell\!\!\!\\&\emptycell\\&\emptycell\\\emptycell\\\emptycell \\\vdotscell%
}}
\unitlength=\cellsize
\begin{picture}(0,0)
\thicklines
{\dottedline{0.1}(-.3,1.3)(2.7,-1.7)}
\drawline(0,-5)(0,-3)(1,-3)(1,-1)(3,-1)(3,0)(5,0)(5,1)(7,1)
\thinlines
\put(0,-4.5){\circle*{0.3}}
\put(0,-3.5){\circle*{0.3}}
\put(1,-2.5){\circle*{0.3}}
\put(1,-1.5){\circle*{0.3}}
\put(3,-0.5){\circle*{0.3}}
\put(5,+0.5){\circle*{0.3}}
{\color{white}
\put(6.5,1){\circle*{0.3}}
\put(5.5,1){\circle*{0.3}}
\put(4.5,0){\circle*{0.3}}
\put(3.5,0){\circle*{0.3}}
\put(2.5,-1){\circle*{0.3}}
\put(1.5,-1){\circle*{0.3}}
\put(0.5,-3){\circle*{0.3}}
}
\put(6.5,1){\circle{0.3}}
\put(5.5,1){\circle{0.3}}
\put(4.5,0){\circle{0.3}}
\put(3.5,0){\circle{0.3}}
\put(2.5,-1){\circle{0.3}}
\put(1.5,-1){\circle{0.3}}
\put(0.5,-3){\circle{0.3}}
\end{picture}
\eeq
The configuration corresponding to the above diagram is illustrated below, with again a dotted line separating sites 0 and 1.
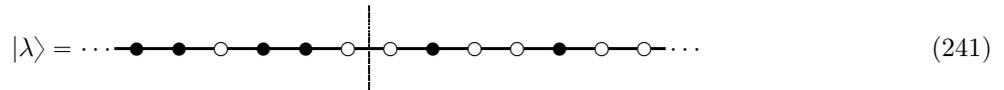
\beq
\label{diagramspins}
\ket{\lambda}=\cdots
\vcenterbox{\unitlength=\cellsize\begin{picture}(13,2)
{\color{gray}\dottedline{0.1}(6,0)(6,2)}%
\linethickness{0.7pt}
\put(0,1){\line(1,0){13}}
\put(.5,1){\circle*{0.3}}
\put(1.5,1){\circle*{0.3}}
\put(3.5,1){\circle*{0.3}}
\put(4.5,1){\circle*{0.3}}
\put(7.5,1){\circle*{0.3}}
\put(10.5,1){\circle*{0.3}}
{\color{white}
\put(2.5,1){\circle*{0.3}}
\put(5.5,1){\circle*{0.3}}
\put(6.5,1){\circle*{0.3}}
\put(8.5,1){\circle*{0.3}}
\put(9.5,1){\circle*{0.3}}
\put(11.5,1){\circle*{0.3}}
\put(12.5,1){\circle*{0.3}}
}
\put(2.5,1){\circle{0.3}}
\put(5.5,1){\circle{0.3}}
\put(6.5,1){\circle{0.3}}
\put(8.5,1){\circle{0.3}}
\put(9.5,1){\circle{0.3}}
\put(11.5,1){\circle{0.3}}
\put(12.5,1){\circle{0.3}}
\put(6,0){\dashbox{0.04}(0,2){}}
%\put(4.87,-0.5){\tiny }
\end{picture}}
\cdots 
\eeq
It is clear from the above association between diagrams and particle-hole configurations that $\lambda$ affects particles from position $j=-\ell(\lambda)+1$ up to $j=\lambda_1$, which means that there is an interval containing $\ell(\lambda)$ particles and $\lambda_1$ holes, the leftmost of which is a hole and the rightmost a particle. In particular, the state $\ket{\lambda}$ has a hole at $j=-\ell(\lambda)+1$ and a particle at $j=\lambda_1$, and the remaining $\ell(\lambda)-1$ particles and $\lambda_1-1$ holes are are distributed over sites $j=\{-\ell(\lambda)+2, -\ell(\lambda)+3,\dots ,\lambda_1-1\}$, which can be seen explicitly for $\lambda = (5,3,1^2)$ above.
\section{Evaluating unitary matrix integrals}
\label{secmatrint}
	We briefly review the evaluation of weighted unitary integrals over Schur polynomials. We start from an absolutely integrable function on the unit circle in $\mathds{C}$,
	\begin{equation}
	\label{fun}
	f(e^{i\theta}) = \sum_{k\in\mathds{Z}} d_k e^{ik\theta}~.
	\end{equation}
We require that $f(e^{i\theta})$ satisfies the assumptions of Szeg\H{o}'s theorem. That is, we write $f(e^{i\theta})$ as
	\beq
	f(e^{i\theta}) = \exp\left(\sum_{k\in\mathds{Z}}c_k e^{ik\theta}\right)~,
	\eeq
	and demand that
	\beq
	\label{szegreq}
	\sum_{k\in \mathds{Z}} \abs{c_k} <\infty ~~,~~~ \sum_{k \in \mathds{Z} } \abs{k}\abs{c_k}^2 < \infty ~.
	\eeq
Writing%	
	\beq
	D_N(f) = \det(T_N(f))= \det \left( d_{j-k}\right)_{j,k=1}^N~,
	\eeq
	the strong Szeg\H{o} limit theorem then states that \cite{szego}, 
	\beq
	\label{szegthm}
	\lim_{N\to \infty}\frac{ D_N(f)}{e^{-Nc_0}} =  \exp\left(\sum_{k=1}^\infty kc_kc_{-k} \right)~.
	\eeq
	The above determinant and various generalization thereof can be related to integrals over the group of $N$ by $N$ unitary matrices with some weight function $f$ with the insertion of Schur polynomials. Take the following matrix,
		\beq
		T^{\lambda,\mu}_N(f) = \left(d_{j-\lambda_j-k+\mu_k}\right)_{j,k=1}^N~.
		\eeq
Then \cite{bd}, \cite{adlervmbk},
	\begin{align}
	\label{toepdn}
	D_N^{\lambda,\mu} (f) 	& \coloneqq \det T_N^{\lambda,\mu}(f)  = \int_{U(N)}s_\lambda(U^{-1]}) s_\mu(U) 	\det f(U) dU \notag \\
	& = \frac{ 1}{N! (2\pi)^N} 	\int_0^{2\pi} s_\lambda(e^{-i\theta_1} ,\dots, e^{-i\theta_M})  s_\mu(e^{i\theta_1} ,\dots, e^{i\theta_M}) \prod_{j=1}^M f(e^{i\theta_j}) \prod_{1\leq j<k\leq N} \abs{e^{i\theta_j} - e^{i\theta_k}}^2 d\theta_j\notag\\
	& = \det\left(d_{j-\lambda_j-k+\mu_k}\right)_{j,k=1}^N~.
	\end{align}
	From the above expression, one can see that $D^{\lambda,\mu}_N(f)$ can be expressed as a minor of $T_{N+k}(f)$, where $k= \max \{ \lambda_1,\mu_1\}$. We remind the reader that a minor of some matrix $M$ is the determinant of a matrix obtained from $M$ by removing some of its rows and columns. In our case, the rows and columns which are removed are specified by $\lambda$ and $\mu$. We write the expectation value $\avg{\dots}$ with respect to the matrix model with weight function $f$ as
	\beq
	\label{schuravg}
	\avg{s_\lambda (U) s_\mu (U^{-1}) } = \frac{	\int s_\lambda(U)s_\mu(U^{-1})\det f(U)dU}{ \int \det f(U)dU} ~.
	\eeq
	We will often neglect to write $U^{\pm1 }$ explicitly, instead writing $\avg{s_\lambda s_\mu}$ or, more generally, $\avg{AB}$ for symmetric polynomials $A$ and $B$. For two functions of the form
	\beq
	a(z)=\sum_{k\leq 0}a_k z^k~~,~~~ 	b(z)=\sum_{k\geq 0}b_k z^k ~,
	\eeq
	the associated Toeplitz matrix satisfies
	\beq
	T(ab) = T(a) T(b)~.
	\eeq
	Let us therefore write $f(e^{i\theta})$ as follows 
	\beq
	\label{fhh}
	f(z)=H(x;z)H(y;z^{-1})~,
	\eeq
	where $H(x;z)$ is the generating function of the homogeneous symmetric polynomials $h_k$ given in \eqref{genfuncsymp} and where we assume that $h_k(x)$ and $h_k(y)$ are square-summable, i.e. $\sum_k \abs{h_k}^2 < \infty$. We can also define our weight function as 
		\beq
	\label{fee}
	f(z)=E(x;z)E(y;z^{-1})~,
	\eeq
	at the cost replacing $h_j$ by $e_j$ everywhere, which is equivalent to transposing all diagrams. We repeat that the CUE corresponds to $f=1$, which corresponds to $x_j=0$ for all $j$ for both $f(z) = H(x;z)H(y;z^{-1})$ and $f(z) = E(x;z)E(y;z^{-1})$. We will consider the case where $f(z) = H(x;z)H(y;z^{-1})$ unless stated otherwise.

	It was shown by Gessel \cite{gessel} that $D_N(f)$ can be expressed in terms of Schur polynomials as
	\beq
	\label{gesselid}
	D_N(f) = \sum_{\ell(\nu)\leq N} s_\nu (x) s_\nu(y)~.
	\eeq
%	. Here, we only consider the case where $y=x \in \mathds{R}$, but the expressions here easily generalize to $x \neq y$ and $x,y \in \mathds{C}$, subject to the assumptions of Szeg\H{o}'s theorem.
Equation \eqref{toepdn} can be similarly expressed in terms of (skew) Schur polynomials \cite{bd}, \cite{GGT1}, \cite{GGT2},
	\beq
	\label{thm5}
	\int s_\lambda(U)s_\mu(U^{-1})\det f(U)dU= \sum_{\ell(\rho)\leq N}s_{\rho/\lambda}(x)s_{\rho/\mu}(y)~.
	\eeq
 We now take the limit $N \to \infty$. From \eqref{thm5} and the fact that [Chapter I.5, example 26 in
	
\cite{mcd}]
	\beq
	\label{mcdschurs}
	\sum_\rho s_{\rho/\lambda}(x) s_{\rho/\mu}(y)=\sum_\nu s_{\mu/\nu}(x) s_{\lambda/\nu}(y) \sum_\eta s_\eta(x) s_\eta(y) ~,
	\eeq
	where the sums run over all partitions, it follows that  \cite{GGT1}, \cite{GGT2} 
	\beq
	\label{largen}
 	\avg{s_\lambda (U) s_\mu (U^{-1}) }= \sum_\nu s_{\lambda/\nu}(y) s_{\mu/\nu}(x)~,
	\eeq
	where the sum runs over all $\nu$ such that $\nu \subseteq \lambda ,\mu  $. As noted at equation \eqref{fee}, we can also define $f(z)= E(x;z)E(y;z^{-1})$ at the cost of transposing all diagrams. In this case, one has 
	\beq
	\label{largen2}
 	\avg{s_\lambda (U) s_\mu (U^{-1}) }= \sum_\nu s_{(\lambda/\nu)^t}(y) s_{(\mu/\nu)^t}(x)~,
	\eeq
	We will consider $f(z) = H(x;z) H(y;z^{-1})$ and apply \eqref{largen} unless stated otherwise. If we take $\mu= \emptyset$ in \eqref{largen}, the only choice for $\nu$ that contributes to the above sum is $\nu=\emptyset$ as well, and we have
	\beq
	\label{singleschur}
	\avg{s_\lambda (U)} = s_\lambda(y)~~,~~~	\avg{s_\mu (U^{-1})} = s_\mu(x)~.
	\eeq
 We also define the connected expectation value ,
		\beq
		\avg{s_\lambda(U) s_\mu(U^{-1})}_c \coloneqq 	\avg{s_\lambda(U) s_\mu(U^{-1})}- 	\avg{s_\lambda(U)} \avg{ s_\mu(U^{-1})}~,
		\eeq
		which corresponds to subtracting the contribution corresponding to $\nu=\emptyset $ in \eqref{largen}. 
\end{appendices}
	\bibliographystyle{unsrt}
	\bibliography{cit}

\end{document}